\documentclass[11pt,a4paper]{article}
\usepackage{jcappub}
\usepackage[normalem]{ulem}
\usepackage{float}
\usepackage{epsfig}
\usepackage{epstopdf}
\usepackage{amsmath,dsfont}
\usepackage{hyperref}
\hypersetup{colorlinks=True,citecolor=blue}
\usepackage{amsfonts}
\usepackage{amssymb}
\usepackage{caption}
\usepackage{bm}
\usepackage{graphicx}
\usepackage{color}
\usepackage{verbatim}
\usepackage{subfigure}
\usepackage{blindtext, multicol}
\usepackage{setspace}
\usepackage{multirow}
 
\usepackage{xcolor}
\usepackage{hyperref}
\usepackage[capitalise]{cleveref}
\usepackage[activate={true,nocompatibility},final,kerning=true,factor=1100,stretch=10,shrink=10]{microtype}
\usepackage{academicons}
\usepackage{xcolor}[dvipsnames]
\usepackage{fontawesome5}

\definecolor{orcidlogocol}{rgb}{0.65, 0.807, 0.223}

\newcommand{\orcid}[1]{$\,$\href{https://orcid.org/#1}{\textcolor{orcidlogocol}{\faOrcid}}}

\usepackage[T1]{fontenc} 

\newcommand{\ber}{\begin{eqnarray}}
\newcommand{\eer}{\end{eqnarray}}

\def\beq{\begin{equation}}
\def\eeq{\end{equation}}
\def\ber{\begin{eqnarray}}
\def\eer{\end{eqnarray}}
\def\benu{\begin{enumerate}}
\def\eenu{\end{enumerate}}

\def\ie{{\em i.e.}}

\def\l{\left}
\def\r{\right}

\def\f{\frac}
\def\mpl{m_{\rm p}}

\def\sq{\lower.25ex\hbox{\large$\Box$}}
\def \lleq {\lower0.9ex\hbox{ $\buildrel < \over \sim$} ~}
\def \ggeq {\lower0.9ex\hbox{ $\buildrel > \over \sim$} ~}

\def\om0{\Omega_{0m}}

\title{Primordial black holes and stochastic inflation beyond slow roll: I - noise matrix elements}

\author[a]{Swagat S. Mishra \orcid{0000-0003-4057-145X},}
\author[a]{Edmund J. Copeland \orcid{0000-0003-3959-6051},}
\author[a]{and Anne M. Green \orcid{0000-0002-7135-1671}}

\affiliation[a]{School of Physics and Astronomy,  University of Nottingham, Nottingham, NG7 2RD, UK.}

\emailAdd{swagat.mishra@nottingham.ac.uk}
\emailAdd{edmund.copeland@nottingham.ac.uk}
\emailAdd{anne.green@nottingham.ac.uk}

\abstract{Primordial Black Holes (PBHs) may form in the early Universe, from the gravitational collapse of large density perturbations, generated by large quantum fluctuations during inflation. Since PBHs form from rare over-densities, their abundance is sensitive to the tail of the primordial probability distribution function (PDF) of the perturbations. It is therefore important to calculate the full PDF of the perturbations, which can be done non-perturbatively using the `stochastic inflation' framework. In single field inflation models generating large enough perturbations to produce an interesting abundance of PBHs requires
violation of slow roll. It is therefore necessary to extend the stochastic inflation formalism beyond slow roll. A crucial ingredient
for this are the stochastic noise matrix elements of the inflaton potential. We carry out 
analytical and numerical calculations of these matrix elements for a potential with a feature which violates slow roll  and produces large, potentially PBH generating, perturbations.   We find that the transition to an ultra slow-roll phase  results in the  momentum induced noise terms becoming larger than the field noise whilst each of them falls exponentially for a few e-folds. The noise terms then start rising with their original order restored, before approaching  constant values   which depend on the nature of the slow roll  parameters in the post transition epoch.
This will significantly impact the quantum diffusion of the coarse-grained inflaton field, and hence the PDF of the perturbations and the PBH mass fraction. } 
\keywords{inflation, primordial black holes,  early universe, dark matter}

\begin{document}
\maketitle

\section{Introduction}
\label{sec:Intro}

   A multitude of cosmological and astrophysical observations indicate that most of the matter in the Universe is non-baryonic cold dark matter~\cite{Bertone:2016nfn,Peebles:2017bzw,Green:2021jrr}. Primordial black holes (PBHs), black holes that might have been formed from over-densities in the early Universe \cite{Zeldovich:1967lct,Hawking:1971ei,Carr:1974nx,Carr:1975qj,Novikov:1979aa,Sasaki:2018dmp}, are a potential dark matter candidate \cite{Hawking:1971ei,Chapline:1975ojl,Ivanov:1994pa,Meszaros:1975ef,Carr:2016drx,Green:2020jor,Carr:2020xqk}. There has been a surge of interest in PBHs in recent years, both their formation and observational probes, following the  detection of gravitational waves from binary black hole mergers \cite{LIGOScientific:2016aoc,Bird:2016dcv,Clesse:2016vqa,Sasaki:2016jop}.

  `Cosmic inflation' (a period of accelerated expansion) has emerged as the leading scenario for the very early Universe, prior to the commencement of the radiation-dominated hot Big Bang \cite{Starobinsky:1980te,Guth:1980zm,Linde:1981mu,Albrecht:1982wi,Linde:1983gd,Baumann:2009ds}. A period of at least $60$-$70$ e-folds of inflation generates natural initial conditions \cite{Mukhanov:1981xt,Guth:1982ec,Starobinsky:1982ee,Hawking:1982cz,Mukhanov:1990me,Baumann:2018muz}. Furthermore quantum fluctuations of the inflaton field can generate the density perturbations from which structure forms.  Observations of the anisotropies in the Cosmic Microwave Background (CMB) radiation \cite{Planck:2018nkj,Planck:2018vyg} provide strong evidence that structure formation on cosmological scales is seeded by almost scale-invariant, nearly Gaussian, adiabatic initial density fluctuations, consistent with the predictions of the simplest single field slow-roll inflation scenario \cite{Tegmark:2004qd,Baumann:2009ds}.

\begin{figure}[hbt]
\begin{center}
\includegraphics[width=0.8\textwidth]{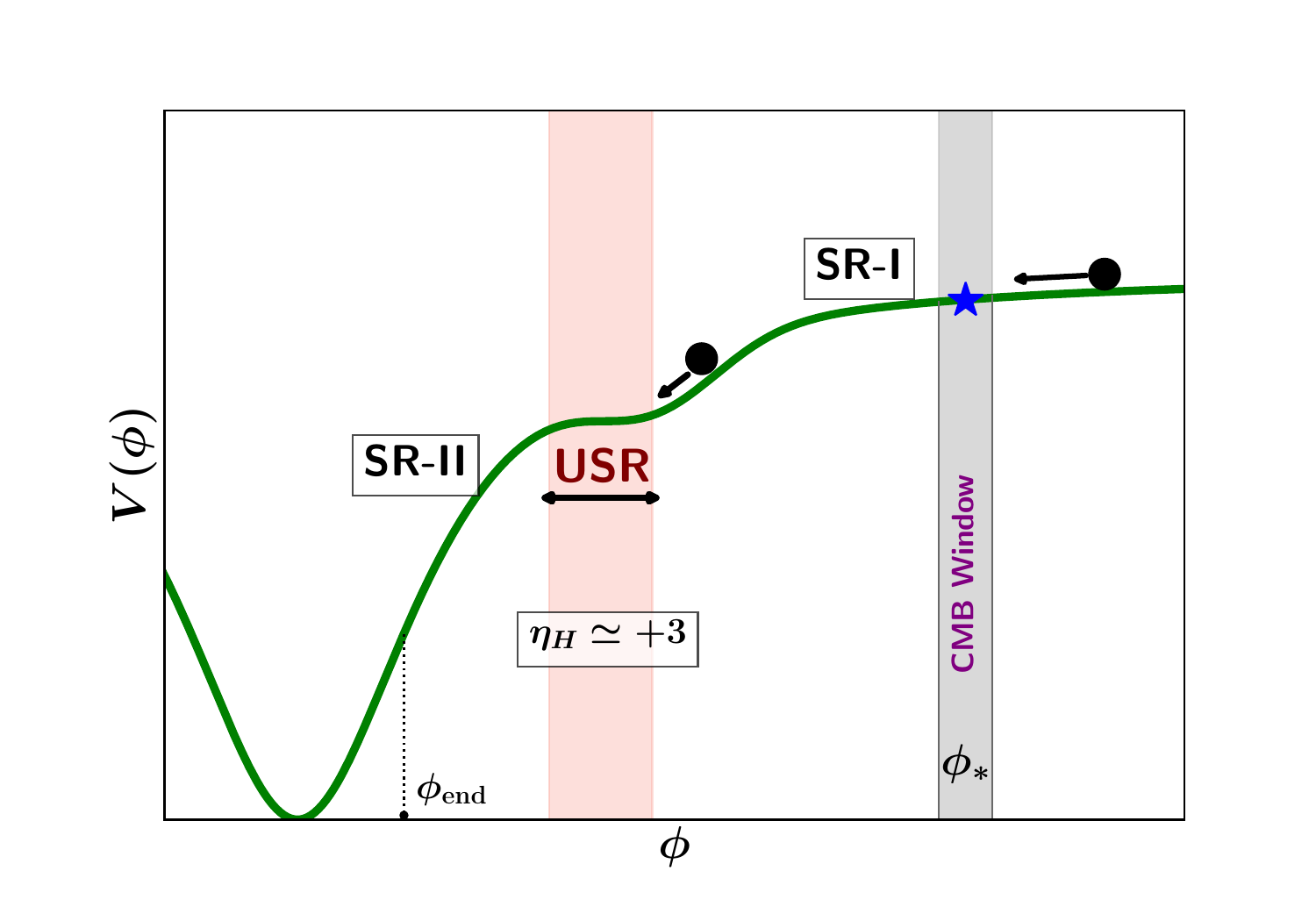}
\caption{A schematic illustration of a plateau potential (solid green line). The `CMB Window' represents field values corresponding to cosmological scales  $k\in \l[0.0005,0.5\r]~{\rm Mpc}^{-1}$ that are probed by CMB observations. The blue star represents the CMB pivot scale $k_* = 0.05 \, {\rm Mpc}^{-1}$. The potential has a flat inflection-point like segment (highlighted with pink shading) which results in ultra slow-roll (USR) inflation. After the first slow-roll phase (SR-I) near the CMB Window, the inflaton enters into an USR phase. During this transient phase of USR, the second slow-roll condition (see Eq.~(\ref{eq:eta_H})) is violated, specifically $\eta_H\simeq +3$. This leads to an enhancement in the primordial perturbations on small scales. Later, the inflaton emerges from the USR into another slow-roll phase (SR-II) before inflation ends at $\phi_{\rm end}$.}
\label{fig:inf_pot_toy_USR_feature}
\end{center}
\end{figure}

CMB observations \cite{Planck:2018jri,BICEP:2021xfz} are consistent with the inflaton field, $\phi$, rolling slowly down an asymptotically flat potential $V(\phi)$ during the epoch when cosmological scales exit the Hubble radius, $50$-$60$ e-foldings before the end of inflation. However the scales probed by CMB and large scale structure (LSS) observations correspond to only $7$-$8$ e-folds of inflation, and hence a relatively small region of the inflaton potential.
On smaller scales, deviations from slow roll may lead to interesting changes in the primordial perturbations. In  particular, if the scalar perturbations are sufficiently large on small scales, then PBHs may form when these modes reenter the Hubble radius during 
the post-inflationary epoch.  PBHs are therefore a powerful probe of the inflaton potential over the full range of field values. Large, PBH forming, fluctuations can be generated by a feature in the inflationary potential, such as a flat inflection point  (see Fig.~\ref{fig:inf_pot_toy_USR_feature}). Such a feature can substantially slow down the already slowly rolling inflaton field, causing the inflaton to enter into an ultra slow-roll (USR) phase, which leads to an enhancement of the power spectrum, ${\cal P}_{\zeta}$, of the primordial curvature perturbation, $\zeta$.

There are several subtleties in calculating the abundance of PBHs formed from inflation models with a feature in the potential. 
Firstly, the sharp drop in the classical drift speed of the inflaton means that the effects of stochastic quantum diffusion on its motion become non-negligible, and potentially even dominant. Even more importantly, since PBHs form from the rare extreme peaks of curvature fluctuations\footnote{Note that while we will be working with the PDF of curvature fluctuations  $P[\zeta]$, the criterion for PBH formation is best specified in terms of the density contrast, $\delta$. Hence, an accurate computation  of the PBH abundance requires the multivariate joint probabilities of the curvature perturbations  \cite{Biagetti:2021eep,DeLuca:2022rfz,Ferrante:2022mui,Gow:2022jfb,Escriva:2022duf}.} their mass fraction at formation, $\beta_{\rm PBH}$, is sensitive  to the tail of the  probability distribution function (PDF) $P[\zeta]$ of the primordial fluctuations. Consequently,  perturbative computations using the  power-spectrum  may  lead to an inaccurate estimation of the PBH mass fraction.  Hence the calculation of the full primordial  PDF, which can be done non-perturbatively using the `\textit{Stochastic Inflation}' (SI) framework   \cite{Starobinsky:1982ee,Starobinsky:1986fx,Salopek:1990re,Salopek:1990jq,Starobinsky:1994bd,Fujita:2013cna,Fujita:2014tja,Vennin:2015hra}, is extremely important (see  Refs.~\cite{Celoria:2021vjw,Cohen:2022clv,Hooshangi:2021ubn,Cai:2021zsp,Ezquiaga:2022qpw,Cai:2022erk,Ferrante:2022mui,Gow:2022jfb,Achucarro:2021pdh}).

There has been a profusion of work on SI in the recent literature in the context of PBH formation \cite{Pattison:2017mbe,Ezquiaga:2018gbw,Firouzjahi:2018vet,Biagetti:2018pjj,Ezquiaga:2019ftu,Pattison:2019hef,Vennin:2020kng,Ando:2020fjm,Ballesteros:2020sre,De:2020hdo,Figueroa:2020jkf,Cruces:2021iwq,Rigopoulos:2021nhv,Pattison:2021oen,Tomberg:2021xxv,Figueroa:2021zah,Tada:2021zzj,Mahbub:2022osb,Jackson:2022unc}, and it has been shown that SI generically predicts a (highly non-Gaussian) exponential tail  in the PDF \cite{Pattison:2017mbe,Ezquiaga:2019ftu}. However, since PBH formation usually requires slow-roll violation, it is important to develop the SI formalism beyond slow-roll \cite{Ezquiaga:2018gbw,Pattison:2019hef,Firouzjahi:2018vet,Ballesteros:2020sre,Figueroa:2020jkf}. In this context,  the validity of various assumptions of the SI formalism have been scrutinised \cite{Pattison:2019hef,Cruces:2021iwq,Rigopoulos:2021nhv}.

  The stochastic inflation formalism is an effective treatment of  the dynamics of the  long-wavelength (IR)  part of the inflaton field  coarse-grained on scales much greater than the  Hubble radius \textit{i.e.}~$k \leq \sigma \, aH$, with  the constant $\sigma \ll 1$.  In this framework, the evolution of the  coarse-grained inflaton field is governed by two first-order  non-linear classical   stochastic differential equations (Langevin equations) which receive constant quantum kicks from  the small scale UV modes that are exiting the Hubble radius due to the accelerated expansion during inflation. Hence the small-scale fluctuations constitute  classical stochastic noise terms in the Langevin equations denoted by $\Sigma_{\phi\phi}$, $\Sigma_{\pi\pi}$, and $\Sigma_{\phi\pi}$ corresponding to the inflaton field noise, momentum noise, and the cross-noise terms (defined in Eq.~(\ref{eq:noise_comp_cor_matrix})). The SI formalism is generally combined  with  the classical $\delta N$ formalism \cite{Starobinsky:1982ee,Starobinsky:1985ibc,Sasaki:1995aw,Lyth:2004gb,Wands:2000dp,Lyth:2005fi}  in order to compute cosmological correlators in this framework.  This leads to the emergence of  the  stochastic $\delta {\cal N}$ formalism\footnote{Note that $N \propto {\rm ln}(a)$, the number of e-folds during inflation, is a deterministic variable, while ${\cal N}$ is a stochastic variable as defined in Sec.~\ref{sec:Langevin}.} \cite{Starobinsky:1986fx,Fujita:2013cna,Fujita:2014tja,Vennin:2015hra,Pattison:2017mbe}.

  The PDF $P[\zeta]$ of the primordial curvature perturbation can then  be determined by using the techniques of first-passage time analysis for the stochastic distribution of the number of e-folds $\cal N$ with fixed boundary conditions on the coarse-grained inflaton field.  A convenient analytic approach to  obtain the distribution of first-passage e-folds (and hence the  PDF $P[\zeta]$) is  to solve  the corresponding Fokker-Planck equation (FPE) with the same boundary conditions \cite{Pattison:2017mbe,Ezquiaga:2019ftu,Pattison:2021oen}. In the analysis of stochastic dynamics, the noise terms in the FPE are assumed \cite{Vennin:2015hra,Pattison:2017mbe,Ezquiaga:2019ftu,Pattison:2021oen} to be of de Sitter-type\footnote{ See Ref.~\cite{Grain:2017dqa} for an analysis of slow-roll noise terms  beyond the de Sitter approximations.}, \textit{i.e.}~$\Sigma_{\pi\pi},\Sigma_{\phi\pi} \simeq 0$, and the field noise, $\Sigma_{\phi\phi} = (H/(2\pi))^2$, is constant  (where $H$ is the Hubble expansion rate during the de-Sitter type phase).  However, in the context of PBH formation, since slow roll is usually violated, it is important to compute the stochastic  noise matrix elements more accurately, otherwise the determination of the PDF becomes inaccurate, which in turn leads to an imprecise estimation of the PBH mass fraction $\beta_{\rm PBH}$.
  
  \bigskip
  
  Our aim is to develop analytical and semi-analytical techniques to  estimate the full PDF  $P[\zeta]$  using the stochastic inflation formalism beyond slow-roll. In the present  paper,   we carry out a thorough  analytical and numerical computation of  the stochastic noise-matrix elements accurately beyond the slow-roll approximations. In a forthcoming paper \cite{MCG_ST_FPE_22} we determine the PDF  $P[\zeta]$ by solving the Fokker-Planck equation beyond slow-roll with  appropriate noise matrix elements and discuss the implications for estimating the mass fraction of PBHs.
 \bigskip

 In what follows, we begin with a brief introduction to the classical inflationary  dynamics in Sec.~\ref{sec:Inf_dyn}, with particular focus on the ultra slow-roll dynamics across a flat segment in the inflaton potential. In Sec.~\ref{sec:SI_formalism}, we discuss the quantum dynamics in the stochastic inflation framework. We introduce the Langevin equation in Sec.~\ref{sec:Langevin}  and emphasise the importance of the noise matrix elements in the adjoint Fokker-Planck equation in Sec.~\ref{sec:Adj_FP_SI}. Section \ref{sec:SI_noise} is dedicated to the computation of the noise matrix elements which is the primary focus of this work. We numerically compute the noise matrix elements for a slow-roll potential as well as a potential with a slow-roll violating feature in Sec.~\ref{sec:SI_noise_numerical} before proceeding to carry out a thorough  analytical treatment in Sec.~\ref{sec:SI_noise_analytical} for instantaneous  transitions between different phases during inflation. We discuss the  potential implications of our results  for the  computation of the PBH mass fraction  and spell out a number of  complexities associated with the computation   in Sec.~\ref{sec:Discussions} before  concluding with   a summary of our main results in Sec.~\ref{sec:Conclusions}. Appendix \ref{app:MS_flat_gauge} provides a derivation of the Mukhanov-Sasaki equation in spatially flat gauge.  Appendix \ref{app:MS_analyt_sol} deals with the  analytical solutions of the Mukhanov-Sasaki equation in the absence of  any transition,  while Appendix~\ref{app:noise_elements_expanded} provides analytical  expressions for the noise matrix elements in the super-Hubble limit. Appendices \ref{app:eta_to_nu} and \ref{app:Sig_2_trans} are dedicated to the dynamics during instantaneous transitions.

\bigskip

We work in natural units with $c=\hbar =1$ and define the reduced Planck mass to be  $\mpl \equiv 1/\sqrt{8\pi G} = 2.43 \times 10^{18}~{\rm GeV}$. We assume the background
 Universe to be described by a spatially flat Friedmann-Lemaitre-Robertson-Walker (FLRW)  metric with signature $(-,+,+,+)$. An overdot $^{(.)}$ denotes derivative with respect to cosmic time $t$, while an overdash $^{(')}$ denotes derivative with respect to the conformal time $\tau$. 

\section{Inflationary dynamics beyond slow roll}
\label{sec:Inf_dyn}

We focus on the inflationary  scenario of a single canonical  scalar field  $\phi$  with a self-interaction potential $V(\phi)$ which is minimally coupled to gravity. The system is described by the action

\beq
S[g_{\mu\nu},\phi] = \int \,  {\rm d}^4x \,  \sqrt{-g} \,  \l[ \, \f{m_p^2}{2} \, R - \f{1}{2}  \, \partial_{\mu}\phi \,  \partial_{\nu}\phi  \, g^{\mu\nu}-V(\phi)\r]~,
\label{eq:Action:GR_inf}
\eeq
where $R$ is the Ricci scalar and $g_{\mu\nu}$ is the metric tensor.  Specializing to the spatially flat FLRW background metric
\beq 
{\rm d}s^2  =  - {\rm d} t^2 + a^2(t)    \l [{\rm d} x^2 + {\rm d} y^2 + {\rm d} z^2 \r] \, ,
\label{eq:FLRW}
\eeq
the evolution equations for the scale factor, $a(t)$, and inflaton, $\phi(t)$, are 

\ber
H^2 \equiv \frac{1}{3m_p^2} \, \rho_{\phi} &=& \frac{1}{3m_p^2} \left[\frac{1}{2}{\dot\phi}^2 +V(\phi)\right] \, ,
\label{eq:friedmann1}\\
\dot{H} \equiv \frac{\ddot{a}}{a}-H^2 &=& -\frac{1}{2m_p^2}\, \dot{\phi}^2 \, ,
\label{eq:friedmann2}\\
{\ddot \phi}+ 3\, H {\dot \phi} + V_{,\phi}(\phi) &=& 0 \,.
\label{eq:phi_EOM}
\eer 
where $H\equiv \dot{a}/a$ and $V_{,\phi} \equiv dV/d\phi$.
 
The slow-roll regime of inflation is usually characterised by the first  two kinematic Hubble slow-roll parameters $\epsilon_H$ and  $\eta_H$, defined by 
\ber
\epsilon_H &=& -\frac{\dot{H}}{H^2}=\frac{1}{2m_p^2} \, \frac{\dot{\phi}^2}{H^2} \,,
\label{eq:epsilon_H}\\
\eta_H &=& -\frac{\ddot{\phi}}{H\dot{\phi}}=\epsilon_H  - \frac{1}{2\epsilon_H} \, \frac{{\rm d}\epsilon_H}{{\rm d} N} \,,
\label{eq:eta_H}
\eer
where  $N = {\rm ln}(a/a_i)$ is the number of e-folds of expansion during inflation with $a_i$ the initial scale factor at some early epoch during inflation before the Hubble-exit of CMB scale fluctuations. The slow-roll conditions correspond to 
\beq
\epsilon_H,~\eta_H\ll 1~.
\label{eq:slow-roll_conditions}
\eeq
 It follows from the definition of the Hubble parameter, $H$, and $\epsilon_H$ in Eq.~(\ref{eq:epsilon_H}), that  
 the condition for the Universe to accelerate, ${\ddot a} > 0$, is $\epsilon_{H} < 1$. Before proceeding further, we  remind the reader of the
distinction between the \textit{quasi-de Sitter} (qdS) and  \textit{slow-roll} (SR)  approximations.   

\begin{itemize}
\item {\bf Quasi-de Sitter} inflation corresponds to the condition $\epsilon_H \ll 1$.
\item {\bf Slow-roll} inflation corresponds to both $\epsilon_H,\, \eta_H \ll 1$. 
\end{itemize}

Hence,  one can deviate from the slow-roll regime by having $|\eta_H| \geq 1$ while still maintaining the qdS expansion by keeping $\epsilon_H \ll 1$, which is exactly what happens during ultra slow-roll (USR) inflation. This distinction will be important for the rest of this paper. Under either of the aforementioned assumptions, the conformal time, $\tau$, is given by
\begin{align}
 -\tau \simeq \f{1}{aH}  \, .
\label{eq:tau_qdS}
\end{align}

 As discussed in Sec.~\ref{sec:Intro}, in order to facilitate PBH formation, we need to  significantly amplify the scalar power at small-scales. This can be achieved with a feature in the inflaton  potential, such as an inflection point-like feature  (as shown in Fig.~\ref{fig:inf_pot_toy_USR_feature}) for which $V_{,\phi} \ll 3 H \dot{\phi}$.  The following criteria need to be satisfied for an
 inflationary potential to be compatible with observations on cosmological scales~\cite{Planck:2018jri} while also generating perturbations on smaller scales that are large enough to form an interesting abundance of PBHs: 

\begin{itemize}
\item
At the CMB pivot scale, $k_*=(aH)_*=0.05~{\rm Mpc}^{-1}$, the amplitude of the
scalar power spectrum is
\beq
{\cal P}_{\zeta}(k_*) = 2.1\times 10^{-9} \,,
\label{eq:CMB_power_constraint}
\eeq
with the scalar spectral index $n_{_S}$ and tensor-to-scalar ratio $r$ satisfying
\beq
n_{_S}(k_{*}) \in \left[0.957,0.975\right]~, \quad r(k_*) \leq 0.036 ~~ {\rm at} ~~ 95\% ~ \mbox{C.L} \, .
\label{eq:CMB_ns_r_constraint}
\eeq

\item

A feature in $V(\phi)$ on a smaller scale $k \gg k_*$ (closer to the end of inflation  $N_e<N_*$) to enhance the 
primordial scalar power spectrum by a factor of roughly $10^7$ with respect to its value at the CMB pivot scale. Here $N_e$ is the number of e-folds before the end of inflation and $N_*$ is the value of $N_e$ when the CMB pivot scale made its Hubble-exit. Typically $N_* \in [50,60]$ depending on the reheating history of the Universe (see \textit{e.g.}~Ref.~\cite{Liddle:2003as}).  Throughout this work we take $N_* = 60$. 

\item
The potential steepens, so that inflation ends. Reheating (and the transition to the subsequent radiation dominated epoch) then occurs as the field oscillates around a minimum in the potential.

\end{itemize}

 Given that PBH formation requires the enhancement of the  inflationary power spectrum  
by a factor of $10^7$ within less than 40 e-folds of expansion (see \textit{e.g.}~Ref.~\cite{Motohashi:2017kbs}), the quantity $\Delta \ln{\epsilon_H}/\Delta N$, and hence 
$|\eta_H|$, must  grow to become of order  unity, thereby violating the second slow-roll condition in Eq.~(\ref{eq:slow-roll_conditions}). In the particular case of a flat plateau region in the potential  at intermediate field values, the inflaton enters a transient  period of ultra slow-roll (USR) (see Refs.~\cite{Tsamis:2003px,Kinney:2005vj,Pattison:2017mbe,Vennin:2020kng,Pattison:2021oen}). Since $V_{,\phi} =0$, from the equation of motion Eq.~(\ref{eq:phi_EOM}), $\ddot{\phi} + 3  H  \dot{\phi}  =0  \Rightarrow - \ddot{\phi}/(H\,\dot{\phi}) = + 3$, leading to 
\beq
\eta_H = +3 ~~~ ~~({\rm during ~ USR}).
\label{eq:eta_USR}
\eeq 
As a consequence, the inflaton speed drops exponentially with the number of e-folds during this USR phase:
\beq
\dot{\phi} = \dot{\phi}_{\rm en} \, e^{-3\, H \, (t-t_{\rm en})} \propto e^{-3\,N}~,
\label{eq:USR_speed}
\eeq
where $\dot{\phi}_{\rm en}$ is the entry velocity to the USR phase at time $t_{\rm en}$. 

\begin{figure}[t]
\begin{center}
\includegraphics[width=0.7\textwidth]{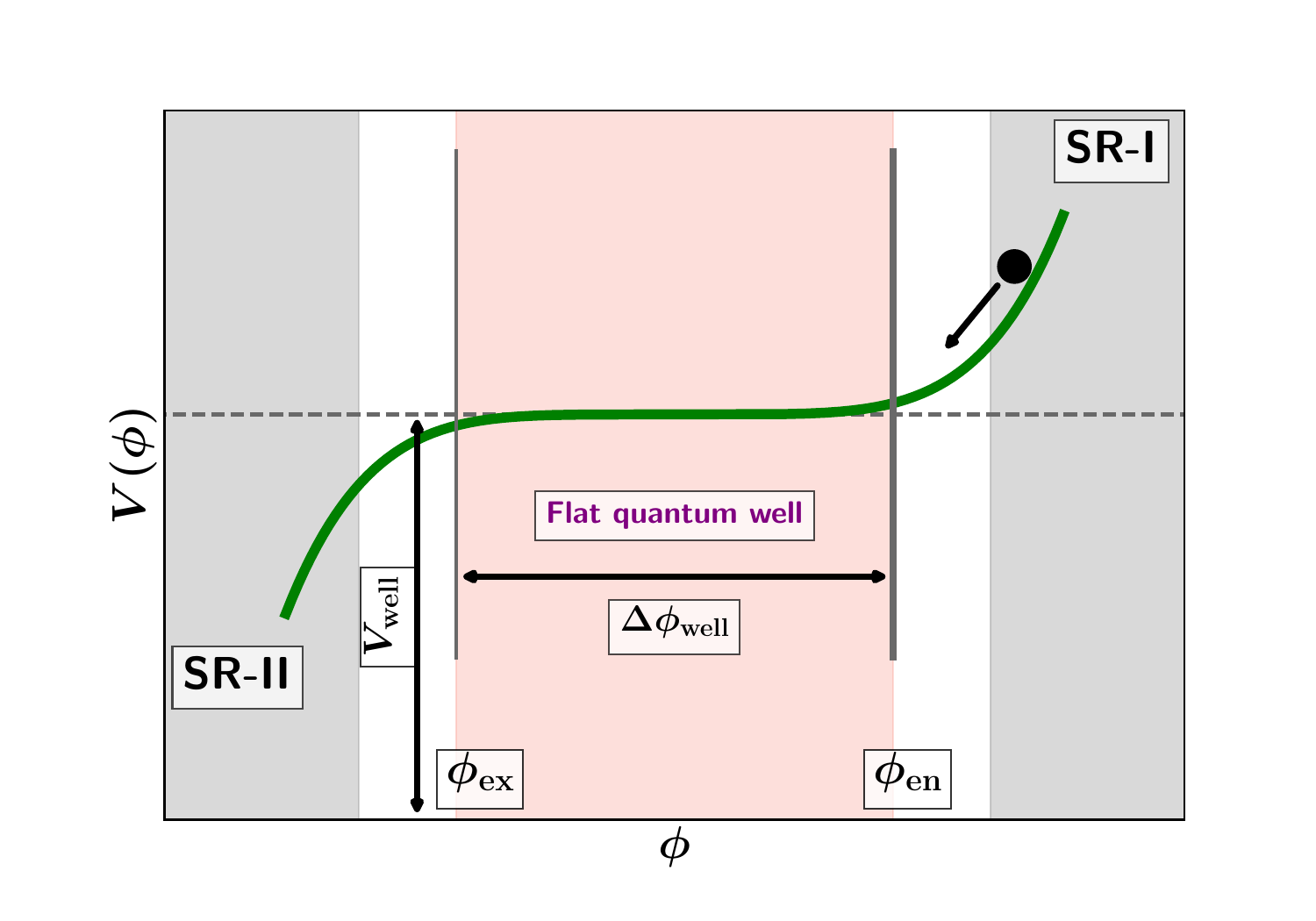}
\caption{A zoomed-in version of Fig.~\ref{fig:inf_pot_toy_USR_feature} in order to schematically illustrate the intermediate flat quantum well feature (highlighted with pink shading) in the inflaton potential.   The height and width of the flat segment are denoted by $V_{\rm well}$ and $\Delta\phi_{\rm well}$ respectively. After exiting the first  slow-roll phase (SR-I) near the CMB window, the inflaton enters the flat region at $\phi = \phi_{\rm en}$ at intermediate field values. During this  USR phase, the effects of quantum diffusion might become significant and hence one should use the stochastic inflation formalism to compute the primordial PDF of $\zeta$.  Later, the inflaton emerges from the USR phase to another slow-roll phase (SR-II) at $\phi = \phi_{\rm ex}$, before the end of inflation. }
\label{fig:inf_flat_Q_well}
\end{center}
\end{figure}

Since USR is a transient non-attractor phase, the inflaton dynamics during this phase are   sensitive to the  initial conditions, in particular to the  speed $\dot{\phi}_{\rm en}$ with which the inflaton enters the plateau. In this context, the  inflaton potential exhibits three important  regimes, namely, the slow-roll SR-I phase for $\phi> \phi_{\rm en}$ around the CMB scale, the USR phase at some intermediate field values $\phi_{\rm ex} \leq \phi \leq \phi_{\rm en}$, succeeded by the final  SR-II phase  for $\phi<\phi_{\rm ex} $ before the end of inflation at $\phi = \phi_{\rm end}$. Figure \ref{fig:inf_flat_Q_well} schematically illustrates the three regimes.  The flat regime (flat quantum well)\footnote{Note that we refer to the flat USR regime as the `flat quantum well' because the inflaton dynamics are usually dominated  by stochastic quantum diffusion, as discussed in the subsequent Sections.} is characterised by  its width $\Delta\phi_{\rm well} = \phi_{\rm en} - \phi_{\rm ex}$, and height, $V_{\rm well}$. During this regime
\beq
 \dot{\phi} = \dot{\phi}_{\rm en} - 3 H \l(\phi - \phi_{\rm en} \r)~.
\label{eq:USR_speed_1}
\eeq

The total number of e-folds of expansion  during the USR period  up to $\phi$, where $\phi_{\rm ex} \leq \phi \leq \phi_{\rm en}$ is given by
\beq
 N_{\rm USR}(\phi) = \f{1}{3}  \, {\rm log} \l( \f{\pi_{\rm en}}{\pi} \r) = \f{1}{3}  \, {\rm log} \l( \f{\pi_{\rm en}}{\pi_{\rm en} - 3 \, \l( \phi - \phi_{\rm en} \r)} \r) ~,
 \label{eq:N_USR_phi}
 \eeq
where 
\beq
\pi = \f{{\rm d}\phi}{{\rm d}N} = \f{\dot{\phi}}{H} \,.\label{eq:USR_mom_1}
\eeq 
 In order to amplify the perturbations sufficiently to generate an interesting abundance of PBHs, the USR phase typically has to last for around $2-3$ e-folds (see Refs.~\cite{Byrnes:2018txb,Karam:2022nym}). The above expression for $N_{\rm USR}(\phi)$ can be used in the `\textit{non-linear classical $\delta$N formalism}' to determine the PDF of primordial fluctuations \cite{Hooshangi:2021ubn}.

From the above expressions, it is clear that the dynamics of inflation during USR is sensitive to the  initial conditions  $\lbrace\phi_{\rm en},\pi_{\rm en}\rbrace$. Let  us define the  critical entry velocity $\dot{\phi}_{\rm cr} $ to be the speed at which the inflaton must enter the flat quantum well in order to come to a halt  at $\phi_{\rm ex}$. From Eqs.~(\ref{eq:USR_speed_1}) and (\ref{eq:USR_mom_1}) it follows that 
\beq
\dot{\phi}_{\rm cr} = - 3\, H \, \Delta\phi_{\rm well} \, ,~~~~
\pi_{\rm cr}= - 3\,\Delta\phi_{\rm well} \, .
\label{eq:speed_cr}
\eeq
If $\dot{\phi}_{\rm en} > \dot{\phi}_{\rm cr}$, then the classical speed of the inflaton is large enough to drive it all the way across the quantum well, while for $\dot{\phi}_{\rm en} < \dot{\phi}_{\rm cr}$, the inflaton comes to a halt at some intermediate point $\phi \in \l( \phi_{\rm ex}, \phi_{\rm en}  \r)$.  Another important constraint comes from requiring inflation to continue,  $\epsilon_H < 1$,  hence  from Eq.~(\ref{eq:epsilon_H}) we get
\beq
 0 \leq \dot{\phi}_{\rm en} < \sqrt{2}\,H\, m_p \, .
\label{eq:entry_speed_bound}
\eeq

\medskip

In this Section, we have discussed the classical dynamics of the inflaton field beyond slow roll, with the specific example of ultra slow-roll inflation across a flat potential well. We now move on to describe the large-scale quantum   dynamics of the inflaton field which is coarse-grained over super-Hubble scales, using the stochastic inflation formalism. This will enable us to study the PDF of the primordial fluctuations generated by  the quantum diffusion of the inflaton.

\section{Quantum dynamics: stochastic inflation formalism}
\label{sec:SI_formalism}

Stochastic inflation is an effective  long wavelength   IR treatment of  inflation in which the inflaton field is coarse-grained over super-Hubble scales $k \leq \sigma \, aH$, with  the constant $\sigma \ll 1$.    On the other hand, the Hubble-exiting   smaller scale UV   modes  are constantly converted into IR modes due to the accelerated expansion during inflation. Hence the coarse-grained inflaton field follows a Langevin-type  stochastic differential equation featuring classical  stochastic noise terms sourced by the smaller scale UV modes,  on top of the classical drift terms  sourced by the gradient of the self-interaction potential $V_{,\phi}(\phi)$.  
   
   \bigskip
   
   We start with the Hamiltonian equations \cite{Grain:2017dqa}  of the system, Eq.~(\ref{eq:Action:GR_inf}), for  Heisenberg operators of  the   inflaton $\hat{\phi}$ and its momentum $\hat{\pi}_{\phi}$
   \ber
\f{{\rm d}\hat{\phi}}{{\rm d}N} &=& \hat{\pi}_\phi ~,\label{eq:phi_H}\\
\f{{\rm d}\hat{\pi}_\phi}{{\rm d}N} &=& - \l( 3 - \epsilon_H  \r) \hat{\pi}_\phi - \f{V_{,\phi}}{H^2} ~,\label{eq:pi_H}
\eer   
where we choose the number of e-folds $N$ as our time evolution variable for $\phi(N,\vec{x})$ and $\pi_{\phi}(N,\vec{x})$ following Refs.~\cite{Vennin:2015hra,Pattison:2017mbe}.

 We split  the inflaton $\hat{\phi}(N,\vec{x})$ and its conjugate momentum $\hat{\pi}_{\phi}(N,\vec{x})$  into the corresponding IR $\lbrace  \hat{\Phi},\hat{\Pi} \rbrace$ and UV $\lbrace  \hat{\varphi},\hat{\pi} \rbrace$ parts:
 
 \beq
\hat{\phi} = \hat{\Phi} + \hat{\varphi}~~,~~ \hat{\pi}_\phi = \hat{\Pi} + \hat{\pi}~,
 \label{eq:phi_pi_split}
 \eeq 
where the UV fields are defined as 
 \ber
\hat{\varphi}(N,\vec{x}) &=& \int \, \f{{\rm d}^3\vec{k}}{\l(2\pi\r)^{\f{3}{2}}} \, W\l(\f{k}{\sigma aH}\r) \, \l[ \phi_k(N)\,  \hat{a}_{\vec{k}} \, e^{-i\vec{k}.\vec{x}} +  \phi_k^*(N) \,  \hat{a}_{\vec{k}}^{\dagger} \,  e^{i\vec{k}.\vec{x}}  \r] \,, \label{eq:UV_phi}\\
\hat{\pi}(N,\vec{x}) &=& \int \, \f{{\rm d}^3\vec{k}}{\l(2\pi\r)^{\f{3}{2}}} \, W\l(\f{k}{\sigma aH}\r) \, \l[ \pi_k(N)\,  \hat{a}_{\vec{k}} \, e^{-i\vec{k}.\vec{x}} +  \pi_k^*(N) \,  \hat{a}_{\vec{k}}^{\dagger} \,  e^{i\vec{k}.\vec{x}}  \r]  \, . \label{eq:UV_pi}
\eer
   Here $W\l(k/\sigma aH\r)$ is the `window function' that selects out  modes with  $k > \sigma aH$. This coarse-graining guarantees that the IR fields $\lbrace  \hat{\Phi},\hat{\Pi} \rbrace$ are  composed of modes of super-Hubble wavelengths and hence can be treated as  classical (stochastic) variables. Substituting the expressions for the UV fields from Eqs.~(\ref{eq:UV_phi}) and (\ref{eq:UV_pi}) into Eqs.~(\ref{eq:phi_H}) and (\ref{eq:pi_H}), the equations for the coarse-grained fields are  \cite{Pattison:2021oen} 
\ber
\f{{\rm d}\hat{\Phi}}{{\rm d}N} &=& \hat{\Pi} + \hat{\xi}_{\phi}(N)~,\label{eq:phi_H_IR}\\
\f{{\rm d}\hat{\Pi}}{{\rm d}N} &=& - \l( 3 - \epsilon_H  \r) \hat{\Pi}- \f{V_{,\phi}(\hat{\Phi})}{H^2} + \hat{\xi}_{\pi}(N) ~,\label{eq:pi_H_IR}
\eer  
where the field and momentum noise operators $\hat{\xi}_{\phi}(N)$ and $\hat{\xi}_{\pi}(N)$ are  given by 
\ber
\hat{\xi}_{\phi}(N) &=& -\int \, \f{{\rm d}^3\vec{k}}{\l(2\pi\r)^{\f{3}{2}}} \, \f{{d}}{{\rm d}N}W\l(\f{k}{\sigma aH}\r) \, \l[ \phi_k(N)\,  \hat{a}_{\vec{k}} \, e^{-i\vec{k}.\vec{x}} +  \phi_k^*(N) \,  \hat{a}_{\vec{k}}^{\dagger} \,  e^{i\vec{k}.\vec{x}}  \r] ~,\label{eq:xi_phi}\\
 \hat{\xi}_{\pi}(N) &=& -  \int \, \f{{\rm d}^3\vec{k}}{\l(2\pi\r)^{\f{3}{2}}} \, \f{{d}}{{\rm d}N}W\l(\f{k}{\sigma aH}\r)  \, \l[ \pi_k(N)\,  \hat{a}_{\vec{k}} \, e^{-i\vec{k}.\vec{x}} +  \pi_k^*(N) \,  \hat{a}_{\vec{k}}^{\dagger} \,  e^{i\vec{k}.\vec{x}}  \r]   ~.\label{eq:xi_pi}
\eer  
We  assume a window function which imposes a sharp cut off\footnote{This assumption  has been questioned, especially because the sharp cut-off window function may not lead to well-behaved coarse-grained field correlators in the  physical space. See Refs.~\cite{Winitzki:1999ve,Andersen:2021lii,Mahbub:2022osb} for more discussion.}  between the IR and UV momentum space modes:
\beq
W\l(\f{k}{\sigma aH}\r) = \Theta\l(\f{k}{\sigma aH} - 1\r) \, .
\label{eq:Window_Theta}
\eeq
 It has the advantage of making the calculation of the noise correlation matrix elements more tractable.

Physically, the noise terms $\hat{\xi}_{\phi}$ and $\hat{\xi}_{\pi}$ in the Langevin Eqs.~(\ref{eq:phi_H_IR}) and (\ref{eq:xi_pi})   are sourced by the constant outflow of UV modes into the IR modes, {\it i.e.} as a UV mode exits  the cut-off scale $k=\sigma aH$ to become part of the  IR field on super-Hubble scales, the IR field receives a `{\it quantum kick}' whose typical amplitude is given by $\sim \sqrt{\langle 0 | \hat{\xi}(N)\hat{\xi}(N') | 0  \rangle}$, where $| 0  \rangle$ is usually taken to be the Bunch-Davies vacuum.  
Given that $\sigma \ll 1$, this happens on ultra super-Hubble scales, where the UV modes must have already become classical fluctuations\footnote{While the quantum-to-classical transition is still an open problem, the treatment of UV noise operators as stochastic noise terms is ensured  to be valid as long as  the  decaying  mode of $\phi_k$ is negligible compared to the non-decaying  mode on super-Hubble scales \cite{Pattison:2019hef}.} due to the rapid  decline of the non-commuting parts of the fields $\lbrace \phi_k,\pi_k \rbrace$ outside the Hubble radius \cite{Polarski:1995jg,Kiefer:1998qe,Kiefer:2008ku}. This leads to the  classical  stochastic description of  the dynamics of the coarse-grained quantum fields  $\hat{\Phi}$, $\hat{\Pi}$ as discussed in the following subsection(s). 

\subsection{Langevin equation}
\label{sec:Langevin}

The Langevin equations corresponding to Eqs.~(\ref{eq:phi_H_IR}) and (\ref{eq:pi_H_IR}) take the compact form
\beq
\f{{\rm d}\Phi_i}{{\rm d}N} = D_i + \xi_i(N) \,
\label{eq:Lang_comp}
\eeq
with coarse-grained IR variables $\Phi_i = \lbrace \Phi, \Pi \rbrace$ and  the  drift terms 
\beq
D_i  = \left\{ \Pi, - \l( 3 - \epsilon_H  \r) \Pi- \f{V_{,\phi}(\Phi)}{H^2}  \right\} , \label{eq:Class-drift-term}
\eeq
 along with the noise operator terms $\xi_i =\lbrace \xi_\phi, \xi_\pi \rbrace$.

In this  compact notation the expressions for the noise operators, Eqs.~(\ref{eq:xi_phi}) and (\ref{eq:xi_pi}), become
\beq
\hat{\xi}_{i}(N) = -\int \, \f{{\rm d}^3\vec{k}}{\l(2\pi\r)^{\f{3}{2}}} \, \f{{d}}{{\rm d}N} W \l(\f{k}{\sigma aH}\r) \, \l[ \phi_{i_k}(N)\,  \hat{a}_{\vec{k}} \, e^{-i\vec{k}.\vec{x}} +  \phi_{i_k}^*(N) \,  \hat{a}_{\vec{k}}^{\dagger} \,  e^{i\vec{k}.\vec{x}}  \r] \, ,
\label{eq:noise_comp}
\eeq
with $\phi_{i_k} = \lbrace \phi_k, \pi_k \rbrace$ being the field and momentum mode functions  respectively. Assuming the sharp $k$-space window function, Eq.~(\ref{eq:Window_Theta}), it is easy to show that the equal-space noise correlators  (auto-correlators) take the form \cite{Grain:2017dqa} 
\beq
 \langle \xi_i(N) \, \xi_j(N') \rangle  = \Sigma_{ij}(N) \, \delta_D(N-N') \, ,
\label{eq:noise_comp_cor}
\eeq
where the noise correlation matrix $\Sigma_{ij}$ has the form 
\beq
 \Sigma_{ij}(N) = (1 - \epsilon_H) \, \f{k^3}{2\pi^2} \, \phi_{i_k}(N)\phi_{j_k}^*(N)\bigg\vert_{k = \sigma aH} ~.
\label{eq:noise_comp_cor_matrix}
\eeq

\bigskip

The stochastic nature of the noise leads to a probabilistic description of the system $\Phi_i = \lbrace \Phi, \Pi \rbrace$. One approach to analyse the system is by solving the Langevin equation, Eq.~(\ref{eq:Lang_comp}), numerically  for many (tens of millions) stochastic  realizations and then proceeding to compute different moments of the  physical (stochastic) variables. This method  is direct, however cumbersome, non-analytical and requires significant computational power. See Refs.~\cite{Fujita:2013cna,Fujita:2014tja} for some  of the earlier attempts in this direction, while for a more concrete analysis beyond slow-roll, see Ref.~\cite{De:2020hdo}, and  for state-of-the-art numerical simulations, relevant for  determining the PDF of primordial fluctuations, see Refs.~\cite{Figueroa:2020jkf,Tomberg:2021xxv,Figueroa:2021zah,Jackson:2022unc}.

 There is also an analytically concrete way to study this system, using the {\em first-passage time} analysis which involves making a transition from the Langevin equations to an equivalent second order partial differential Fokker-Planck equation (FPE) \cite{SDE_Gardiner,SDE_Evans,Starobinsky:1986fx,Starobinsky:1994bd}, that describes the time evolution of the PDF of the stochastic variables $\lbrace \Phi, \Pi \rbrace$, subject to appropriate boundary conditions. Given our primary goal of computing the full PDF $P[\zeta]$,  we take this route following Refs.~\cite{Pattison:2017mbe,Ezquiaga:2019ftu,Pattison:2021oen}.

The FPE corresponding to the Langevin equation,  Eq.~(\ref{eq:Lang_comp}), takes the form
\beq
\f{\partial}{\partial N} \, P(\Phi_i;N) = {\cal L}_{\rm FP}(\Phi_i) \,  . \,  P(\Phi_i;N) \, ,
\label{eq:FP_fQwell_xy}
\eeq
 where ${\cal L}_{\rm FP}(\Phi_i)$ is the second-order  Fokker-Planck differential operator and  $P(\Phi_i;N)$ is the probability density function of the stochastic process that is related to the probability of finding the phase-space variables at a given value  $\Phi_i = \lbrace \Phi,\Pi\rbrace$  at some time $N$. However such a quantity is not of primary concern  to us since we are not interested in studying the phase-space dynamics of the inflaton\footnote{This would have been our primary goal if we were studying the initial conditions for inflation, or exit from eternal inflation to a SR classical regime \cite{Creminelli:2008es,Rudelius:2019cfh}.}. Rather, we are interested in finding  the probability distribution $P_{\Phi_i}({\cal N})$ of the number of e-folds ${\cal N}$. Note the important difference between  our time variable $N$  and the stochastic variable ${\cal N}$. $N$ denotes the background expansion of the Universe, while ${\cal N}$ is the number of e-folds of expansion obtained from the Langevin equations with fixed boundary conditions in the IR field space, $\phi_{\rm en}$ and $\phi_{\rm ex}$.  The coarse-grained curvature perturbation  $\zeta_{\rm cg}$  is related to the stochastic number of e-folds ${\cal N}$ via the stochastic $\delta{\cal N}$ formalism \cite{Starobinsky:1986fx,Vennin:2015hra,Pattison:2017mbe,Ezquiaga:2019ftu,Pattison:2021oen} 
 \beq
\zeta_{\rm cg} \equiv \zeta(\Phi_i) = {\cal N} - \langle {\cal N}(\Phi_i) \rangle \, ,
\label{eq:zeta_deltaN_stoc}
\eeq
with
\beq
\langle {\cal N}(\Phi_i) \rangle  = \int_0^{\infty} \,  {\cal N} \, P_{\Phi_i}({\cal N}) \, {\rm d}{\cal N} \, ,
\label{eq:N_avg}
\eeq
where  the PDF $P_{\Phi_i}({\cal N})$ of the number of e-folds  satisfies the adjoint FPE which we discuss below in Sec.~\ref{sec:Adj_FP_SI}.  Note that   ${\cal N}(\Phi_i)$ and $P_{\Phi_i}({\cal N})$ correspond to  ${\cal N}(\Phi,\Pi)$ and  $P_{(\Phi,\Pi)}({\cal N})$ respectively.

\subsection{Adjoint Fokker-Planck equation and first-passage time analysis}
\label{sec:Adj_FP_SI} 

  The adjoint FPE for the PDF $P_{\Phi_i}({\cal N})$ corresponding to the general Langevin equation, Eq.~(\ref{eq:Lang_comp}), is given by
\beq
 \f{\partial}{\partial {\cal N}} \, P_{\Phi_i}({\cal N}) =   \l[ D_i \, \f{\partial}{\partial \Phi_i} + \f{1}{2} \,  \Sigma_{ij} \, \f{\partial^2}{\partial \Phi_i \partial \Phi_j} \r]   P_{\Phi_i}({\cal N})  \, .
\label{eq:Adj_FP_Comp_gen}
\eeq

Our primary goal is to solve Eq.~(\ref{eq:Adj_FP_Comp_gen}), with appropriate  boundary conditions  for $\Phi_i\equiv \lbrace \Phi, \Pi \rbrace$  in order to compute the PDF  $P_{\Phi_i}({\cal N}) \equiv P_{\Phi,\Pi}({\cal N})$. A physically well-motivated set of boundary conditions  includes  an absorbing boundary   at smaller field values $\phi^{(\rm A)}$ closer to the end of inflation and a reflecting boundary  at a larger field value $\phi^{(\rm R)}$ closer to the CMB scale. The PDF at the boundaries satisfies
 
 \begin{enumerate}
\item Absorbing  boundary at $\phi^{(\rm A)}$
\beq
P_{\Phi=\phi^{(\rm A)},\Pi}({\cal N}) = \delta_D({\cal N})~,
\label{eq:BC_IR_PDF}
\eeq

\item Reflecting  boundary at $\phi^{(\rm R)}$ 
\beq
\f{\partial}{\partial \Phi}P_{\Phi=\phi^{(\rm R)},\Pi}({\cal N}) = 0~.
\label{eq:BC_UV_PDF}
\eeq
\end{enumerate}

The absorbing boundary condition ensures that for  $\Phi < \phi^{(\rm A)}$, the dynamics is heavily drift dominated and quantum diffusion effects are negligible. Similarly, the reflecting boundary condition arises from assuming that the potential is  steep  enough  in the region $\Phi > \phi^{(\rm R)}$ so that a freely diffusing inflaton can not climb back to a region  of the potential beyond $\phi^{(\rm R)}$. Both the boundary conditions play a crucial role in  determining  the functional form of the  PDF, thus affecting the PBH mass fraction.

\bigskip

A convenient method for determining  the PDF, as discussed in Ref.~\cite{Pattison:2017mbe}, involves  considering the `characteristic function' (CF) $\chi_{\cal N}(q;\Phi_i)$, given by \footnote{The subscript ${\cal N}$ in  $\chi_{\cal N}(q;\Phi_i)$ denotes that the characteristic function is obtained by taking the Fourier transformation of the PDF with respect to ${\cal N}$, and hence  $\chi_{\cal N}(q;\Phi_i)$ is in fact independent of ${\cal N}$ and  only a function of $q, \, \Phi$. } 
\beq        
\chi_{\cal N}(q;\Phi_i) \equiv \langle  e^{i \, q \, {\cal N}}\rangle = \int_{-\infty}^{\infty}  e^{i \, q \, {\cal N}} P_{\Phi_i}({\cal N}) \, {\rm d}{\cal N}  \,,
 \label{eq:CF_gen}
\eeq 
 which is the Fourier transform of the PDF $P_{\Phi_i}({\cal N})$ \textit{w.r.t} the dummy variable $q$ (which is a complex number in general). Hence the PDF   is the inverse Fourier transform of the CF:
  \beq 
   P_{\Phi_i}({\cal N}) = \f{1}{2\pi}   \int_{-\infty}^{\infty}  e^{- i \, q \, {\cal N}}  \chi_{\cal N}(q;\Phi_i) \, {\rm d}q    \,.
\label{eq:PDF_CF_gen}
\eeq 

Since  the PDF satisfies the adjoint FPE, Eq.~(\ref{eq:Adj_FP_Comp_gen}), the CF satisfies 
\beq
  \l[ D_i \, \f{\partial}{\partial \Phi_i} + \f{1}{2} \,  \Sigma_{ij} \, \f{\partial^2}{\partial \Phi_i \partial \Phi_j}   + i q \r] \chi_{\cal N}(q;\Phi_i) = 0 \,,
\label{eq:CF_USR_eq}
\eeq
which is a partial differential equation with one less dynamical variable than the adjoint FPE. The corresponding  boundary conditions, Eqs.~(\ref{eq:BC_IR_PDF}) and (\ref{eq:BC_UV_PDF}), for the characteristic function are given by
\beq
 \chi_{\cal N}(q;\phi^{(\rm A)},\Pi) = 1\, , ~ ~~ \f{\partial}{\partial \Phi} \, \chi_{\cal N}(q;\phi^{(\rm R)},\Pi) = 0 ~.
\label{eq:CF_BCs_gen}
\eeq

The characteristic equation, Eq.~(\ref{eq:CF_USR_eq}), corresponding to a potential $V(\phi)$ in a general situation is quite difficult to solve. In practice, one has to make crucial approximations regarding the classical drift  $D_i$ and the quantum noise $\xi_i$. The most common approximation used in the literature assumes that  the noise matrix elements $\Sigma_{ij}$ in Eq.~(\ref{eq:noise_comp_cor_matrix}) are of the de Sitter-type, \textit{i.e.}  (see Sec.~\ref{sec:SI_noise})
\beq
\Sigma_{\phi\phi} = \l( \f{H}{2\pi} \r)^2 \, , ~~~~ \Sigma_{\phi\pi}, \,  \Sigma_{\pi\pi} \simeq 0 \, .
\label{eq:Sigma_dS_type}
\eeq

We now specialise to the case of quantum diffusion across a flat segment  of the inflaton potential, as discussed in Sec.~\ref{sec:Inf_dyn} and shown in Fig.~\ref{fig:inf_flat_Q_well}. It is helpful to make a change of variables 
\beq
f = \f{\Phi - \phi_{\rm ex} }{\Delta\phi_{\rm well}} \, , ~~~ y = \f{\Pi}{\pi_{\rm cr}} \, ,
\label{eq:f_y_dimless}
\eeq
where $f$ is the fraction of the flat well which remains to be traversed and $y$ is the momentum relative to the critical momentum defined in Eq.~(\ref{eq:speed_cr}), the initial momentum for which the fields comes to a halt at $\phi_{\rm ex}$. The CF, Eq.~(\ref{eq:CF_USR_eq}), then becomes (see Ref.~\cite{Pattison:2021oen})
\beq \l[  \f{1}{\mu^2} \f{\partial^2}{\partial f^2} - 3y \l(\f{\partial}{\partial f} + \f{\partial}{\partial y} \r) + i q \r] \chi_{\cal N}(q;f,y) = 0 \, ,\label{CFfy}
\eeq
where 
\beq
\mu^2  \simeq \f{\Delta\phi^2_{\rm well}}{ m_p^2} \, \f{1}{v_{\rm well} \,} \,,\label{mudef}
\eeq 
with $v_{\rm well} = V_{\rm well}/m_p^4$, where $V_{\rm well}$ is the height of  the flat quantum well. The corresponding boundary conditions  now become 
\beq \chi_{\cal N}(q;0,y) = 1 \, , ~ ~~ \f{\partial}{\partial f} \, \chi_{\cal N}(q;1,y) = 0 \, ,\label{fybdcond}
\eeq
Such a system has been solved \cite{Pattison:2021oen} in two distinct limits, namely 

\begin{itemize}
\item \underline{Free stochastic diffusion} for which  $\pi_{\rm en} \ll \pi_{\rm cr} \Rightarrow   y_{\rm en}\ll 1$,  implying that 
the classical drift term, Eq.~(\ref{eq:Class-drift-term}), can be safely ignored, in which case the PDF takes the form (see Refs.~\cite{Pattison:2017mbe,Ezquiaga:2019ftu})
\beq
 P_f({\cal N})  =  \sum_{n=0}^{\infty} A_n \, \sin{\l[ \l( 2n+1 \r) \, \f{\pi}{2} \, f\r]} \, {\rm e}^{-\Lambda_n \, {\cal N}} ~,
\label{eq:sep_var_SR_gen_sol_x}
\eeq
where 
\beq
A_n =  \l(2n+1\r) \f{\pi}{\mu^2}~,~~~~~ \Lambda_n = \l(2n+1\r)^2\f{\pi^2}{4}\f{1}{\mu^2}\,.
\label{eq:sep_var_SR_fQW_LA}
\eeq

\begin{figure}[H]
\begin{center}
\subfigure[][]{
\includegraphics[width=0.487\textwidth]{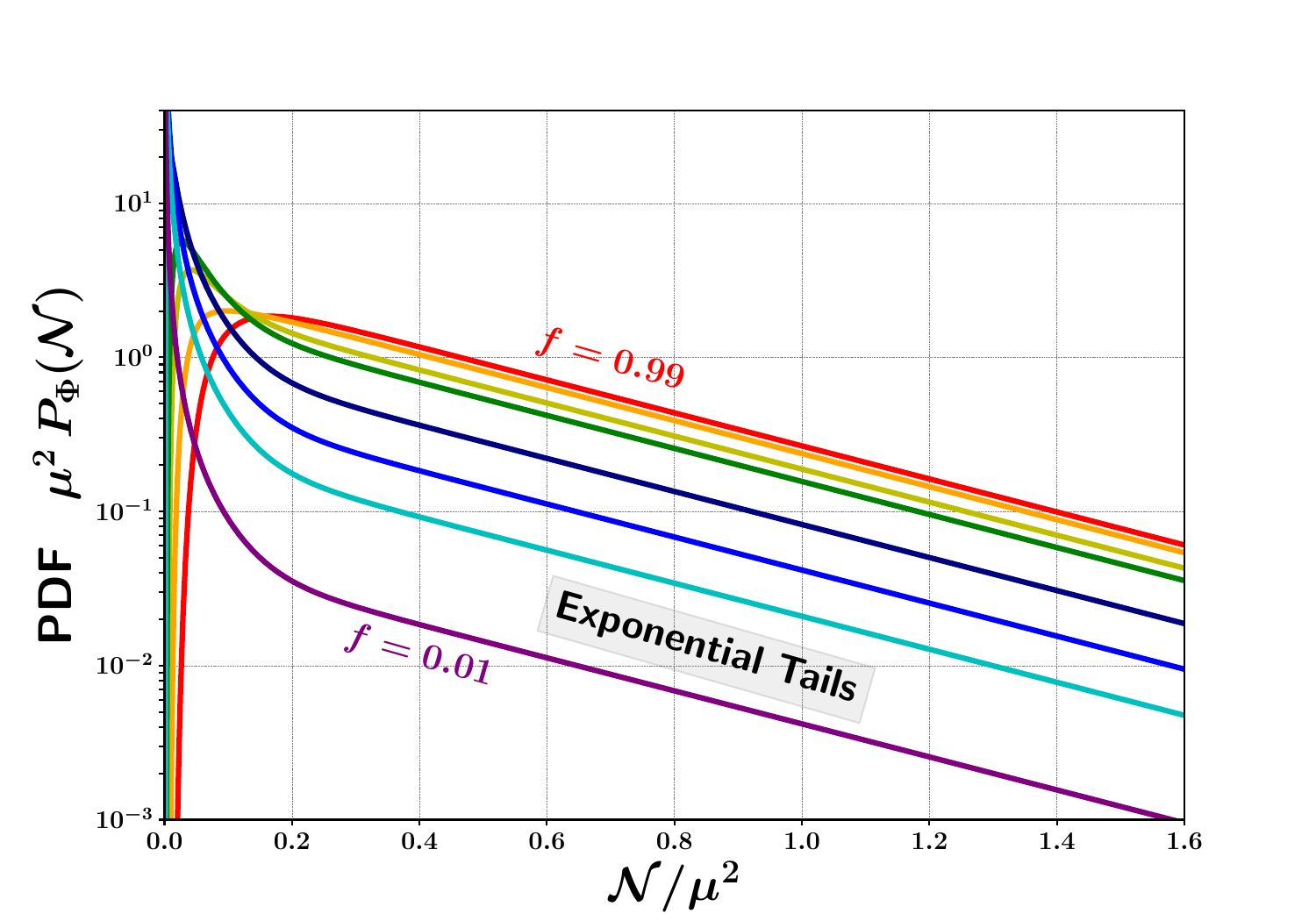}}
\subfigure[][]{
\includegraphics[width=0.487\textwidth]{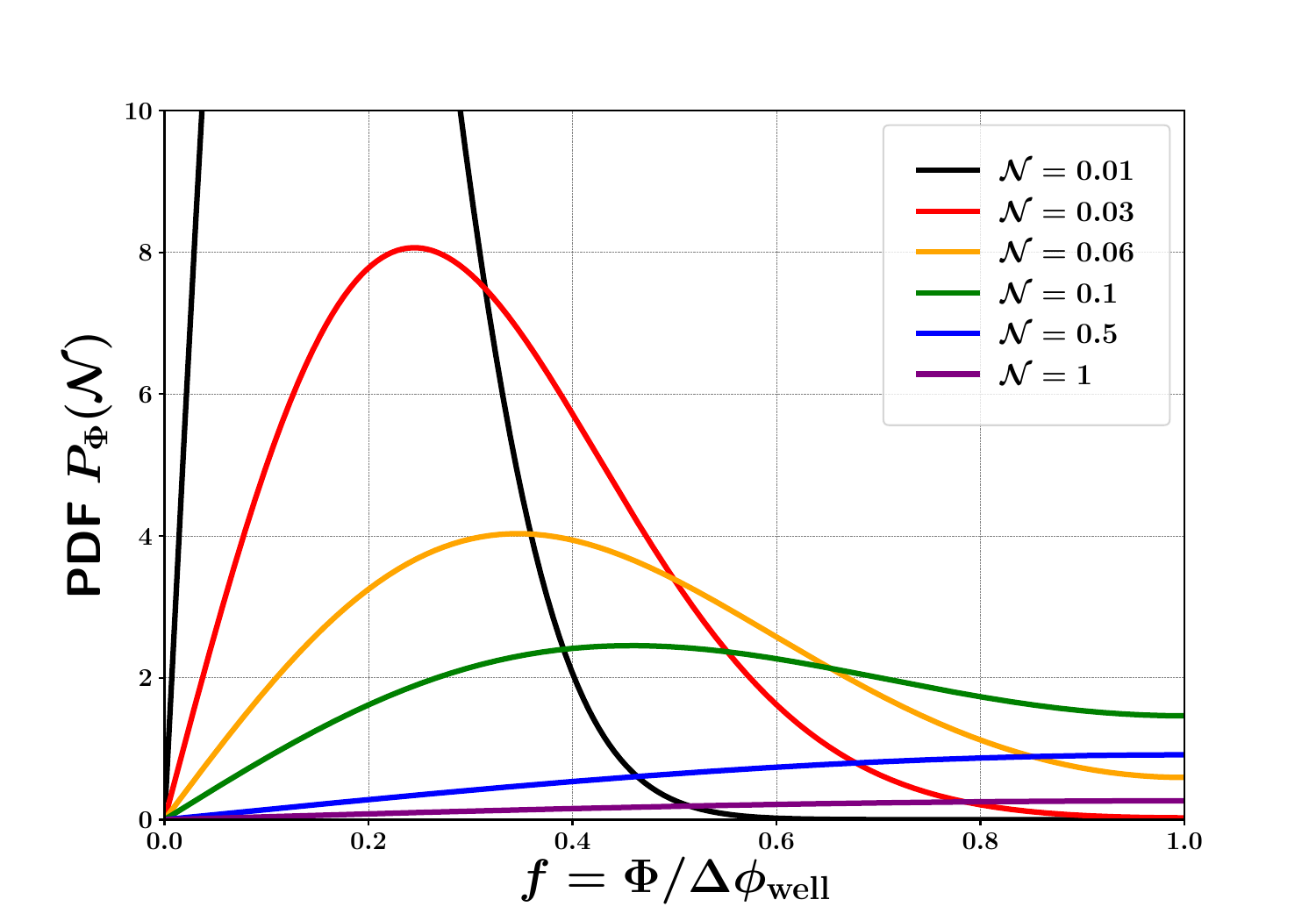}}
\caption{ {\bf Left panel:} the full PDF for the flat quantum well as a function of ${\cal N}$ for different values of the initial condition $\Phi = \phi_i$, expressed in terms of $f = \Phi/ \Delta\phi_{\rm well}$ (for simplicity, we assume $\phi_{\rm ex} =0$ here).  {\bf Right panel:} the full  PDF as a function of initial field value $\Phi$ for 
realizations which have different values of $\cal N$.}
\label{fig:PDF_SR_fQwell}
\end{center}
\end{figure}
The full PDF as a function of ${\cal N}$ is plotted in the left panel of Fig.~\ref{fig:PDF_SR_fQwell}. In the  limit ${\cal N} \gg 1$,  the PDF  exhibits an `{\em exponential tail}' of the form  
$$P_\Phi({\cal N}) \simeq  A_0 \, e^{-\Lambda_0 {\cal N}} \,. $$ 
It follows from Eq.~(\ref{eq:sep_var_SR_fQW_LA}) that the amplitude, $A_{0}$ and coefficient of the exponential, $\Lambda_{0}$, are given by 
 $$A_0 = \f{\pi}{\mu^2} \, , ~~\Lambda_0 = \f{1}{4} \, \f{\pi^2}{\mu^2} \,,$$
(note that $\Lambda_0$ is independent of $\Phi$), so that 
 $$P_\Phi({\cal N}) \simeq \f{\pi}{\mu^2}\, \cos{\l[\f{\pi}{2} \l(\f{\Phi}{\Delta\phi_{\rm well}} -1\r)\r]} \, e^{-\f{1}{4} \, \f{\pi^2}{\mu^2} \, {\cal N}} \, .$$
 
In fact, the exponential tail was shown in Ref.~\cite{Ezquiaga:2019ftu} to be a universal feature of the PDF for quantum diffusion across a generic slow-roll potential with absorbing  and reflecting  boundary conditions,  Eqs.~(\ref{eq:BC_IR_PDF}) and (\ref{eq:BC_UV_PDF}). Larger values of $f$ correspond to  more quantum diffusion before exiting the flat quantum well and hence result in more prominent exponential tails. Notice that the PDFs saturate towards $f=1$ which is a consequence of the reflective  boundary condition given in Eq.~(\ref{eq:BC_UV_PDF}). 

The full  PDF as a function of initial field value $\Phi$  is plotted in the right panel of Fig.~\ref{fig:PDF_SR_fQwell} for realizations which have different values of $\cal N$. It is clear that for ${\cal N} \rightarrow 0$, starting from $f = \Phi/\Delta\phi_{\rm well} \simeq 0$ yields a sharply peaked  distribution, in accordance with the absorbing boundary condition given in Eq.~(\ref{eq:BC_IR_PDF}).

\item  \underline{Large classical drift} where $\pi_{\rm en} \gg \pi_{\rm cr} \Rightarrow   y_{\rm en}\gg 1$. In this case the dynamics of the inflaton is primarily governed by its classical drift and hence the PDF is approximately Gaussian even for ${\cal N} \gg 1$ (see Ref.~\cite{Pattison:2021oen}). 
\end{itemize}

\bigskip

However, in cases where the power spectrum is amplified sufficiently to form an interesting abundance of PBHs, the inflaton typically enters the intermediate flat USR segment from the CMB scale SR-I phase (see Fig.~\ref{fig:inf_flat_Q_well}), with speed of the order $\pi_{\rm en} \simeq \pi_{\rm cr} \Rightarrow   y_{\rm en} \simeq  1$. In this case, both classical drift and stochastic diffusion become important (at least initially during the entry into the USR segment) and hence the aforementioned approximations will not be valid. 
Furthermore, the de Sitter approximations  for the noise matrix elements, Eq.~(\ref{eq:Sigma_dS_type}), might breakdown \cite{Ahmadi:2022lsm} during the transition into the USR phase.   Consequently, it becomes important to estimate the noise matrix elements more accurately. 

\medskip

We conclude that in order to properly use the stochastic $\delta{\cal N}$ formalism to estimate the abundance of PBHs,  one must correctly determine the PDF $P_{\Phi_i}({\cal N})$ from the adjoint FPE Eq.~(\ref{eq:Adj_FP_Comp_gen}) with appropriate boundary conditions. As discussed above,  this can be carried out in  two important steps:

\begin{enumerate}
 
 \item  Calculate  the noise matrix elements $\Sigma_{ij}$ from Eq.~(\ref{eq:noise_comp_cor_matrix}) accurately for the transitions between the CMB scale slow roll and subsequent slow-roll violating phases.
 
 \item Determine the form of the PDF $P_{\Phi,\Pi}({\cal N})$, taking into account the initial momentum with which the inflaton enters the USR segment.

\end{enumerate}

\medskip

In the rest of this paper, we carry out the first task of accurately computing the noise matrix elements, first numerically in Sec.~\ref{sec:SI_noise_numerical} for a potential with a slow-roll violating feature, and then analytically in Sec.~\ref{sec:SI_noise_analytical} for  the case of  instantaneous  transitions between different phases during inflation. We reserve the second task to an upcoming paper \cite{MCG_ST_FPE_22}.

\section{Noise matrix elements in stochastic inflation}
\label{sec:SI_noise}

In this section we calculate the expressions for the noise matrix elements $\Sigma_{ij}$, \textit{i.e.}~the correlators  of the field  and momentum noise operators $\hat{\xi}_i = \lbrace \hat{\xi}_{\phi}, \hat{\xi}_{\pi} \rbrace$. We do this  initially for standard slow-roll inflation, and compare the estimates for $\Sigma_{ij}$ computed using the pure de Sitter approximation to those obtained using the slow-roll approximations.

The key equations that we use are the following: the definition of the noise operators, Eq.~(\ref{eq:noise_comp}), which along with a step-like $k$-space window function,
Eq.~(\ref{eq:Window_Theta}),  leads to the noise correlators of Eq.~(\ref{eq:noise_comp_cor}), with the noise correlation matrix $\Sigma_{ij}$ being given by Eq.~(\ref{eq:noise_comp_cor_matrix}).
It is important to note that these UV-noise mode functions are to be computed, not at Hubble crossing, but at $k=\sigma aH$, where they chronologically  become  part of the coarse-grained IR field and momentum, and provide  quantum kicks. Hence, in order to compute the elements of the noise  matrix $\Sigma_{ij}$, we  need to compute the mode functions $\phi_{i_k} = \lbrace \phi_k, \pi_k \rbrace$. This can be done by solving the Mukhanov-Sasaki (MS)  equation  in terms of conformal time $\tau$ defined in Eq.~(\ref{eq:tau_qdS}) \footnote{Note that depending upon the situation, the MS equation, Eq.~(\ref{eq:MS_modes}), written  in terms of the number of e-folds $N\sim {\rm ln}(a)$ as $$  \f{{\rm d}^2v_k}{{\rm d}N^2}  + \l( 1-\epsilon_H \r)\f{{\rm d}v_k}{{\rm d}N} + \l[ \l( \f{k}{aH} \r)^2 - \f{z''}{z} \r] v_k = 0    \, $$  might be more useful.  We note that in terms of $N$, the MS equation features a friction term, and both the terms inside the square bracket evolve with time. However, in terms of conformal time, $\tau$, it is a simple harmonic oscillator equation with time dependent mass terms $(aH)^{-2} z''/z$, while the comoving mode frequency $k$ is fixed, which is why we choose to work with conformal time.}    \beq
 v_k'' + \l( k^2 - \f{z''}{z} \r) v_k = 0 ~,
\label{eq:MS_modes}
\eeq
 where 
\ber
z &=& am_p\sqrt{2\epsilon_H}~, \label{eq:MS_z} \\
\f{z''}{z} &=& (aH)^2 \l[ 2 + 2 \epsilon_H - 3 \eta_H + 2 \epsilon_H^2 + \eta_H^2 - 3 \epsilon_H \eta_H  - \f{1}{aH} \, \eta'_H \r]~, \label{eq:MS_zdd}
\eer
with appropriate initial conditions. The expressions for the  mode functions  $\phi_{i_k}$ in the spatially flat gauge\footnote{In this work we compute the mode functions $\lbrace \phi_k, \pi_k \rbrace$, and hence the noise matrix elements, $\Sigma_{ij}$, in the spatially flat gauge, while  the Langevin equations are written in the uniform-$N$ gauge. This introduces small corrections to the noise terms which we assume to be negligible \cite{Pattison:2019hef}. We discuss this further in Sec.~\ref{sec:Discussions}.}  are given by (see App.~\ref{app:MS_flat_gauge})
\beq
\phi_k = \f{v_k}{a}~,~~ \pi_k = \f{{\rm d}}{{\rm d}N} \l( \f{v_k}{a} \r) \, .
\label{eq:modef_gen}
\eeq

From Sec.~\ref{sec:Adj_FP_SI}, it is clear that in order to accurately  compute the noise matrix elements $\Sigma_{ij}$ appearing in the adjoint FPE, Eq.~(\ref{eq:Adj_FP_Comp_gen}), we need to solve the MS equation as accurately as possible. For slow-roll inflation, all relevant scales were sub-Hubble at early times, and hence we impose the Bunch-Davies \cite{Bunch:1978yq} initial conditions 
\beq
\lim_{k\tau \longrightarrow -\infty } v_k(\tau) =  \f{1}{\sqrt{2k}} \, e^{-ik\tau}  ~.
\label{eq:MS_BD}
\eeq

We introduce  a convenient new time variable, $T$, defined as 
\beq
 T = -k\tau = \f{k}{a H} \, .
 \label{eq:T_tau}
\eeq
During quasi-dS expansion, the conformal time $\tau$ runs from 
$-\infty$ to $0$, so $T$ runs from $\infty$ to $0$.
Modes undergo Hubble-exit at $T \equiv k/(aH) = 1$, 
and the sub- and super-Hubble regimes correspond to $T \gg 1$ and $T \ll 1$ respectively. 

In terms of $T$ the MS equation, Eq.~(\ref{eq:MS_modes}), takes the form 
\beq
\f{{\rm d}^2 v_k}{{\rm d} T^2} + \l( \, 1 - \f{\nu^2 - \f{1}{4}}{T^2}  \, \r) v_k = 0 \, ,
\label{eq:MS_T_nu} 
\eeq
where
\beq
\nu^2 = \f{1}{(aH)^2} \, \f{z''}{z} + \f{1}{4} \, .  
\label{eq:z_nu_rel}
\eeq
For slow-roll inflation, $\nu^2$ is greater than or equal to $9/4$ at early times and  increases monotonically towards the end of inflation. In the limit where $\nu$ is a constant, the MS Eq.~(\ref{eq:MS_T_nu}) can be converted to a Bessel equation as shown in App.~\ref{app:MS_analyt_sol}.     In what follows, we start with the computation of the noise-matrix elements for the case of  featureless slow-roll potentials, before proceeding to discuss the case of  potentials possessing a slow-roll violating feature. 

\subsection{Featureless potentials}
\label{sec:Sig_SR}

 In the case of a featureless  potential for which slow roll is a good approximation up until the end of inflation, the effective mass term $(aH)^{-2} z''/z$ in the MS Eq.~(\ref{eq:MS_modes}) is almost a constant and evolves monotonically.  Hence the MS Eq.~(\ref{eq:MS_T_nu}) can be solved analytically by approximating $\nu$ in Eq.~(\ref{eq:z_nu_rel}) to be a constant.

 Let us first demonstrate this calculation for the case of the pure de Sitter limit which is usually employed in the computation of noise matrices in the stochastic formalism. In the pure dS limit, both $\epsilon_H, \, \eta_H =0 $, leading to $z''/z = 2 a^2 H^2$ and $\nu^2 = 9/4$. Since $a(\tau) = -1/(H\tau) $ in the pure dS  approximation, 
\beq
a(T) = \f{k}{HT}~,
\label{eq:a_T}
\eeq
 and hence the  MS Eq.~(\ref{eq:MS_T_nu}) reduces to the familiar form

 \beq
\f{{\rm d}^2 v_k}{{\rm d} T^2} + \l(\, 1 - \f{2}{T^2}  \, \r) v_k = 0 \, .
\label{eq:MS_modes_dS}
\eeq

The general solution of this equation is given by 

\beq
v_k(T) =  \f{1}{\sqrt{2k}} \, \l[  \, \alpha_k \, \l( 1 + \f{i}{T} \r) \, e^{i \, T}  \, +   \,  \beta_k \, \l( 1 - \f{i}{T} \r) \, e^{-i \, T}  \, \r] \, ,
\label{eq:MS_sol_gen_qdS}
\eeq
where the positive and negative frequency Bogolyubov coefficients satisfy the canonical normalisation (Wronskian) condition
\beq
|\alpha_k|^2  - |\beta_k|^2 = 1~.
\label{eq:MS_norm_can}
\eeq

Imposing the  Bunch-Davies boundary conditions given in Eq.~(\ref{eq:MS_BD}) 

\beq
\lim_{T \gg 1} v_k(T) =  \f{1}{\sqrt{2k}} \, e^{iT}  \, ,
\label{eq:MS_BD_T}
\eeq
selects out the positive frequency solution only, \textit{i.e} $\alpha_k =1$, $\beta_k = 0$, resulting in the final expression for the MS mode functions
 
\beq
 v_k(T) =  \f{1}{\sqrt{2k}} \, \l( 1 + \f{i}{T} \r) \, e^{i \, T}  \, ,
\label{eq:MS_sol_qdS}
\eeq 
 from which we find the field and momentum mode functions from Eq.~(\ref{eq:modef_gen})  to be 
\ber
\phi_k (T) &=& \f{H}{\sqrt{2k^3}} \, \l( T + i \r)   \, e^{i \, T}  \, ,\label{eq:phi_k_qdS} \\
 \pi_k (T) &=&  - \f{H}{\sqrt{2k^3}} \, i\, T^2   \, e^{i \, T}  \, . \label{eq:pi_k_qdS} 
\eer 

  Using the above expressions for the mode functions, we derive exact expressions\footnote{Note that we have dropped the imaginary part of the off-diagonal terms of the noise correlator matrix  since they do not correspond to classical noise sources \cite{Grain:2017dqa}.}  for the noise  matrix  elements, Eq.~(\ref{eq:noise_comp_cor_matrix}), in the form (recall they are evaluated at $k=\sigma aH$, hence when $T=\sigma$) 
\ber
\Sigma_{\phi\phi}  &=&  \l( 1+\sigma^2 \r) \,  \l(\f{H}{2\pi}\r)^2  ~, \label{eq:Sig_phiphi_dS} \\
{\rm Re}(\Sigma_{\phi\pi})  &=& -\sigma^2 \, \l(\f{H}{2\pi}\r)^2 ~, \label{eq:Sig_phipi_dS} \\
\Sigma_{\pi\pi}  &=& \sigma^4 \, \l(\f{H}{2\pi}\r)^2  \label{eq:Sig_pipi_dS}~.
\eer  
 
 In the stochastic inflation formalism the field and momentum variables are coarse-grained on ultra-Hubble scales, where  $\sigma \ll 1$. For example, taking $\sigma = 0.01$, we get ${\rm Re}(\Sigma_{\phi\pi})
 = 10^{-4} \, \Sigma_{\phi\phi}$ and $\Sigma_{\pi\pi}  = 10^{-8} \, \Sigma_{\phi\phi}$  under the pure de Sitter approximation.  This motivates the usual practice of dropping the momentum-induced noise terms $\Sigma_{\phi\pi}$ and $\Sigma_{\pi\pi}$ from the adjoint FPE, Eq.~(\ref{eq:Adj_FP_Comp_gen}).

 Turning now to slow-roll inflation,  even though  $\epsilon_H, \, \eta_H \ll 1$, the slow-roll parameters do not exactly vanish unlike in pure dS space.  Nevertheless, as long as the quasi-de Sitter expansion is valid (which is justified since $\epsilon_H \ll 1$),  the expression for the scale factor in terms of $T$   is  still given by Eq.~(\ref{eq:a_T}). Under the slow-roll approximations, the MS equation takes the general form Eq.~(\ref{eq:MS_T_nu}) with $\nu \neq 3/2$. In fact for realistic SR potentials, $\nu$ is roughly equal to $3/2$ and evolves slowly and monotonically. 
Assuming $\nu$ to be a constant,  and imposing Bunch-Davies initial conditions, the expression for $v_k$ takes the form (see App.~\ref{app:MS_analyt_sol})

\beq
 v_k(T) =   e^{i \l(  \nu + \f{1}{2}  \r) \f{\pi}{2}}  \,  \sqrt{\f{\pi}{2}}  \, \f{1}{\sqrt{2k}} \,  \sqrt{T} \,  H_{\nu}^{(1)}(T)   \, ,
\label{eq:MS_sol_SR}
\eeq 
 where $ H_{\nu}^{(1)}(T)$ is the Hankel function of the first kind.   For $\nu \neq 3/2$, using the expression for the super-Hubble limit of the Hankel function\footnote{ Expressions for $\Sigma_{ij}$ which are valid for any value of  $\nu$ in the super-Hubble limit  are provided in App.~\ref{app:noise_elements_expanded} (also see Ref.~\cite{Grain:2017dqa}).}  given in Eq.~(\ref{eq:hankel1_super}), we obtain expressions for the field and momentum mode functions 
 
\ber
\phi_k(T) &=&   e^{i \l(  \nu - \f{1}{2}  \r) \f{\pi}{2}} \,  2^{\nu - \f{3}{2}} \,  \f{\Gamma(\nu)}{\Gamma(3/2)} \,  \f{H}{\sqrt{2\,k^3}} \,  T^{-\nu + \f{3}{2}} \, ,\\
\pi_k(T) &=& - e^{i \l(  \nu - \f{1}{2}  \r) \f{\pi}{2}} \,  2^{\nu - \f{3}{2}} \,  \f{\Gamma(\nu)}{\Gamma(3/2)} \,  \f{H}{\sqrt{2\,k^3}} \, \l( -\nu + \f{3}{2} \r) \, T^{-\nu + \f{3}{2}} \, ,
\eer
 which leads to the following expressions for the noise matrix elements $\Sigma_{ij}$ on super-Hubble scales 
 \ber
 \Sigma_{\phi\phi} &=&   2^{2 \l( \nu - \f{3}{2} \r)} \, \l[ \f{\Gamma(\nu)}{\Gamma(3/2)} \r]^2 \, \l(\f{H}{2\pi} \r)^2 \, T^{2\l( -\nu + \f{3}{2}  \r)} \, , \label{eq:Sig_phiphi_SR_supH} \\
 {\rm Re}\l(\Sigma_{\phi\pi}\r) &=&   - 2^{2 \l( \nu - \f{3}{2} \r)} \, \l[ \f{\Gamma(\nu)}{\Gamma(3/2)} \r]^2 \, \l(\f{H}{2\pi} \r)^2 \, \l( -\nu + \f{3}{2} \r) \, T^{2\l( -\nu + \f{3}{2}  \r)}   \, , \label{eq:Sig_phipi_SR_supH} \\
 \Sigma_{\pi\pi} &=&   2^{2 \l( \nu - \f{3}{2} \r)} \, \l[ \f{\Gamma(\nu)}{\Gamma(3/2)} \r]^2 \, \l(\f{H}{2\pi} \r)^2 \, \l( -\nu + \f{3}{2} \r)^2 \, T^{2\l( -\nu + \f{3}{2}  \r)}  \, . \label{eq:Sig_pipi_SR_supH} 
 \eer

 \begin{figure}[hbt]
\begin{center}
\includegraphics[width=0.8\textwidth]{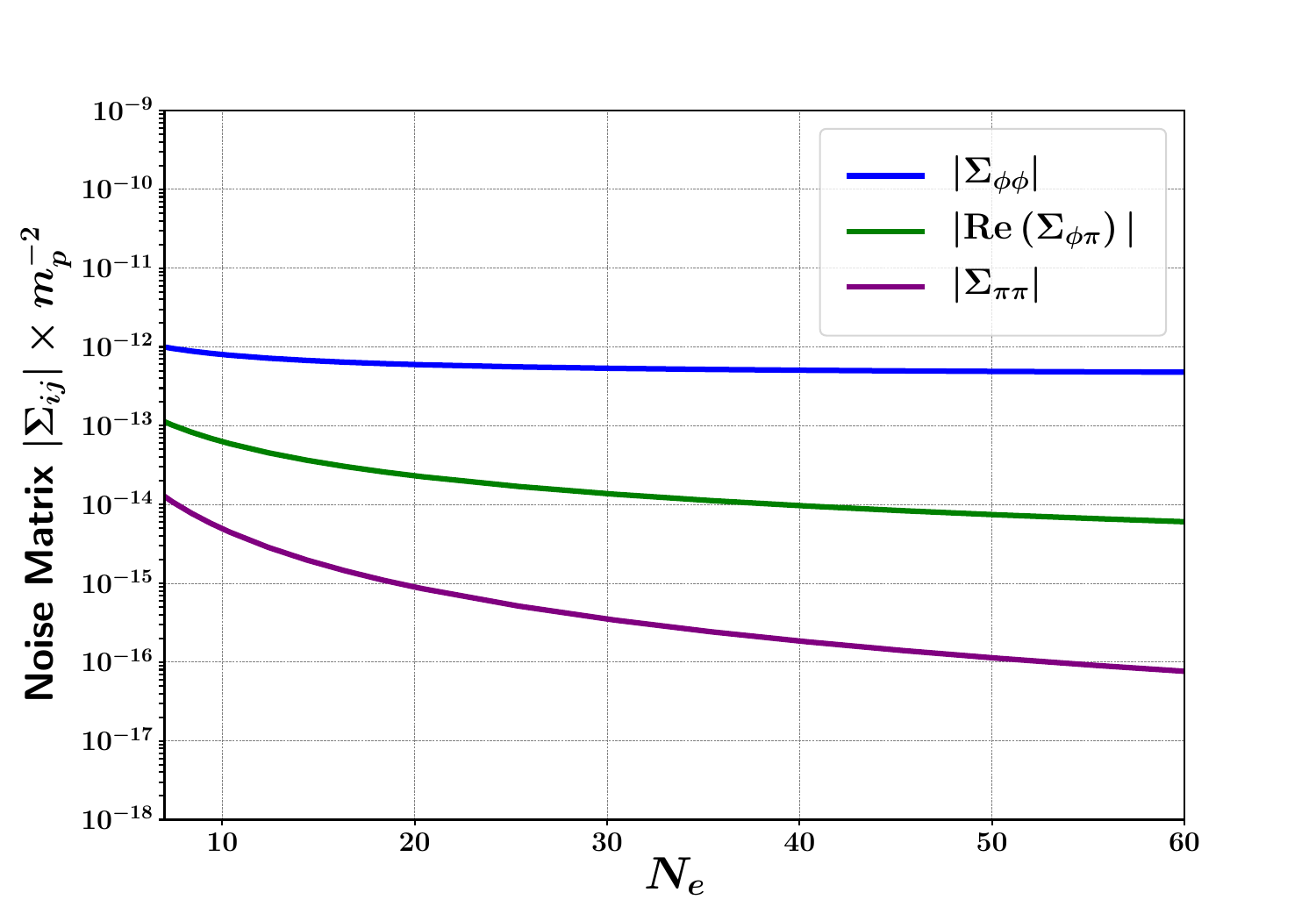}
\caption{The numerically determined noise matrix elements for the slow-roll D-brane  KKLT potential given in Eq.~(\ref{eq:KKLT_base}) for $M=0.5 \, m_p$, in the absence of any features with $\sigma=0.01$, as a function of number of e-folds from the end of inflation $N_{e}$: from top to bottom $|\Sigma_{\phi \phi}|$, $|{\rm Re}(\Sigma_{\phi \pi})|$ and $|\Sigma_{\pi \pi}|$, blue, green and purple lines respectively. We find  significant differences between the numerical calculation and the analytical estimation under the assumption of a de Sitter expansion, Eqs.~(\ref{eq:Sig_phiphi_dS})-(\ref{eq:Sig_pipi_dS}), in which case the ratio of $\Sigma_{\phi\phi}:|{\rm Re(\Sigma_{\phi\pi}})|:\Sigma_{\pi\pi}$ is $1:10^{-4}:10^{-8}$.} 
\label{fig:Sig_SR_KKLT_Num}
\end{center}
\end{figure}
 
 Recalling the definition of $T$ in Eq.~(\ref{eq:T_tau}) and the fact super-Hubble scales correspond to $k=\sigma a H$, hence $T=\sigma$, it follows that the 
above expressions demonstrate that all three noise terms scale as $\Sigma_{ij} \propto \sigma^{2\l( -\nu + 3/2  \r)}$ on super-Hubble scales. This  is  in contrast to the pure dS limit where the three noise terms in Eqs.~(\ref{eq:Sig_phiphi_dS})-(\ref{eq:Sig_pipi_dS}) behave differently, namely, $\Sigma_{\phi\phi} = {\rm const.}$, $\Sigma_{\phi\pi} \propto \sigma^2$ and $\Sigma_{\pi\pi} \propto \sigma^4$. Hence, during  SR inflation for which $\nu \simeq 3/2$, even though the momentum-induced noise terms $\Sigma_{\phi\pi}$ and $\Sigma_{\pi\pi}$ are small  compared to the field noise $\Sigma_{\phi\phi}$,  they may not be negligible, depending upon the value of $(\nu -3/2)$. As mentioned previously, for  most slow-roll potentials, $\nu$ evolves slowly and monotonically. The numerically determined noise matrix elements, $\Sigma_{ij}$, are shown in Fig.~\ref{fig:Sig_SR_KKLT_Num} for an example asymptotically flat SR potential, which we choose to be the D-Brane KKLT potential \cite{KKLT,KKLMMT,Kallosh_Linde_CMB_targets1,inf_encyclo} which has the form
 \beq
V(\phi) = V_0 \, \f{\phi^2}{M^2+\phi^2} \, ,
\label{eq:KKLT_base}
\eeq 
where $M$ is the mass scale in the KKLT model  which we have chosen to be $M=0.5 \, m_p$. We have chosen $\sigma=0.01$ as is the standard practice (see Ref.~\cite{Grain:2017dqa}).  We notice that the momentum induced noise  terms  $\Sigma_{\phi\pi}$ and $\Sigma_{\pi\pi}$ are  much higher  than their corresponding values   in the pure de Sitter limit.  In particular, we find the ratio\footnote{The ratio of the noise terms is not strongly dependent on the value of the KKLT mass scale, $M$. Smaller values of $M$ result in smaller values of $\epsilon_H$, without changing the value of $\eta_H$ significantly. Therefore, in the quasi-dS limit $\epsilon_H \ll 1$, the value of $\nu^2$ from  Eq.~(\ref{eq:z_nu_rel}) and hence the noise matrix elements, Eqs.~(\ref{eq:Sig_phiphi_SR_supH})-(\ref{eq:Sig_pipi_SR_supH}), do not change significantly. Such weak dependence of the ratio of noise terms on the parameters of the  potential is also the case for $\alpha$-attractors as well as for a number of other asymptotically flat potentials (see Ref.~\cite{Kallosh_Linde_CMB_targets1,Mishra:2022ijb}). However, in general, if a change in the parameters of the potential changes the value of $\nu$, it will change the ratio of the noise terms.}  of $\Sigma_{\phi\phi}:|{\rm Re(\Sigma_{\phi\pi}})|:\Sigma_{\pi\pi}$  to be  $1:2\times 10^{-2}:4\times 10^{-4}$  for large $N_e$  as opposed to the de Sitter analytic estimate of $1:10^{-4}:10^{-8}$.   Additionally, the momentum induced noise terms scale approximately in the same way as the field noise $\Sigma_{\phi\phi}$  in accordance with Eqs.~(\ref{eq:Sig_phiphi_SR_supH})-(\ref{eq:Sig_pipi_SR_supH}) at early times during inflation when $\nu \simeq {\rm const.}$. Towards the end of inflation, since $\nu$ starts to evolve  faster, our analytical results based on $\nu \simeq {\rm const.}$ are no longer applicable.

\subsection{Potentials with a slow-roll violating feature}
\label{sec:Sig_USR}
 Potentials possessing a feature that generates large, PBH-forming, perturbations, typically exhibit slow-roll violation, during which the quasi-dS approximation is still valid ($\epsilon_H \ll 1$), while $\eta_H \geq 1$ (see Ref.~\cite{Motohashi:2017kbs}). In particular, $\eta_H \simeq 3$ during an ultra slow-roll phase  as discussed in Sec.~\ref{sec:Inf_dyn}. From  Eq.~(\ref{eq:MS_zdd}), the expression for the effective mass term  $z''/z$ under the quasi-dS approximation becomes 
 \beq
\f{1}{(aH)^2} \, \f{z''}{z}  \simeq 2 - 3\, \eta_H + \eta_H^2 + \tau \, \f{{\rm d}\eta_H}{{\rm d}\tau} \, .
\label{eq:meff_qdS}
\eeq
In this case, the inflationary dynamics undergoes transitions between a number of phases  driven by the behaviour of $\eta_H$.  In single field models in which perturbations grow sufficiently to produce an interesting abundance of PBHs, the inflaton typically undergoes two important transitions (see Ref.~\cite{Karam:2022nym}). The first transition T-I occurs from the CMB scale SR-I to  a near-USR phase, followed by a second transition T-II, from the  near-USR phase to the subsequent second slow-roll phase, SR-II, before the end of inflation. For some class of features (see Refs.~\cite{Karam:2022nym,Bhatt:2022mmn}), the second transition T-II also leads to an intermediate constant-roll (CR) phase \cite{Motohashi:2014ppa}  during which $\eta_H$ is negative, almost constant, and of order unity.
 
\begin{figure}[htb]
\begin{center}
\includegraphics[width=0.8\textwidth]{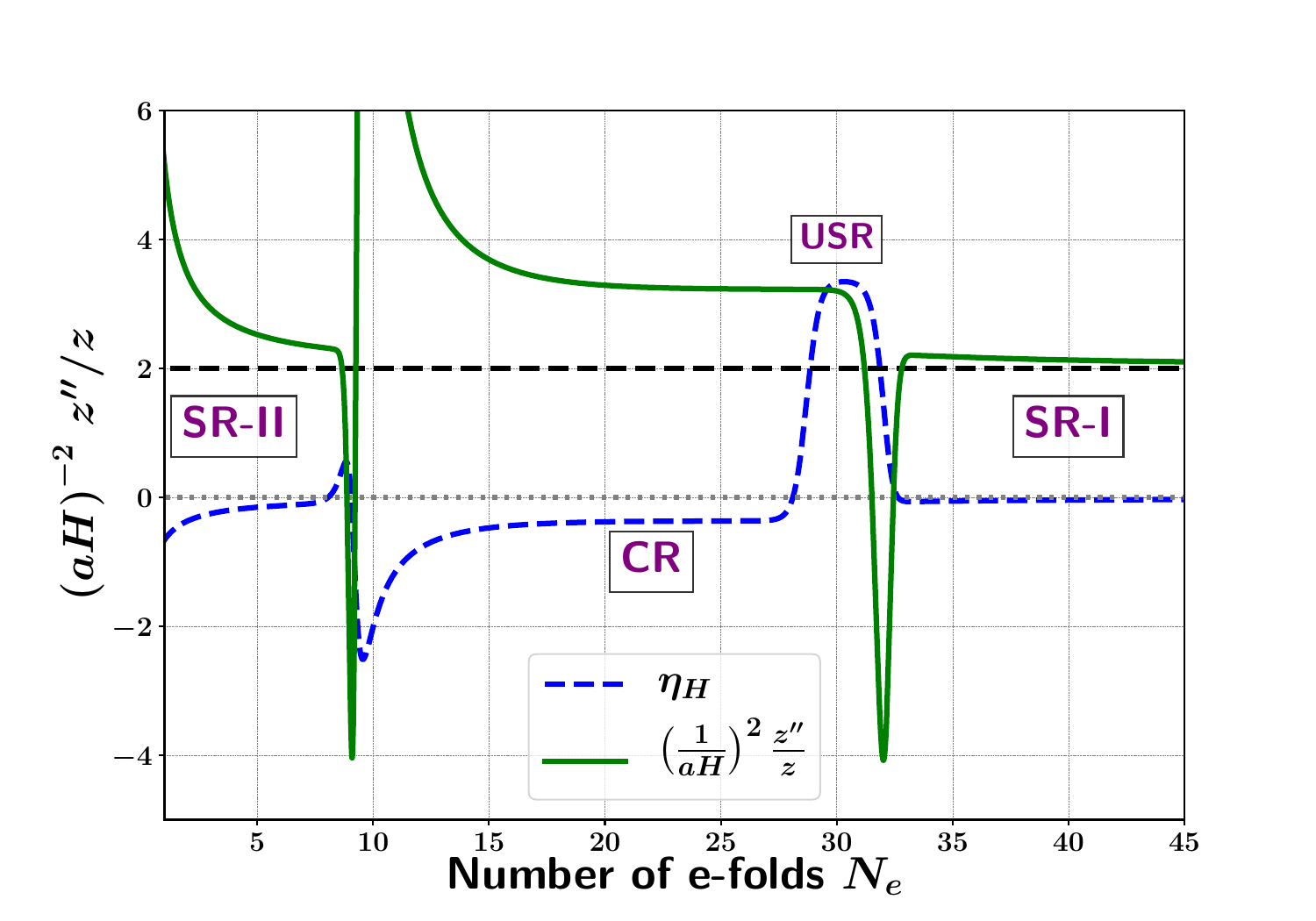}
\caption{Evolution of the effective mass term  $(1/a H)^2 z''/z$  in the Mukhanov-Sasaki equation Eq.~(\ref{eq:MS_modes}) (solid green curve) and $\eta_H$ (dashed blue curve)  for the  modified KKLT potential  featuring a tiny Gaussian bump as given in Eq.~(\ref{eq:KKLT_gaussian_bump}).  The black-dashed line is $(1/a H)^2 z''/z$ for a de Sitter expansion (Eq.~\ref{eq:z_nu_rel}), namely $\nu=3/2$. In the  modified KKLT case,  $(1/a H)^2 z''/z$ makes a sharp yet smooth dip around the transition from the CMB scale SR-I to the subsequent  near-USR phase, after  which it remains almost constant throughout the USR and constant-roll (CR) phases  (but with $\nu > 3/2$), until the inflaton enters into another slow-roll phase, SR-II, before the end of inflation. The  corresponding noise matrix  elements  associated with this potential are shown in Fig.~\ref{fig:Sig_SR_USR_KKLT_GB}.} 
\label{fig:inf_meff_GB}
\end{center}
\end{figure}

 As a specific example, we consider a modified KKLT potential with an additional tiny Gaussian bump-like feature \cite{Mishra:2019pzq}:
 \beq
V_{\rm b}(\phi) = V_0 \, \f{\phi^2}{M^2+\phi^2} \, \l[1 + A \, \exp{\l({-\f{1}{2} \, \f{(\phi-\phi_0)^2}{\tilde{\sigma}^2}}\r)}\r] ~,
\label{eq:KKLT_gaussian_bump}
\eeq 
where $A$, $\tilde{\sigma}$  and $\phi_0$ represent the height,  width  and position of the  bump respectively. The evolution of $\eta_H$ and $z''/z$  for this potential is shown  in Fig.~\ref{fig:inf_meff_GB}.   Following Ref.~\cite{Mishra:2019pzq}, we fix $M=0.5 \, m_p$, and take  the bump parameters to be  $A =  1.87 \times 10^{-3}$, $\tilde{\sigma} = 1.993 \times 10^{-2} \, m_p$ and $\phi_0 =  2.005 \, m_p$.  These bump parameter values lead to a peak in the scalar power-spectrum of ${\cal P}_\zeta \sim 10^{-2}$ at a $k$ value corresponding to $\sim 10^{17} \, {\rm g}$ PBHs, \ie~at the lower end of the asteroid mass window where PBHs can make up all of the dark matter (see e.g.~Refs.~\cite{Green:2020jor,Carr:2020xqk}).

 The inflationary dynamics in this case display the aforementioned three  key phases, namely SR-I, USR and CR with $\eta_H$ making sharp (yet smooth) transitions between them, as shown by the dashed blue curve in Fig.~\ref{fig:inf_meff_GB}. However, during the second transition from USR to the CR phase  ($15 \leq N_e \leq 30$), the effective mass term  $(aH)^{-2} z''/z$ remains nearly   constant\footnote{
  The effective mass term remains almost constant during the second transition because of the upward step-like evolution of $\eta_H$ as a function of $N_e$. In the quasi-dS approximation, $\epsilon_H \ll 1$, the effective mass term, Eq.~(\ref{eq:meff_qdS}), becomes $(aH)^{-2} z''/z \simeq 2-3\eta_H+\eta_H^2+\tau \eta_H'$. If $\eta_H$ has the form  $$\eta_H = \f{3}{2} + C \, \tanh{\l[C\l(N_e - \tilde{N_e}\r)\r]} \,$$ where $C$ is a constant and $\tilde{N_e}$ is the value of $N_e$ at which $\eta_H = 3/2$, \ie~an upward step, then $(aH)^{-2} z''/z = C^2-1/4 = {\rm const.} \, $. Note that the effective mass term is only constant for an upwards step in $\eta_{H}$, and not for a downward step, as occurs at the first transition.}, as emphasized in Ref.~\cite{Karam:2022nym}. The evolution of the mode functions (and hence the noise matrix elements)  is determined by  $(aH)^{-2} z''/z$   through the MS Eq.~(\ref{eq:MS_modes}). The expression for the mode functions therefore remains the same in the subsequent CR phase through the second transition  because of the duality first noticed by Wands (see Ref.~\cite{Wands:1998yp}). Hence  it is only necessary to follow the evolution   through the first transition, T-I, from SR-I to the near-USR phase.

  In what follows, we will first describe how to compute the noise matrix elements $\Sigma_{ij}$ numerically for  the potential Eq.~(\ref{eq:KKLT_gaussian_bump}), before  finding accurate analytic solutions for them. Note that we use this particular model to demonstrate our numerical framework because of its mathematical simplicity and efficiency. However,  the results we present are representative of models with a broad range of features, including inflection point-like behaviour. This is because, as shown in Ref.~\cite{Karam:2022nym}, the behaviour of the effective mass term $z''/z$ is similar across these large class of models, hence our primary conclusions will apply to all of them, and not just this modified KKLT model.
 
\subsubsection{Numerical analysis}
\label{sec:SI_noise_numerical}

In order to numerically compute the noise matrix elements for the  potential Eq.~(\ref{eq:KKLT_gaussian_bump}), our strategy is to split the mode functions $\phi_k$, $\pi_k$ and $v_k$ into their real and imaginary parts (see Ref.~\cite{Bhatt:2022mmn})
 
 \beq
 v_k = v^{(R)}_k + i \, v^{(I)}_k~,~~~ \phi_k = \phi^{(R)}_k + i \, \phi^{(I)}_k~,~~~ \pi_k = \pi^{(R)}_k + i \, \pi^{(I)}_k~.
 \label{eq:phik_pik_vk}
 \eeq
We impose  Bunch-Davies initial conditions Eq.~(\ref{eq:MS_BD}) deep in the sub-Hubble regime, $k \gg aH$, to obtain  $v^{(R)}_k $, $v^{(I)}_k $. Using  Eq.~(\ref{eq:modef_gen}), we then find the real and imaginary parts of $\phi_k$ and $\pi_k$:
\ber
\phi_k \equiv \phi^{(R)}_k + i \, \phi^{(I)}_k &=& \f{1}{a} \, \l[ \, v^{(R)}_k + i \, v^{(I)}_k  \, \r]~, \label{eq:phi_k_RI} \\
\pi_k \equiv \pi^{(R)}_k + i \, \pi^{(I)}_k &=& \f{1}{a} \, \l[ \, \l( \, \f{v'^{(R)}_k}{aH} - v^{(R)}_k  \, \r)  + i \, \l( \,  \f{v'^{(I)}_k}{aH} - v^{(I)}_k  \, \r)  \, \r]~.
\label{eq:pi_k_RI}
\eer

\begin{figure}[t]
\begin{center}
\includegraphics[width=0.8\textwidth]{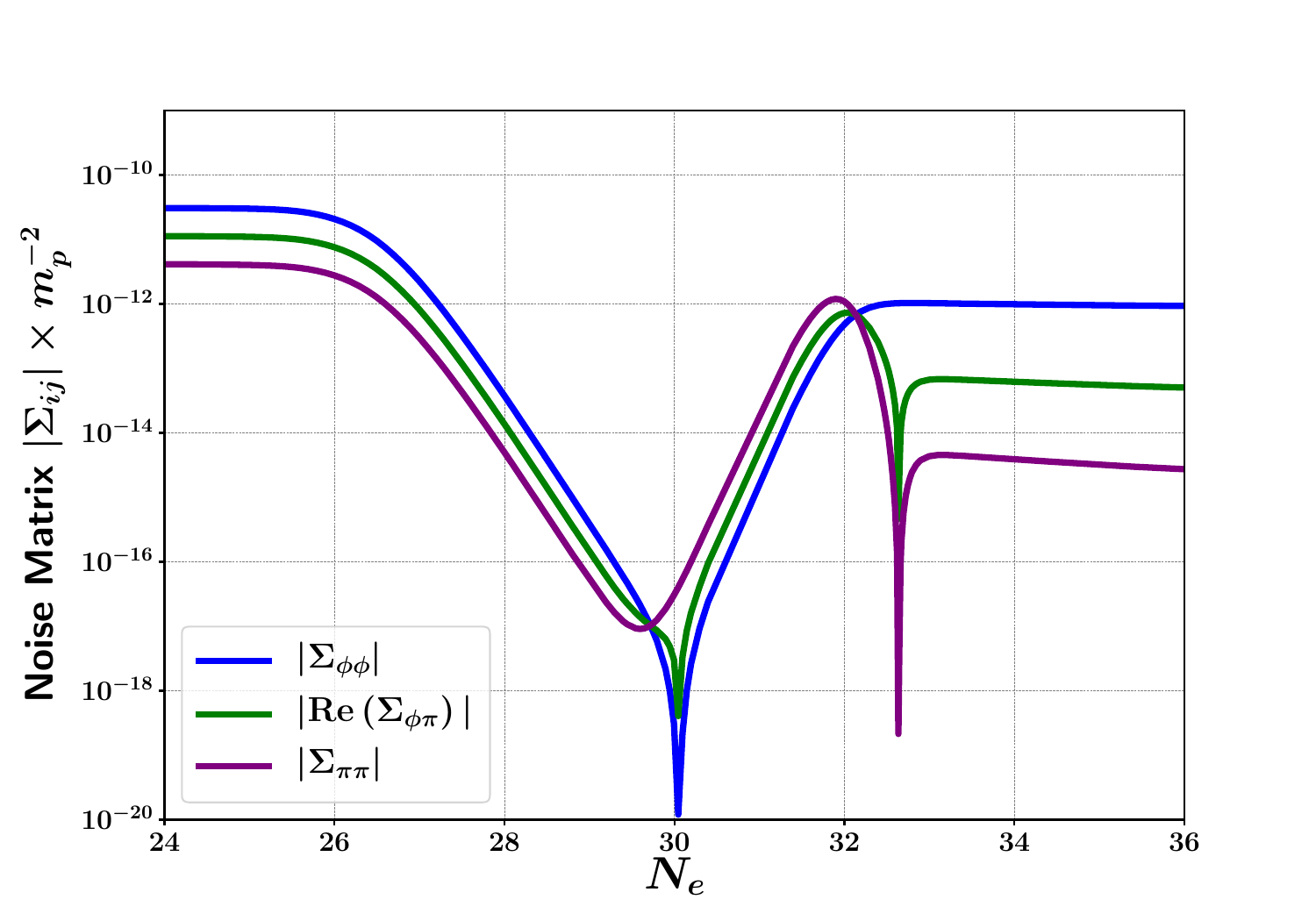}
\caption{The numerically determined  noise matrix elements, $\Sigma_{ij}$, with $\sigma =0.01$, for the  modified KKLT potential with a tiny Gaussian bump, Eq.~(\ref{eq:KKLT_gaussian_bump}), leading to a realistic smooth transition from SR-I to a near-USR phase. (Note that the plot  shows the behaviour of $\Sigma_{ij}$ only in the vicinity of the USR regime.)  The transition leads to an enhancement of the  momentum induced noise terms, $\Sigma_{\phi \pi}$ and $\Sigma_{\pi \pi}$,  relative to the field noise, $\Sigma_{\phi \phi}$,  in the USR epoch.}
\label{fig:Sig_SR_USR_KKLT_GB}
\end{center}
\end{figure}
Substituting  Eqs.~(\ref{eq:phik_pik_vk})-(\ref{eq:pi_k_RI}) into Eq.~(\ref{eq:noise_comp_cor_matrix}), we derive the following compact expressions for the noise matrix elements  $\Sigma_{ij}$
\ber
 \Sigma_{\phi \phi} &=& \l( 1- \epsilon_H \r)\f{k^3}{2\pi^2} \times \f{1}{a^2} \times \l[ \, \l( v^{(R)}_k \r)^2 + \l( v^{(I)}_k \r)^2   \, \r] \Bigg\vert_{k = \sigma aH} ~~  \label{eq:Sig_phiphi_RI} \\
{\rm Re} \l( \Sigma_{\pi \phi} \r) &=&  {\rm Re} \l( \Sigma_{\phi \pi} \r) = \l( 1- \epsilon_H \r) \f{k^3}{2\pi^2} \times \f{1}{a^2} \times \l[   v^{(R)}_k \l(  \f{v'^{(R)}_k}{aH} - v^{(R)}_k  \r) +   v^{(I)}_k \l(  \f{v'^{(I)}_k}{aH} - v^{(I)}_k  \r)  \r] \Bigg\vert_{k = \sigma aH}  ~~\label{eq:Sig_phipi_RI} \\
 \Sigma_{\pi \pi} &=& \l( 1- \epsilon_H \r) \f{k^3}{2\pi^2} \times \f{1}{a^2} \times \l[ \,  \l(  \f{v'^{(R)}_k}{aH} - v^{(R)}_k  \r)^2 +  \l(  \f{v'^{(I)}_k}{aH} - v^{(I)}_k  \r)^2    \, \r] \Bigg\vert_{k = \sigma aH} \, . \label{eq:Sig_pipi_RI} 
\eer

\begin{figure}[htb]
\begin{center}
\includegraphics[width=0.8\textwidth]{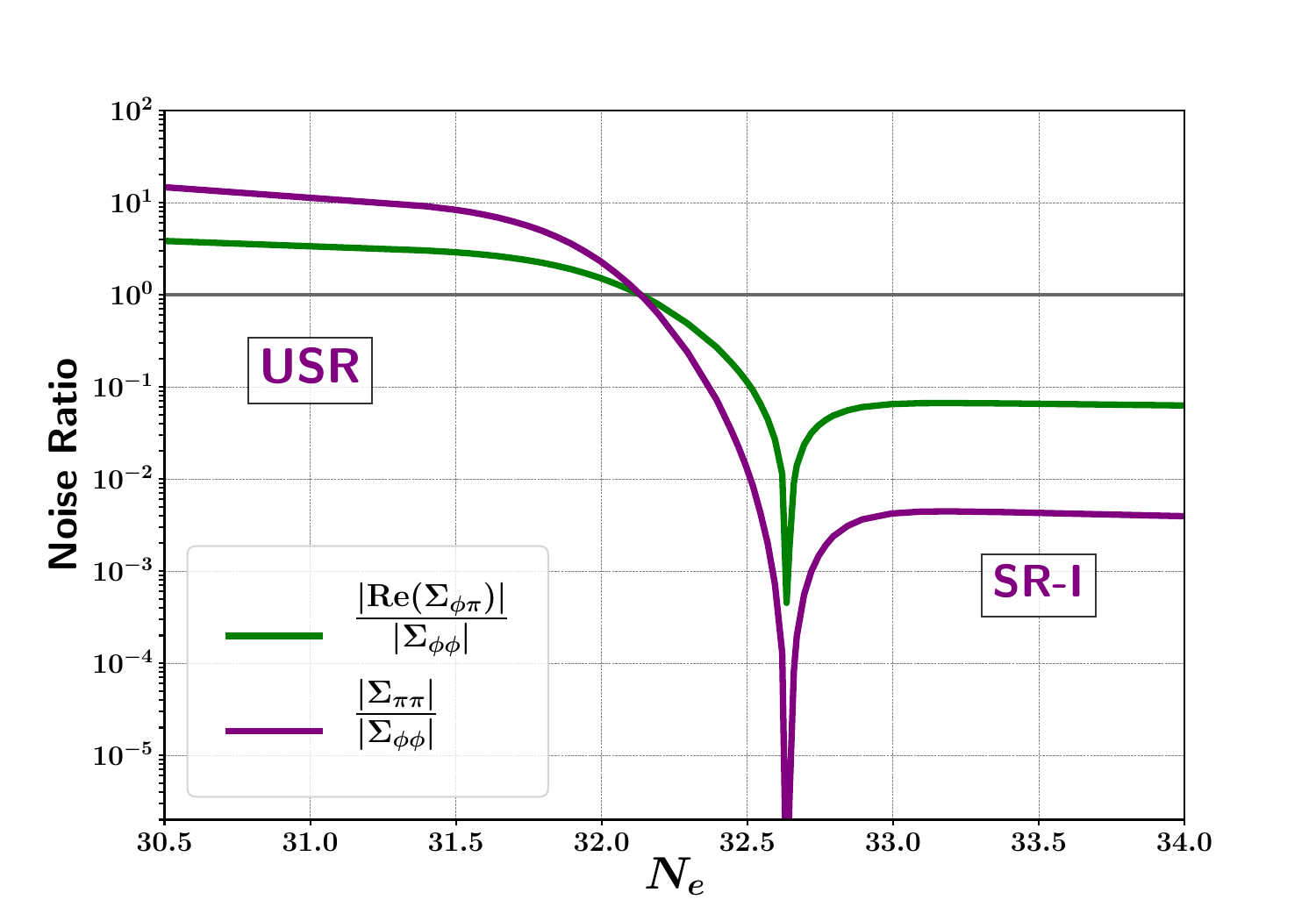}
\caption{The ratios of the momentum-induced noise terms and the field noise, $|{\rm Re}(\Sigma_{\phi \pi})|/ |\Sigma_{\phi \phi}|$ in green
and $|\Sigma_{\pi \pi}|/ |\Sigma_{\phi \phi}|$ in purple,  with $\sigma=0.01$, for the potential Eq.~(\ref{eq:KKLT_gaussian_bump}) with a tiny Gaussian bump as a function of $N_{\rm e}$ around the SR-I to USR transition.
The transition from SR-I to USR leads to an enhancement of the  momentum induced noise terms, $\Sigma_{\phi \pi}$ and $\Sigma_{\pi \pi}$,  relative to the field noise, $\Sigma_{\phi \phi}$, in the USR epoch.}
\label{fig:Sig_SR_USR_KKLT_GB_zoom}
\end{center}
\end{figure}

 As we mentioned earlier, the imaginary part of the off-diagonal term
$\Sigma_{\pi \phi}$ does not correspond to a stochastic classical noise source \cite{Grain:2017dqa}, hence we only need consider its real part in Eq.~(\ref{eq:Sig_phipi_RI}). The evolution of the absolute values of $\Sigma_{\phi \phi}$, Re($\Sigma_{\phi \pi}$) and $\Sigma_{\pi \pi}$ for the potential Eq.~(\ref{eq:KKLT_gaussian_bump}) are plotted in Fig.~\ref{fig:Sig_SR_USR_KKLT_GB} for  $\sigma=0.01$,   while Fig.~\ref{fig:Sig_SR_USR_KKLT_GB_zoom}  shows the ratios between the momentum-induced noise terms and the field noise,  $|{\rm Re}(\Sigma_{\phi\pi})|/|\Sigma_{\phi \phi}|$ and $|\Sigma_{\pi \pi}|/|\Sigma_{\phi \phi}|$ around the transition epoch. The transition leads to an enhancement of the  momentum induced noise terms   relative to the field noise with $\Sigma_{\pi \pi} > |{\rm Re}(\Sigma_{\phi \pi})|
>  \Sigma_{\phi \phi}$. This is followed by  a near-exponential   fall of each $\Sigma_{ij}$ during USR, since the slope of $\Sigma_{ij}$ is almost constant during this epoch.    We see that $|\Sigma_{\pi \pi}|/|\Sigma_{\phi \phi}| \gtrsim 3 \times |{\rm Re}(\Sigma_{\phi\pi})|/|\Sigma_{\phi \phi}|$. At late times the noise matrix elements begin to  rise again and asymptote to constant values, and the   hierarchy between the noise terms gets reversed back to  $\Sigma_{\pi \pi} < |{\rm Re}(\Sigma_{\phi \pi})| <  \Sigma_{\phi \phi}$. We also notice that the asymptotic value of each $\Sigma_{ij}$ at late times is greater than its corresponding value in the SR-I phase. 

 From Figs.~\ref{fig:Sig_SR_USR_KKLT_GB} and \ref{fig:Sig_SR_USR_KKLT_GB_zoom}, we conclude that the noise matrix elements for a potential with a PBH-forming feature evolve in a more complicated way than for pure de Sitter or pure slow-roll. We next show that  the aforementioned  interesting features  of the noise terms  across different epochs, such as during SR-I, immediately after the transition from SR-I to USR, as well as the late time asymptote, can be understood by making appropriate analytical approximations. In the following subsection, we compute the noise matrix elements analytically by assuming the transition T-I from SR-I to the near-USR phase to be instantaneous.  We will also demonstrate that in the quasi-dS limit, $\epsilon_H \ll 1$, the   noise terms are completely determined by  the second slow-roll parameter $\eta_H$.

\subsubsection{Analytical treatment for instantaneous transitions}
\label{sec:SI_noise_analytical}

 In order to compute $\Sigma_{ij}$ analytically, we  consider an approach which captures the key features of the full numerical evolution, namely solving  the MS Eq.~(\ref{eq:MS_T_nu}) under the following assumptions.
 
 \begin{enumerate}
 \item We assume  the second slow-roll parameter $\eta_H$ to be a piece-wise constant function which makes an  instantaneous (yet finite) transition, $\eta_H: \, \eta_1 \to \eta_2$ at time $\tau = \tau_1$, given by
 \beq
\eta_H (\tau) = \eta_1 + \l( \eta_2 - \eta_1 \r) \, \Theta(\tau-\tau_1)~,
\label{eq:eta_T1}
\eeq
where $\Theta$ is the Heaviside  step function:
\beq
\Theta(\tau-\tau_1) =     \begin{cases}  ~ 0 \, , & \tau < \tau_1 ~ ,\\ ~ 1 \, ,   &  \tau  > \tau_1 ~. \end{cases}  
\label{eq:Theta}
\eeq
\item The corresponding expression for   $\nu$  given in Eq.~(\ref{eq:z_nu_rel}) is then determined using Eq.~(\ref{eq:MS_zdd}), which under the quasi-dS approximation, $\epsilon_H \simeq 0$, becomes Eq.~(\ref{eq:meff_qdS}). Using the expression for $\eta_H$ from Eq.~(\ref{eq:eta_T1}) in Eq.~(\ref{eq:meff_qdS}), we obtain 

\beq
\nu^2 -\f{1}{4} \equiv \f{z''}{z} \, \tau^2 =  {\cal A} \, \tau \, \delta_D(\tau-\tau_1) +  \nu_1^2 - \f{1}{4} + \l( \nu_2^2 - \nu_1^2 \r) \, \Theta(\tau-\tau_1) \, ,
\label{eq:meff_qdS_eta_T1}
\eeq 
where 
\beq
{\cal A} = \eta_2 - \eta_1 \, ,~~~ \nu_{1,2}^2 - \f{1}{4} = 2 - 3 \, \eta_{1,2} + \eta_{1,2}^2 \, .
\label{eq:A_nu_etaH_T1}
\eeq
Hence the piece-wise constant $\eta_H$ in Eq.~(\ref{eq:eta_T1}) results in  a piece-wise constant $\nu$ in Eq.~(\ref{eq:meff_qdS_eta_T1}). We notice that the effective mass term $z''/z$ contains a Dirac delta-function  arising from the derivative of the $\Theta$ function in Eq.~(\ref{eq:eta_T1}). Note that for $\eta_2 > \eta_1$ (which is the case for the SR-I $\to$ USR transition in Fig.~\ref{fig:inf_meff_GB}) we have ${\cal A} > 0$ and hence the term containing the Dirac delta-function in Eq.~(\ref{eq:meff_qdS_eta_T1}) is negative (since $\tau < 0$ during inflation). This delta-function dip for an instantaneous transition analytically represents the observed dip of finite width and depth for potentials with a smooth feature, as seen in $ (1/ aH)^2  z''/z$ in Fig.~\ref{fig:inf_meff_GB}  (around $N_e \sim 32.5$). 

 \item We impose Bunch-Davies initial conditions, Eq.~(\ref{eq:MS_BD}), only for modes that become super-Hubble at early times before the transition \textit{i.e.} $\tau < \tau_1$.
 
 \item General solutions to the MS equation  in different piece-wise constant $\nu$ regimes are matched during the transition $\tau = \tau_1$ by using the Israel Junction conditions \cite{Deruelle:1995kd,Byrnes:2018txb,Ahmadi:2022lsm}
 \ber
v_k^L(\tau_1) &=&  v_k^E(\tau_1) ~~~~~~~~~~~~~~~~~ ({\rm Continuity})   ~, \label{eq:Israel_USR1} \\
\f{{\rm d}}{{\rm d}\tau} v_k^L\bigg\vert_{\tau_1^+} -  \f{{\rm d}}{{\rm d}\tau} v_k^E\bigg\vert_{\tau_1^-}  &=&   \int_{\tau_1^-}^{\tau_1^+} {\rm d}\tau \f{z''}{z}\,v_k^L(\tau)  ~~~ ({\rm Differentiability}) \, , \label{eq:Israel_USR2} 
\eer  
where $v_k^E(\tau)$ and $v_k^L(\tau)$ are  the mode functions before and after the transition respectively, represented  by 
\begin{align}
  v_k (\tau)  \, &=  \,  \begin{cases} ~  v_k^{E}(\tau)  \, , & \tau < \tau_1 ~ ,\\ 
                                     ~  v_k^{L}(\tau)  \, , &  \tau > \tau_1 ~. 
                                          \end{cases} 
\label{eq:vk_jump_gen}
\end{align}

 \end{enumerate}

We would ultimately  like to derive expressions for the  noise matrix elements which can be  expressed  in terms of the mode functions $v_k$ in the following compact form 
\ber
 \Sigma_{\phi \phi} &=&   \l( \f{H}{2\pi} \r)^2 \, T^2 \,  \Big\vert  \, \sqrt{2k} \, v_k (T) \, \Big\vert ^2  \bigg\vert_{T   =   \sigma }  ,  \label{eq:Sigma_phi_phi_gen_T} \\
 {\rm Re} \l( \Sigma_{\pi \phi} \r)  &=& -   \l( \f{H}{2\pi} \r)^2  T^2     {\rm Re} \l( \sqrt{2k}  v_k^* (T)  \l[  T  \f{{\rm d}}{{\rm d}T}  \l( \sqrt{2k} v_k(T) \r) + \sqrt{2k}  v_k (T) \r] \r) \bigg\vert_{T   =   \sigma }   ,  \label{eq:Sigma_pi_phi_gen_T} \\
 \Sigma_{\pi \pi} &=&   \l( \f{H}{2\pi} \r)^2 \, T^2 \,  \Big\vert  \, T \, \f{{\rm d}}{{\rm d}T} \l( \sqrt{2k} \,v_k(T) \r) + \sqrt{2k} \, v_k (T)  \Big\vert ^2   \bigg\vert_{T   =   \sigma }  ,
  \label{eq:Sigma_pi_pi_gen_T}
\eer
where we  take $\sigma = 0.01$ as discussed earlier. We start with the computation of noise matrix elements for an instantaneous transition in the pure dS limit where $\nu_1 =  \nu_2 = 3/2$, before moving on to a general transition between constant values of $\nu: \, \nu_1 \to \nu_2$, with $\nu_2 > \nu_1$.

\medskip

\begin{center}
{\bf {\large Case 1:} Instantaneous transition in the pure dS limit with $\nu_1 = \nu_2 = \f{3}{2}$}
\end{center}
\medskip

In the case of an instantaneous transition at $\tau=\tau_1$ in the pure dS limit (first considered in Ref.~\cite{Starobinsky:1992ts}), we have $\eta_1=0$ and $\eta_2=3$ and the system makes a transition from a SR to an exact USR phase. Accordingly, the effective mass term in the MS equation takes the form
\beq
 \f{z''}{z} \, \tau^2=  {\cal A} \, \tau \,  \delta_D(\tau-\tau_1) \, +  \, 2 \, ,
\label{eq:meff_jump_dS_dS}
 \eeq
where  the transition strength is ${\cal A}= 3$.  The expressions for the  mode functions, obtained by solving Eq.~(\ref{eq:MS_modes}) are given (in terms of $T = -k \tau$) by 
\begin{flalign}
v_k (T)  \equiv  \begin{cases} ~  v_k^{E}(T) = \f{1}{\sqrt{2k}} \, \l( 1 + \f{i}{T} \r) \, e^{i \, T}  \, , & T > T_1 ~ ,\\
 v_k^{L}(T) = \f{1}{\sqrt{2k}} \, \l[  \, \alpha_k \, \l( 1 + \f{i}{T} \r) \, e^{i \, T}  \, +   \,  \beta_k \, \l( 1 - \f{i}{T} \r) \, e^{-i \, T}  \, \r] \, , &  T < T_1 \,, \end{cases}
\label{eq:vk_jump_dS_dS}
\end{flalign}
 where $\alpha_k$ and $\beta_k$ are constants of integration (to be determined from  the Israel junction conditions given in Eqs.~(\ref{eq:Israel_USR1}) and (\ref{eq:Israel_USR2})), 
while their derivatives  are given by 
\begin{flalign}
 \f{{\rm d}v_k}{{\rm d}T} \equiv  \begin{cases} ~  \f{{\rm d}v_k^E}{{\rm d}T}  = \f{1}{\sqrt{2k}} \, \l[ -\f{1}{T} + i \, \l( 1 - \f{1}{T^2}  \r) \r] \, e^{i \, T}  \, , & T > T_1 ~ ,\\ 
 \f{{\rm d}v_k^L}{{\rm d}T}  = \f{1}{\sqrt{2k}}  \l[   \alpha_k  \l[ -\f{1}{T} + i  \l( 1 - \f{1}{T^2}  \r) \r]  e^{i \, T}   +     \beta_k  \l[ -\f{1}{T} - i  \l( 1 - \f{1}{T^2}  \r) \r]  e^{-i \, T}   \r] \, , &  T < T_1 \, , \end{cases}
\label{eq:vkT_jump_dS_dS}
\end{flalign}
 where recall $T>T_1$  corresponds to the epoch before the transition  and $T<T_1$ to the epoch after the transition. Note that we have  imposed Bunch-Davies initial conditions on the mode function $v_k^{E}(T)$ before the transition $T>T_1$, in accordance with our  third assumption as discussed above. 
The corresponding Fourier modes of the field fluctuations are  obtained from Eq.~(\ref{eq:modef_gen})  
\begin{flalign}
\phi_k(T) \, =  \,  \begin{cases} ~  \phi_k^E(T)  = \f{H}{\sqrt{2k^3}} \, \l( T + i \r)   \, e^{i \, T}  \, , & T > T_1 ~ ,\\ 
\phi_k^L(T) =  \f{H}{\sqrt{2k^3}} \, \l[  \, \alpha_k \, \l( T + i \r)   \, e^{i \, T}  \, +   \,  \beta_k \, \l( T - i \r)   \, e^{-i \, T}  \, \r] \, , &  T < T_1 ~. \end{cases} 
\label{eq:phikT_jump_dS_dS}
\end{flalign}
 as are those of the field momentum fluctuations
\begin{flalign}
\pi_k(T) \, =  \,  \begin{cases} ~  \pi_k^E(T)  =  - \f{H}{\sqrt{2k^3}} \, i\, T^2   \, e^{i \, T}  \, , & T > T_1 ~ ,\\ 
\pi_k^L(T) = -  \f{H}{\sqrt{2k^3}} \,  i\, T^2   \, \l[  \, \alpha_k  \,   e^{i \, T}  -  \beta_k \,  e^{-i \, T}  \, \r] \, , &  T < T_1 ~. \end{cases}
\label{eq:pikT_jump_dS_dS}
\end{flalign}

\begin{figure}[H]
\begin{center}
\hspace{-0.35in}
\subfigure[][]{
\includegraphics[width=0.54\textwidth]{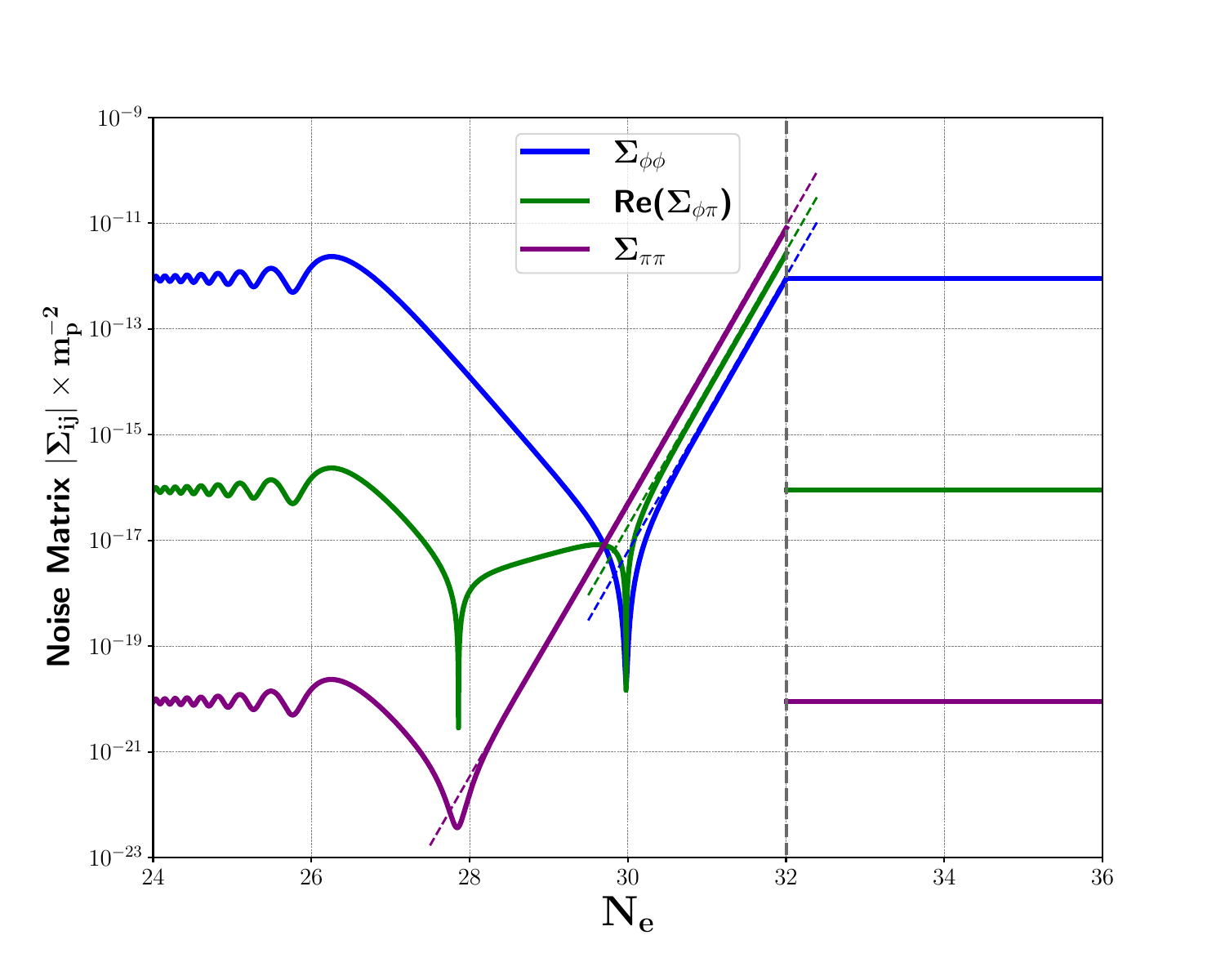}}
\hspace{-0.4in}
\subfigure[][]{
\includegraphics[width=0.54\textwidth]{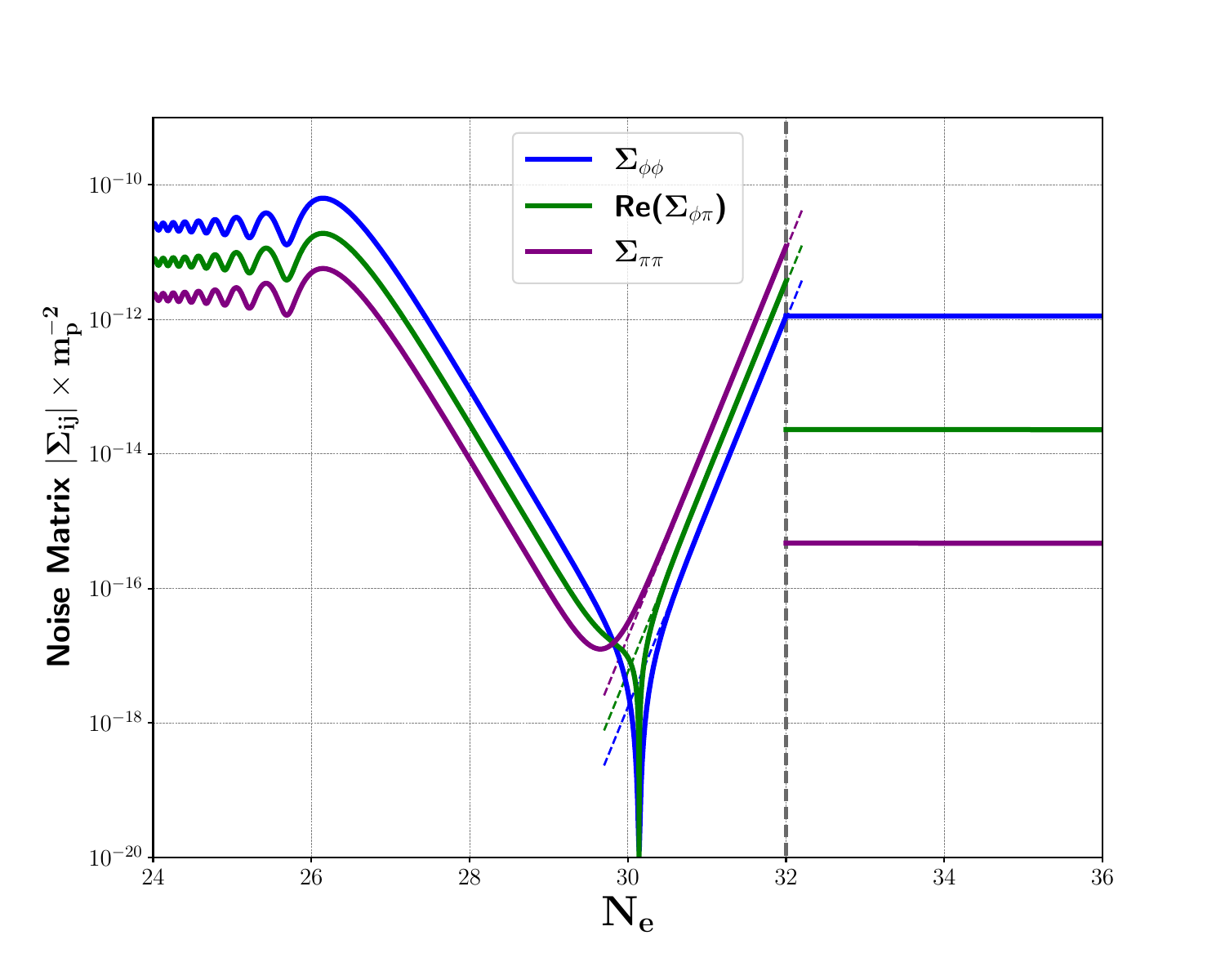}}
\caption{The noise matrix elements $\Sigma_{ij}$ computed analytically using Eqs.~(\ref{eq:Sigma_phi_phi_gen_T})-(\ref{eq:Sigma_pi_pi_gen_T}), for $\sigma = 0.01$. {\bf Left panel}: an instantaneous transition from SR to USR using the pure de Sitter limit, $\nu_1 = \nu_2 = 3/2$. {\bf Right panel}: an instantaneous transition from  a SR phase with $\nu_1 =1.52$ to a near-USR phase with  $\nu_2 = 1.8$. In both cases the thin dashed lines show the analytical asymptotes immediately after the transition, $\Sigma_{ij} \sim e^{2{\cal A}N_e}$.}
\label{fig:Sig_singleT_asymp}
\end{center}
\end{figure}

The Bogolyubov  coefficients $\alpha_k$ and $\beta_k$,  determined by implementing the Israel junction conditions, Eqs.~(\ref{eq:Israel_USR1}) and  (\ref{eq:Israel_USR2}), are given by
\begin{align}
 a_1 \, \alpha_k \, + \,  b_1 \, \beta_k \,   &=  \, d_1 & \\
 a_2 \,  \alpha_k  \, +  \, b_2 \, \beta_k  \,  &=  \, d_2 ~, 
 \end{align}
 which yields 
\begin{align}
  \alpha_k = \f{ d_1 \, b_2 - d_2 \, b_1  }{ a_1 \, b_2 - a_2 \, b_1   } ~,~~~   \beta_k = \f{ d_2 \, a_1 - d_1 \, a_2  }{ a_1 \, b_2 - a_2 \, b_1   } ~, 
\end{align}
where 
\begin{align}
a_1 &=   T_1 + i   & \\
b_1 &=   \l( T_1 - i \r) \, e^{ -i \, 2T_1 }   & \\
a_2 &=  \l( 1 + {\cal A} \r) \, T_1 + i \,   \l( 1 + {\cal A} - T_1^2 \r) & \\
b_2 &=  \l( 1 + {\cal A} \r) \, T_1 - i \,   \l( 1 + {\cal A} - T_1^2 \r)  \, e^{ -i \, 2T_1 } & \\ 
d_1 &=    T_1 + i  & \\  
d_2 &=   T_1 + i \, \l( 1 - T_1^2 \r)
\end{align}
With a little bit of algebra, we obtain\footnote{Eqs.~(\ref{eq:alphak_dS_dS}) and (\ref{eq:betak_dS_dS}) agree with the results obtained in Refs.~\cite{Starobinsky:1992ts,Ahmadi:2022lsm} for an inflaton potential consisting of two linear regimes $V(\phi) \propto \phi$ and slopes $\alpha$, $\beta$ that are joined at a point $\phi = \phi_1$  in which case  $T_1$ is related to $\phi_1$ and ${\cal A} = 3(\alpha-\beta)/\alpha$.} 
\ber
 \alpha_k &=& 1 - i \, \f{{\cal A}}{2} \, \f{1}{T_1} \, \l( 1 + \f{1}{T_1^2} \r)  \, , \label{eq:alphak_dS_dS} \\
\beta_k &=& i \, \f{{\cal A}}{2} \, \f{1}{T_1} \, \l( 1 + \f{i}{T_1} \r)^2 \, e^{ i \, 2T_1 }  \, , \label{eq:betak_dS_dS}
\eer
 From the above expressions, we can compute the noise matrix elements  $\Sigma_{\phi\phi}, \, \Sigma_{\phi\pi}, \, \Sigma_{\pi\pi}$ in Eqs.~(\ref{eq:Sigma_phi_phi_gen_T}) - (\ref{eq:Sigma_pi_pi_gen_T}).  The instantaneous transition leads to a mixing between the positive and negative frequency solutions for the mode functions. Our analytical results are shown in the left panel of Fig.~\ref{fig:Sig_singleT_asymp} for ${\cal A} = +3$ \footnote{ The noise matrix elements for the case of an instantaneous transition  were also estimated in Ref.~\cite{Ahmadi:2022lsm} under the  pure dS approximation for the Starobinsky model~\cite{Starobinsky:1992ts}.  Using our calculations, with a transition strength in the range ${\cal A} \leq 3$,  one can obtain the noise matrix elements for   the entire parameter space of the Starobinsky model.}.  The  asymptotic behaviour of $\Sigma_{ij}$ at different epochs  can be inferred from the scale dependence of  the Bogolyubov coefficients, given by Eqs.~(\ref{eq:alphak_dS_dS}) and (\ref{eq:betak_dS_dS}). Note that different comoving modes contribute to the noise terms at different times. Immediately after the transition,  when $T \lesssim T_1 \ll 1$, the noise matrix elements are due to modes joining the coarse-graining scale $T=\sigma$ at this epoch, for which,  using Eqs.~(\ref{eq:alphak_dS_dS}) and (\ref{eq:betak_dS_dS}), $\alpha_k, \, \beta_k  \simeq -3i/(2T_1^{3})$. Consequently, the noise terms from Eqs.~(\ref{eq:Sigma_phi_phi_gen_T})-(\ref{eq:Sigma_pi_pi_gen_T}) behave as $\Sigma_{ij} \propto T_1^{-6}$. Since $T_1 = k/(a_1H_1)$,  $\Sigma_{ij} \propto k^{-6} \propto  e^{6 N_e}$, 
 and the ratio $\Sigma_{\phi\phi}:|{\rm Re(\Sigma_{\phi\pi}})|:\Sigma_{\pi\pi}$ is given by $1:3:9$, to leading order in $T=\sigma$.  Following this epoch, the noise terms begin to rise exponentially, and the hierarchy between the field and momentum induced terms  gets reversed back to $\Sigma_{\phi\phi} > |{\rm Re(\Sigma_{\phi\pi}})| > \Sigma_{\pi\pi}$.  At sufficiently late times, when $T \ll  1 \ll T_1$, the noise matrix elements are due to modes for which  $\alpha_k \to 1$,  while  $\beta_k \simeq 3i/(2T_1) \, e^{i 2T_1}$ decays to zero with oscillations. Hence the noise matrix elements   asymptote to their corresponding  pre-transition (constant) values given by Eqs.~(\ref{eq:Sig_phiphi_dS})-(\ref{eq:Sig_pipi_dS}).

 Comparing the analytical results for an instantaneous transition in the pure dS limit, shown in the left panel of Fig.~\ref{fig:Sig_singleT_asymp}, with  the numerical results for a potential with a PBH forming feature shown in Fig.~\ref{fig:Sig_SR_USR_KKLT_GB}, we conclude that the former fails to capture\footnote{Note, however, that the pure dS-transition is a good approximation   for potentials featuring   a  `flat' segment  (as shown in Fig.~\ref{fig:inf_flat_Q_well}) rather than a bump. In this case  Eqs.~(\ref{eq:vk_jump_dS_dS})-(\ref{eq:betak_dS_dS}) can be used to compute the noise terms (see Ref.~\cite{Ahmadi:2022lsm}).} the late time asymptotic properties of $\Sigma_{ij}$. Therefore, in the following, we will compute $\Sigma_{ij}$ relaxing the pure dS approximation.

\medskip

\begin{center}
{\bf {\large Case 2:}  Instantaneous transition between two constant values of $\nu$: $\nu_1 \to \nu_2$}
\end{center}

\medskip

In the case of an instantaneous transition at $\tau=\tau_1$  where  $\nu$ makes a jump\footnote{ Earlier  work on  instantaneous transition during inflation beyond the pure de Sitter approximation (when $\nu \neq 3/2$) can be found in Refs.~\cite{Joy:2007na,Hazra:2014goa,Hazra:2021eqk}.} between the constant values $\nu_1 \to \nu_2$, once again by solving Eq.~(\ref{eq:MS_modes}), the expressions for the  mode functions are given (in terms of $T = -k \tau$) by 
\begin{flalign}
 v_k (T)  \, =  \,  \begin{cases} ~  \f{1}{\sqrt{2k}} \, \sqrt{\f{\pi}{2}}  \, e^{i \l(  \nu_1 + \f{1}{2}  \r) \f{\pi}{2}}  \, \sqrt{T} \,  H_{\nu_1}^{(1)}(T)   \, , & T > T_1 ~ ,\\ ~    \sqrt{T} \l[  \, C_1^{L} \, H_{\nu_2}^{(1)}(T) \, +   \, C_2^{L} \, H_{\nu_2}^{(2)}(T) \, \r] \, , &  T < T_1 ~, \end{cases} 
 \label{eq:vk_jump_nu1_nu2}
\end{flalign}
and their derivatives are given by 

\begin{flalign}
\f{{\rm d}v_k}{{\rm d}T}    =    \begin{cases} ~  \f{1}{\sqrt{2k}}  \sqrt{\f{\pi}{2}}   e^{i \l(  \nu_1 + \f{1}{2}  \r) \f{\pi}{2}}   \f{1}{\sqrt{T}}  \l[  \l( \f{1}{2} - \nu_1 \r)   H_{\nu_1}^{(1)}(T) + T   H_{\nu_1-1}^{(1)}(T)  \r] \, , & T > T_1 \\ ~    \f{1}{\sqrt{T}}  \l[  C_1^{L}  \l(  \l( \f{1}{2} - \nu_2 \r)   H_{\nu_2}^{(1)}(T) + T   H_{\nu_2-1}^{(1)}(T)  \r) +    C_2^{L} \l(  \l( \f{1}{2} - \nu_2 \r)   H_{\nu_2}^{(2)}(T) + T   H_{\nu_2-1}^{(2)}(T)  \r)  \r] \, , &  T < T_1   \end{cases}  
 \label{eq:vkT_jump_nu1_nu2}
\end{flalign}

By implementing the Israel junction conditions, Eqs.~(\ref{eq:Israel_USR1}) and  (\ref{eq:Israel_USR2}), the constant coefficients of integration $C_1^{L}$ and $C_2^{L}$ can be shown to satisfy the algebraic equations
\begin{align}
 a_1 \,  C_1^{L} \, + \,  b_1 \, C_2^{L} \,   &=  \, d_1 & \\
 a_2 \,  C_1^{L}  \, +  \, b_2 \, C_2^{L}  \,  &=  \, d_2 ~, 
 \end{align}
which yields 
\begin{align}
   C_1^{L} = \f{ d_1 \, b_2 - d_2 \, b_1  }{ a_1 \, b_2 - a_2 \, b_1   } ~,~~~   C_2^{L} = \f{ d_2 \, a_1 - d_1 \, a_2  }{ a_1 \, b_2 - a_2 \, b_1   } ~, 
\end{align}
where 
\begin{align}
a_1 &=   H_{\nu_2}^{(1)}(T_1)   &\\
b_1 &=   H_{\nu_2}^{(2)}(T_1)   &\\
a_2 &=  \l( \, \f{1}{2} - \nu_2    - {\cal A} \, \r)  \,  H_{\nu_2}^{(1)}(T_1) + T_1 \,  H_{\nu_2 - 1}^{(1)}(T_1) & \\
b_2 &=  \l( \, \f{1}{2} - \nu_2    - {\cal A} \, \r)  \,  H_{\nu_2}^{(2)}(T_1) + T_1 \,  H_{\nu_2 - 1}^{(2)}(T_1)  & \\ 
d_1 &=  \f{1}{\sqrt{2k}} \, \sqrt{\f{\pi}{2}}  \,   e^{i \l(  \nu_1 + \f{1}{2}  \r) \f{\pi}{2}} \, H_{\nu_1}^{(1)}(T_1)  & \\  
d_2 &=    \f{1}{\sqrt{2k}} \, \sqrt{\f{\pi}{2}}  \,   e^{i \l(  \nu_1 + \f{1}{2}  \r) \f{\pi}{2}} \, \l[ \, \l( \f{1}{2} - \nu_1 \r)  \,  H_{\nu_1}^{(1)}(T_1) + T_1 \,  H_{\nu_1 - 1}^{(1)}(T_1)   \, \r]  
\end{align}

 The resulting noise matrix elements, computed  using  Eqs.~(\ref{eq:Sigma_phi_phi_gen_T})-(\ref{eq:Sigma_pi_pi_gen_T}), are shown in the right panel of Fig.~\ref{fig:Sig_singleT_asymp}.  In order to compare our results with the numerical calculation in Fig.~\ref{fig:Sig_SR_USR_KKLT_GB}, we choose $\nu_1 = 1.52$ and $\nu_2 =  1.8$. These values correspond to $\eta_1 = -0.02$ and $\eta_2 = 3.3$ respectively, to match the values of $\eta_H$ during  the SR-I and the near-USR epochs for the modified KKLT potential with a Gaussian bump used for the numerical calculation in Fig.~\ref{fig:Sig_SR_USR_KKLT_GB}.

As in the pure dS case, immediately after the transition,  when $T \lesssim T_1 \ll 1$, the noise matrix elements fall nearly-exponentially  with $\Sigma_{ij} \sim e^{2{\cal A}N_e}$.  The ratio $\Sigma_{\phi\phi}:|{\rm Re(\Sigma_{\phi\pi}})|:\Sigma_{\pi\pi}$ is approximately $ 1:{\cal A}:{\cal A}^2$ (where ${\cal A}\equiv \eta_2 - \eta_1 = 3.32$ from Eq.~(\ref{eq:A_nu_etaH_T1})), and nearly constant. However, following this epoch the noise terms begin to rise and the hierarchy between the field and momentum induced terms is reversed back to $\Sigma_{\phi\phi} > |{\rm Re}(\Sigma_{\phi\pi})| > \Sigma_{\pi\pi}$. At sufficiently late times, $T \ll  1 \ll T_1 $, the coefficient of the negative frequency solution  $C_2^L$ becomes negligible, and the  behaviour of $\Sigma_{ij}$ can be understood  from the constant $\nu$ expressions for the noise terms, Eqs.~(\ref{eq:Sig_phiphi_SR_supH})-(\ref{eq:Sig_pipi_SR_supH}). The late time ratio of noise terms is given by $\Sigma_{\phi\phi}:|{\rm Re(\Sigma_{\phi\pi}})|:\Sigma_{\pi\pi} \to  1:\l(\nu_2 - \f{3}{2}\r): \l(\nu_2 - \f{3}{2}\r)^2$ \ie~the values of $\Sigma_{ij}$ are higher than their pre-transition counterparts in the SR-I phase. This matches the behaviour of the numerically calculated noise matrix elements for the modified KKLT potential with a Gaussian bump shown in Fig.~\ref{fig:Sig_SR_USR_KKLT_GB}. 

\vspace{0.5cm}

The key results of our analytical calculations for an instantaneous transition are:
\begin{enumerate}
\item 
The expressions for the noise matrix elements in the pre-transition epoch are given by Eqs.~(\ref{eq:Sig_phiphi_SR_supH})-(\ref{eq:Sig_pipi_SR_supH}) with $\nu =  \nu_1$, resulting in the ratios  
\beq
\Sigma_{\phi\phi}:|{\rm Re(\Sigma_{\phi\pi}})|:\Sigma_{\pi\pi} \to  1:\l(\nu_1 - \f{3}{2}\r): \l(\nu_1 - \f{3}{2}\r)^2 \, .
\label{eq:Sig_pre_trans_gen}
\eeq
\item 
Immediately after the transition, $\Sigma_{ij} \propto e^{2{\cal A}N_e}$, and
\beq
\Sigma_{\phi\phi}:|{\rm Re(\Sigma_{\phi\pi}})|:\Sigma_{\pi\pi} \to  1:{\cal A}:{\cal A}^2 \,.
\label{eq:Sig_post_trans_gen}
\eeq
\item
At sufficiently late times, the noise terms are again given by Eqs.(\ref{eq:Sig_phiphi_SR_supH})-(\ref{eq:Sig_pipi_SR_supH}), but with $\nu =  \nu_2$, yielding the ratios 
\beq
\Sigma_{\phi\phi}:|{\rm Re(\Sigma_{\phi\pi}})|:\Sigma_{\pi\pi} \to  1:\l(\nu_2 - \f{3}{2}\r): \l(\nu_2 - \f{3}{2}\r)^2 \,.
\label{eq:Sig_post_trans_gen_asymp}
\eeq
\end{enumerate}

 Comparing Fig.~\ref{fig:Sig_singleT_asymp} with Fig.~\ref{fig:Sig_SR_USR_KKLT_GB}, we see  that the analytical treatment assuming an instantaneous transition  between two constant values of $\nu$, $\nu_1$ and  $\nu_2$ (Case 2) captures most of the  asymptotic properties of $\Sigma_{ij}$ for a potential with a PBH forming feature. This is in contrast to the pure dS transition (Case 1) which was not able to  capture the late-time asymptote accurately, due to the assumption that $\nu_1 = \nu_2 = 3/2$.  Furthermore, the pure dS transition also underestimates  the momentum induced noise terms $\Sigma_{\phi\pi}$ and $\Sigma_{\pi\pi}$ in the SR-I phase, as discussed in Sec.~\ref{sec:Sig_SR}.
 
\medskip

\medskip

  We conclude  this Section by briefly commenting on the degree of correlation between the field and momentum noise terms, $ \xi_{\phi}$ and $ \xi_{\pi}$, which can be quantified in terms of the ratio\footnote{We thank David Wands for pointing out the relevance of calculating $\gamma$.} 
 \beq
\gamma = \f{|{\rm Re}(\Sigma_{\phi\pi})|}{\sqrt{\Sigma_{\phi\phi}\Sigma_{\pi\pi}}} \, ,
\label{eq:R_noise_corr}
 \eeq
  where $\gamma$ is related to the determinant of the noise matrix by $\gamma^2 = 1 - {\rm det(\Sigma_{ij})}/(\Sigma_{\phi\phi}\Sigma_{\pi\pi})$. The noise terms are maximally correlated if  $\gamma=1$, while $\gamma=0$ implies that $\xi_\phi$ and $\xi_\pi$ are independent (see Ref.~\cite{Grain:2017dqa}). For featureless potentials, we find that $\gamma \simeq 1$  under the pure dS approximation,  using Eqs.~(\ref{eq:Sig_phiphi_dS})-(\ref{eq:Sig_pipi_dS}), as well as under the SR approximations,   using Eqs.~(\ref{eq:Sig_phiphi_SR_supH})-(\ref{eq:Sig_pipi_SR_supH}).  For potentials with a PBH forming feature, we also find that $\gamma \simeq 1$ throughout the three asymptotic regimes\footnote{We find $\gamma \ll 1$ only for brief transient periods when the noise terms begin to rise after their  exponential fall  post transition.}  described by  Eqs.~(\ref{eq:Sig_pre_trans_gen})-(\ref{eq:Sig_post_trans_gen_asymp}). This property of maximal correlation between the noise terms is  a direct consequence of the fact that for $\sigma \ll 1$, the super-Hubble UV  mode functions, $\phi_k$, are frozen  by the time they join the coarse-graining scale $k = \sigma aH$ (see Refs.~\cite{Figueroa:2021zah,Tomberg:2022mkt} for a  detailed discussion on the freezing behaviour of the UV modes). Consequently, we conclude  that quantum diffusion  can be assumed to be sourced by a single random noise term.   Hence our analysis suggests that the  dynamics during the three asymptotic regimes given in Eqs.~(\ref{eq:Sig_phiphi_dS})-(\ref{eq:Sig_pipi_dS}) can be described by a system with a single stochastic degree of freedom as suggested in Refs.~\cite{Rigopoulos:2021nhv,Ahmadi:2022lsm,Tomberg:2023kli}. This will be  discussed in our forthcoming paper~\cite{MCG_ST_FPE_22}.

\section{Discussion}
\label{sec:Discussions}

 In Sec.~\ref{sec:SI_noise} we accurately calculated the stochastic noise matrix elements  for a sharp transition from SR to USR, using both analytical and numerical techniques. Our ultimate aim is to determine the PDF of the number of e-folds, $P_{\Phi,\Pi}({\cal N})$, by solving the adjoint Fokker-Planck Eq.~(\ref{eq:Adj_FP_Comp_gen}) (using appropriate boundary conditions) and then calculate the mass fraction of PBHs  $\beta_{\rm PBH}$. Using  the Press-Schechter formalism~\cite{Press:1973iz}, the PBH mass fraction is usually estimated by integrating the probability distribution of the coarse grained curvature perturbation, $P(\zeta_{cg})$, above the threshold for PBH formation,
$\zeta_{c}$. The PBH mass fraction in  the Stochastic formalism is given as (see Refs.~\cite{Pattison:2017mbe,Ezquiaga:2018gbw})  
  \beq
\beta(\Phi,\Pi) \equiv \int_{\zeta_c}^{\infty} \,  P(\zeta_{\rm cg}) \, {\rm d}\zeta_{\rm cg}=  \int_{\zeta_c + \langle {\cal N}(\Phi,\Pi) \rangle}^{\infty} \, P_{\Phi,\Pi}({\cal N}) \, {\rm d}{\cal N} \, ,
\label{eq:USR_beta_SI_form}
\eeq
where the average number of e-folds, $\langle {\cal N}(\Phi,\Pi) \rangle$, can be obtained from Eq.~(\ref{eq:N_avg}).

 While this task is reserved for our upcoming paper, we expect that the sharp decline of the noise terms after the transition will decrease the amount of quantum diffusion of the IR fields across the PBH-forming feature. Therefore we expect the tail of the  PDF to decline  more rapidly than what is usually found using the pure dS approximation without any transitions. Indeed such behaviour of the PDF was found in Ref.~\cite{Ahmadi:2022lsm} which focused on a sharp transition in pure dS space using the linear potential model of Starobinsky \cite{Starobinsky:1992ts}. Numerical simulations carried out in Refs.~\cite{Figueroa:2020jkf,Tomberg:2021xxv,Figueroa:2021zah} show that the canonical computation based on the pure dS  noise terms without any transition  typically leads to inaccurate estimates of the PBH abundance (over-estimates for some potentials and under-estimates for other potentials).  However, it is important  to study the relative contributions of the noise terms, $\Sigma_{ij}$, the potential, $V(\phi)$, and the boundary conditions to  the PDF separately. An analytical approach is well-suited to this, and this is one of the primary goals of our upcoming paper.

\bigskip

In the following, we overview  the outstanding complexities in accurately calculating the PBH mass fraction.

\begin{itemize}

\item \textit{Curvature perturbation vs density contrast:}  While the PBH mass fraction is often calculated  from  the PDF of the curvature perturbation using Eq.~(\ref{eq:USR_beta_SI_form}), the criterion for PBH formation is most accurately formulated in terms of the non-linear density contrast $\delta_l$,  see \textit{e.g.} Ref.~\cite{Germani:2018jgr,Tada:2021zzj,Biagetti:2021eep,DeLuca:2022rfz,Escriva:2022duf}.   An accurate computation of the PDF of the density contrast needs the
knowledge of all the higher order  ($n$-point) connected correlators of the curvature perturbation, $\zeta$. Therefore a high-precision calculation of the PBH mass fraction requires the  joint probabilities rather than  the one-point PDF, $P[\zeta]$ (see Ref.~\cite{Tada:2021zzj,DeLuca:2022rfz} for discussion of this issue).

\item \textit{Gauge corrections:} In this work we compute the mode functions $\lbrace \phi_k, \pi_k \rbrace$, and hence the noise correlators of $\lbrace \hat{\xi}_{\phi}, \hat{\xi}_{\pi} \rbrace$, in the spatially flat gauge, however the Langevin equations are written in the uniform-$N$ gauge. This induces corrections to the noise terms that could be non-negligible when the slow-roll approximations are violated \cite{Rigopoulos:2021nhv,Cruces:2021iwq,Pattison:2019hef}. However Refs.~\cite{Pattison:2019hef,Figueroa:2021zah} showed that the gauge corrections are negligible for $\sigma \ll 1$. 

\item   \textit{Choice of coarse-graining parameter:} In Sec.~\ref{sec:SI_formalism}, we mentioned that the coarse-graining parameter $\sigma$ needs to be small enough, $\sigma \ll 1$,  to ensure that the short-wavelength quantum fluctuations $\lbrace \xi_\phi, \xi_\pi \rbrace$  act as  classical noise on the dynamics of the coarse-grained fields $\lbrace \Phi, \Pi \rbrace$.  In fact, the physical results are expected to be independent of $\sigma$  as long as $\sigma \gg e^{-1/(3\epsilon_H)}$ (see Refs.~\cite{Starobinsky:1994bd,Vennin:2015hra}).  In our analysis, we  have considered $\sigma = 0.01$ in order to account for  substantial  non-linearity in the evolution by including as  many modes into the long-wavelength regime as possible without violating the stochastic nature of the noise terms (see Refs.~\cite{Figueroa:2021zah,Tomberg:2022mkt}). Nevertheless, our results given in  Eqs.~(\ref{eq:Sig_pre_trans_gen})-(\ref{eq:Sig_post_trans_gen_asymp}) demonstrate that the ratio of the noise terms are rather insensitive to the choice of $\sigma$. Numerical simulations of the stochastic dynamics carried out in Refs.~\cite{Figueroa:2021zah,Tomberg:2022mkt} also indicate that the mass fraction of PBHs  do not depend upon the particular choice for the value of $\sigma$, as long as it is not arbitrarily small. 

\item \textit{Effects of backreaction:} We have calculated the noise matrix elements by treating the mode functions $\lbrace \phi_k, \pi_k \rbrace$ as linear perturbations in a deterministic (non-stochastic) inflationary background, as is the usual practice in perturbation theory. In stochastic inflation, the noise terms should in principle be evaluated in the stochastically evolving  background of the coarse-grained IR fields $\lbrace \Phi, \Pi \rbrace$. However,  numerical simulations demonstrate that  such non-Markovian corrections due to the backreaction effects of the stochastic IR background are negligible for  single field inflationary potentials with  a   large class of PBH forming features, such as a flat segment, an  inflection point, or a  bump/dip and hence can be safely ignored (see Refs.~\cite{Figueroa:2021zah,Tomberg:2022mkt}).

\item \textit{Classical $\delta N$ formalism:} The PDF of the comoving curvature perturbation can also be computed using the classical (non-linear) $\delta N$ formalism developed in Refs.~\cite{Starobinsky:1982ee,Starobinsky:1985ibc,Sasaki:1995aw,Lyth:2004gb,Wands:2000dp}. For potentials with a broad class of features, the non-perturbative non-Gaussianity induced by  stochastic effects is usually expected to  be dominant (see Ref.~\cite{Figueroa:2021zah}).   The relative significance  of the stochastic effects can be  inferred from a \textit{classicality} criterion, expressed in terms of the parameter $j_{\rm cl} = | V(\phi) \eta_H/(24\pi^2 m_p^4 \, \epsilon_H )|$, obtained from a saddle-point approximation of the stochastic integrals~\cite{Vennin:2015hra}.  For  a potential with $j_{\rm cl} \ll 1$, stochastic effects can  safely be ignored (except in the far tail of the PDF). Hence, it is possible  to construct  potentials  for which the stochastic effects are specifically  negligible by design, while the classical non-linearities can be significant (see Ref.~\cite{Hooshangi:2021ubn}). In such cases the classical $\delta N$ formalism can be successfully used to compute the PDF (see also Refs.~\cite{Cai:2021zsp,Cai:2022erk}).

\item \textit {Loop corrections:} As outlined in Sec.~\ref{sec:Inf_dyn}, to generate a non-negligible abundance of PBHs, the power spectrum of the primordial scalar perturbations on small scales has to be roughly seven orders of magnitude larger than its measured value on CMB scales, \textit{i.e} ${\cal P}_\zeta (k) \simeq 10^{-2}$. Therefore it is crucial  to ask  whether such a large enhancement of power at smaller scales might induce non-negligible loop corrections to the CMB scale power spectrum at higher orders in perturbation theory. Recently such  calculations were carried out perturbatively in  Refs.~\cite{Kristiano:2022maq,Inomata:2022yte,Choudhury:2023vuj,Choudhury:2023jlt,Kristiano:2023scm}. These papers find that the  one-loop corrections to the CMB scale power spectrum can become significant if ${\cal P}_\zeta (k) \gtrsim {\cal O}(10^{-2})$. This appears to rule out the formation of an interesting abundance of PBHs in single field inflationary models. However this conclusion is currently the subject of debate. It has been argued in Refs.~\cite{Riotto:2023hoz,Riotto:2023gpm,Firouzjahi:2023aum} that the  loop corrections are negligible if the transition from  USR to the subsequent attractor phase is smooth enough. Nevertheless we stress that the amplitude of the small scale power spectrum required to form an interesting abundance of PBHs depends on the PDF of the perturbations. The standard value, ${\cal P}_\zeta (k) \simeq 10^{-2}$, assumes the PDF is Gaussian. This amplitude, and therefore the size of the one-loop corrections to the CMB scale power spectrum,  will be different for the non-Gaussian tail usually generated by stochastic effects.

\end{itemize}

\section{Conclusions}
\label{sec:Conclusions}

 PBHs can form due to the gravitational collapse of large fluctuations, in the non-perturbative tail of the PDF. An accurate calculation of the full PDF of the perturbations is therefore crucial 
to calculate their abundance. Stochastic inflation is a powerful framework for computing the cosmological correlators non-perturbatively. Using  the stochastic $\delta{\cal N}$ formalism, the full PDF  can be calculated from the first-passage statistics of the  number of e-folds, ${\cal N}$, during inflation.  However to correctly account for the back-reaction effect of small scale (UV) fluctuations, $\varphi_k$, on the long wavelength coarse-grained (IR) field, $\Phi$, it is essential to compute the noise matrix elements accurately. Since most single field inflationary potentials with a PBH-forming feature  violate the slow-roll conditions, a precise calculation of the stochastic noise matrix elements beyond slow roll is required. In this paper we have done this, both analytically and numerically. 

After a brief overview of single-field inflationary dynamics beyond slow roll in Sec.~\ref{sec:Inf_dyn}, we set up the relevant equations underlying the stochastic inflation formalism in Sec.~\ref{sec:SI_formalism}. 
There are two key steps to using the stochastic inflation formalism to calculate the full PDF of fluctuations in slow-roll violating, PBH-producing, models:
\begin{enumerate}
\item compute the statistics of both field and momentum-induced  noise terms  $\lbrace \xi_\phi, \xi_\pi \rbrace$,  
\item set up the  Langevin equations (or, the corresponding adjoint Fokker-Planck equation)  without ignoring the inflaton IR momentum $\Pi$,
\end{enumerate}
We have addressed the first issue here and will focus on the second in a forthcoming paper \cite{MCG_ST_FPE_22}.

  In Sec.~\ref{sec:SI_noise} we computed the matrix elements, $\Sigma_{ij}$, defined in Eq.~(\ref{eq:noise_comp_cor_matrix}), which characterise the statistics of the field and momentum noise terms. First, in Sec.~\ref{sec:Sig_SR}, we derived expressions for  $\Sigma_{ij}$ for featureless potentials where the slow-roll conditions  $\epsilon_H, \, \eta_H \ll 1$ remain valid  until almost the end of inflation. We compared the results of our analytical calculations,  Eqs.~(\ref{eq:Sig_phiphi_SR_supH})-(\ref{eq:Sig_pipi_SR_supH}), and numerical calculations (shown in Fig.~\ref{fig:Sig_SR_KKLT_Num}) for the  KKLT potential, Eq.~(\ref{eq:KKLT_base}), with the corresponding estimates  under the  pure de Sitter approximation, Eqs.~(\ref{eq:Sig_phiphi_dS})-(\ref{eq:Sig_pipi_dS}). We found that the dS approximation underestimates the momentum induced noise terms, $\Sigma_{\phi\pi}$ and $\Sigma_{\pi\pi}$, by several orders of magnitude, even for a slow-roll potential.

  In Sec.~\ref{sec:Sig_USR}, we calculated the noise matrix elements for single field inflationary potentials with a slow-roll violating, PBH-forming feature. For the numerical calculations we used the modified KKLT potential featuring a tiny Gaussian bump, 
  Eq.~(\ref{eq:KKLT_gaussian_bump}), as a proto-typical single-field PBH-forming potential. This potential has a sharp transition from the CMB scale SR-I phase to the subsequent near-USR phase, as shown in Fig.~\ref{fig:inf_meff_GB}. Our results, plotted in Figs.~\ref{fig:Sig_SR_USR_KKLT_GB} and \ref{fig:Sig_SR_USR_KKLT_GB_zoom}, show that following the transition,  $\Sigma_{ij}$ falls exponentially and  the momentum induced noise terms dominate the field noise with the hierarchy $\Sigma_{\pi \pi} > |{\rm Re}(\Sigma_{\phi \pi})| >  \Sigma_{\phi \phi}$. Subsequently, the noise terms return back to their original hierarchy, before growing and tending to constant values. 
  
  To understand the asymptotic behaviour of the noise terms, we calculated the noise matrix elements analytically using several approximations.  Firstly we treated the sharp transition between the SR-I phase and the subsequent near-USR phase as instantaneous, and assumed  the second slow-roll parameter $\eta_H$ to be piece-wise constant. By solving the Mukhanov-Sasaki equation, Eq.~(\ref{eq:MS_T_nu}), analytically for a constant $\eta_H$ (and hence a  constant $\nu$), and applying the Israel junction matching conditions across the transition, we  computed  the noise matrix elements shown in the right panel of Fig.~\ref{fig:Sig_singleT_asymp}. We found that the behaviour of the noise terms post transition is governed by a single parameter, namely the transition strength, ${\cal A}$, which is defined as the difference between the values of the second slow-roll parameter $\eta_H$ post- and pre-transition as given in Eq.~(\ref{eq:A_nu_etaH_T1}). This analytical computation based on an instantaneous transition $\nu : \nu_1 \to \nu_2$  captures the key features of the noise matrix elements  for potentials with a smooth feature (see Eqs.~(\ref{eq:Sig_pre_trans_gen})-(\ref{eq:Sig_post_trans_gen_asymp})).

  We also compared our calculations with those for an instantaneous transition using the pure dS approximation, {\ie} $\nu_1 = \nu_2 = 3/2$, which was carried out in Ref.~\cite{Ahmadi:2022lsm} for the Starobinsky model \cite{Starobinsky:1992ts}, see the left panel of Fig.~\ref{fig:Sig_singleT_asymp}.  We found that the dS approximation  underestimates the noise terms not only in the SR-I phase (as mentioned before), but also a long time after the transition.  However, the pure dS-transition estimates are a good approximation to the behaviour of the noise terms immediately after the transition. Furthermore, for potentials with  a pure `flat' feature (as shown in Fig.~\ref{fig:inf_flat_Q_well}) rather than a bump, the dS-transition approximations work quite well.

   In our analytical solutions of the MS equation, we focused on a single sharp transition, T-I. In this case the effective mass term $z''/z$ remains almost constant throughout the USR, T-I and CR phases, as can be seen in  Fig.~\ref{fig:inf_meff_GB}. Therefore the expression for the mode functions remains the same after the second transition,  due to the Wands duality as discussed in Sec.~\ref{sec:Sig_USR}. This is a common characteristic  of a broad class of single field inflationary models with a PBH forming feature (see Ref.~\cite{Karam:2022nym}).  Our analytical scheme can be extended to  situations where the effective mass term $z''/z$  undergoes two or more sharp transitions. We provide the relevant analytical expressions for the mode functions in this case in App.~\ref{app:Sig_2_trans}.

  \bigskip
  
  We conclude that in order to accurately determine the PDF of the curvature perturbation, $P[\zeta]$, beyond slow roll, one must solve the adjoint Fokker-Planck Equation (\ref{eq:Adj_FP_Comp_gen}) using the correct asymptotic forms of the noise matrix elements given in Eqs.~(\ref{eq:Sig_pre_trans_gen})-(\ref{eq:Sig_post_trans_gen_asymp}).  Our upcoming paper \cite{MCG_ST_FPE_22} will be dedicated to developing analytical and semi-analytical techniques to solve the adjoint FPE with the knowledge of $\Sigma_{ij}$ obtained here.   While  numerical simulations of the Langevin equations  can be carried out in full generality, they are often quite time-consuming, and demand large computational resources. Furthermore, the analytical approach will allow us to calculate the asymptotic behaviour of the PDF and study the effects of the  noise terms, $\Sigma_{ij}$, the potential, $V(\phi)$, and the boundary conditions on the PDF separately. It is therefore  complementary to the fully numerical simulations of the Langevin equations discussed in Ref.~\cite{Figueroa:2020jkf,Tomberg:2021xxv,Figueroa:2021zah,Jackson:2022unc}.

\section*{Acknowledgements} 
 We would like to thank David Wands  and Antonio Riotto  for helpful comments. SSM, EJC and AMG are supported by a STFC Consolidated Grant [Grant No. ST/T000732/1], and EJC by a Leverhulme Research Fellowship [RF-2021 312]. 
For the purpose of open access, the authors have applied a CC BY public copyright licence to any Author Accepted Manuscript version arising. \\

{\bf Data Availability Statement:} This work is entirely theoretical and has no associated data.

\appendix

\section{Mukhanov-Sasaki equation in spatially flat gauge}
\label{app:MS_flat_gauge}
 In this Appendix we outline the derivation of the Mukhanov-Sasaki equation in a spatially flat gauge, which we solved for in Sec.~\ref{sec:SI_noise}. Starting from the action of a canonical scalar field minimally coupled to gravity 
$$ S[g_{\mu\nu},\phi] = \int \,  {\rm d}^4x \,  \sqrt{-g} \,  \l( \, \f{m_p^2}{2} \, R - \f{1}{2}  \, \partial_{\mu}\phi \,  \partial_{\nu}\phi  \, g^{\mu\nu}-V(\phi)\r)~, $$
we consider linear field fluctuations $\phi(t,\vec{x}) = \bar{\phi}(t) + \delta\phi(t,\vec{x})$ and the linearly perturbed  ADM metric in the spatially flat gauge 
$$ {\rm d}s^2 = -  \alpha^2 \, {\rm d}t^2 + a^2(t) \, \delta_{ij} \l( {\rm d}x^i + \beta^i \, {\rm d}t \r) \l( {\rm d}x^j + \beta^j \, {\rm d}t \r) \, ,$$
where $\alpha = 1 + \delta\alpha$ and $\beta^i$ are the lapse and shift functions.  Imposing the GR momentum and Hamiltonian constraints, one obtains (see Ref.~\cite{Maldacena:2002vr})
$$ \delta\alpha = \sqrt{\f{\epsilon_H}{2}} \, \f{\delta\phi}{m_p} \, , ~~~~ \partial_i \beta^i = - \epsilon_H \, \f{{\rm d}}{{\rm d}t} \l( \f{1}{\sqrt{2\epsilon_H}} \, \f{\delta\phi}{m_p}\r)~.$$

Incorporating the above expressions into the action, expanding around the background, the quadratic action for $\delta\phi$ fluctuations (in the spatially flat gauge) is \cite{Maldacena:2002vr,Baumann:2018muz}
\beq
S[\delta\phi] = \f{1}{2} \, \int \,  {\rm d}t {\rm d}^3{\vec x} ~ a^3  \l[ \dot{\delta\phi}^2  - \f{(\partial_i\delta\phi)^2}{a^2} - \l( \f{{\rm d}^2 V(\phi)}{{\rm d} \phi^2} + 2\, \epsilon_H \, H^2 \, \l( 2 \, \eta_H - \epsilon_H - 3 \r) \r)  \delta\phi^2 \r] \, .
\label{eq:Quad_action_spat_flat}
\eeq
The Euler-Lagrange equation for $\delta\phi$ is given by
\beq
\ddot{\delta\phi} + 3 \, H \, \dot{\delta\phi} - \f{\nabla^2}{a^2} \, \delta\phi - a^2H^2 \l( 2 - \epsilon_H - \f{1}{H^2} \, \f{{\rm d}^2 V(\phi)}{{\rm d} \phi^2}  - \f{2 \dot{\bar{\phi}}}{m_p^2 H^3} \, \f{{\rm d} V(\phi)}{{\rm d} \phi}  - \f{\dot{\bar{\phi}}^2}{m_p^4 H^4}  \, V \r)  \, \delta\phi = 0 ~.
\label{eq:El_deltaphi_spat_flat}
\eeq

With the change of variables $v = a \, \delta\phi$, the Fourier modes $v_k$ satisfy the Mukhanov-Sasaki equation 
$$v_k^{''} + \l( k^2 - \f{z''}{z} \ \r) \, v_k = 0 ~,$$
with 
\begin{align}
 \f{z''}{z} &\equiv a^2H^2 \l( 2 - \epsilon_H - \f{1}{H^2} \, \f{{\rm d}^2 V(\phi)}{{\rm d} \phi^2}  - \f{2 \dot{\bar{\phi}}}{m_p^2 H^3} \, \f{{\rm d} V(\phi)}{{\rm d} \phi}  - \f{\dot{\bar{\phi}}^2}{m_p^4 H^4}  \, V \r) \,, \\
&= (aH)^2 \l[ 2 + 2 \epsilon_H - 3 \eta_H + 2 \epsilon_H^2 + \eta_H^2 - 3 \epsilon_H \eta_H  - \f{1}{aH} \, \eta'_H \r] \, .
\end{align}

\section{Analytical solution of the Mukhanov-Sasaki equation}
\label{app:MS_analyt_sol}

 For the featureless slow-roll potentials that we study in Sec.~\ref{sec:Sig_SR}, $\nu^2$ is greater than or equal to $9/4$ and effectively constant. In 
this case the Mukhanov-Sasaki (MS) Eq.~(\ref{eq:MS_modes}) can be written as a  Bessel equation with constant $\nu$, which can be solved analytically (see Refs.~\cite{book_nist_gov,Birrell:1982ix}). In this Appendix we present this solution in terms of both Hankel functions (App.~\ref{sec:app:MS_analyt_sol_hankel}) and Bessel functions (App.~\ref{sec:app:MS_analyt_sol_bessel}).

For the analytical treatment, assuming  $(aH)^{-2} \, z''/z = \nu^2 - 1/4$ to be a constant,   the Mukhanov-Sasaki (MS) Eq.~(\ref{eq:MS_modes}) can be written as 

\beq
 \f{{\rm d}^2 v_k}{{\rm d} \tau^2} + \l[ \, k^2 - \f{\nu^2 - \f{1}{4}}{\tau^2}  \, \r] v_k = 0 ~,
\label{eq:MS_modes_nu_tau}
\eeq
using  the new time variable, $T$, defined in Eq.~(\ref{eq:T_tau})
 $$ T = -k\tau = \f{k}{a \, H} \, .$$
All modes undergo Hubble-exit at $T  = 1$, with sub (super)-Hubble scales corresponding to $T \gg (\ll) 1$. In terms of this new time variable, the MS equation takes the form

$$\f{{\rm d}^2 v_k}{{\rm d} T^2} + \l[ \, 1 - \f{\nu^2 - \f{1}{4}}{T^2}  \, \r] v_k = 0 \, .$$

Using the variable redefinition $F =   v_k/\sqrt{T}$, this equation can be transformed into the more familiar  Bessel equation:
\beq
\f{{\rm d}^2 F}{{\rm d} T^2} +   \f{1}{T} \, \f{{\rm d} F}{{\rm d} T}  + \l[ \, 1 - \f{\nu^2}{T^2}  \, \r] F = 0 ~.
\label{eq:MS_modes_nu_Bessel}
\eeq

The general solution to Eq.~(\ref{eq:MS_modes_nu_Bessel}) (when $\nu$ is not an integer) can be written either as a linear combination of Hankel functions of the first and second kind $\lbrace  H_\nu^{(1)}(T), \,  H_\nu^{(2)}(T)   \rbrace$ or as a linear combination of positive and negative order ($\pm \nu$) Bessel functions of the first kind  $ \lbrace J_{-\nu}(T), \, J_{\nu}(T) \rbrace$. The functions are related by~\cite{book_nist_gov}

\beq
H_\nu^{(1,2)}(T)  = \f{ \pm  J_{-\nu}(T) \mp e^{\mp i\pi\nu} \, J_{\nu}(T) }{ i\sin{(\pi\nu)}  }    \,. \label{eq:Hank_Bess12} 
\eeq

\subsection{In terms of Hankel functions}
\label{sec:app:MS_analyt_sol_hankel}

The general solution to the Bessel Eq.~(\ref{eq:MS_modes_nu_Bessel}) in terms of the  Hankel functions is given by

\beq
F(T) = C_1 \, H_\nu^{(1)}(T) \, +   \, C_2 \, H_\nu^{(2)}(T)~,
\label{eq:Bessel_sol_Hankel}
\eeq
where the coefficients $C_1$ and $C_2$ are fixed by initial/boundary conditions. Hence the solution to the MS  equation  can be written as 

\beq
 v_k (T) = \sqrt{T} \, \l[  \, C_1 \, H_\nu^{(1)}(T) \, +   \, C_2 \, H_\nu^{(2)}(T) \, \r] \,.
\label{eq:MS_sol_Hankel}
\eeq

In the sub-Hubble limit, $T \gg 1$, the Hankel functions take the form 
\ber
 H_\nu^{(1)}(T) \bigg \vert_{T \to \infty} \simeq \sqrt{\f{2}{\pi}} \, \f{1}{\sqrt{T}}  \, e^{ i T}  \, e^{-i \l(  \nu + \f{1}{2}  \r) \f{\pi}{2}} \, , \label{eq:hankel1_sub} \\
H_\nu^{(2)}(T) \bigg \vert_{T \to \infty} \simeq \sqrt{\f{2}{\pi}} \, \f{1}{\sqrt{T}}  \, e^{ - i T} \, e^{i \l(  \nu + \f{1}{2}  \r) \f{\pi}{2}} \, , \label{eq:hankel2_sub}
\eer
while in the super-Hubble limit, $T \ll 1$, the Hankel functions take the form 
\ber
 H_\nu^{(1)}(T) \bigg \vert_{T \to 0} \simeq \sqrt{\f{2}{\pi}} \, e^{- i  \f{\pi}{2}} \,  2^{\nu - \f{3}{2}}   \, \f{\Gamma(\nu)}{\Gamma(\f{3}{2})} \,  T^{-\nu} \, , \label{eq:hankel1_super} \\
H_\nu^{(2)}(T) \bigg \vert_{T \to 0} \simeq ~ - \sqrt{\f{2}{\pi}} \, e^{- i  \f{\pi}{2}} \,  2^{\nu - \f{3}{2}}   \, \f{\Gamma(\nu)}{\Gamma(\f{3}{2})} \,  T^{-\nu} \, . \label{eq:hankel2_super}
\eer

The Bunch-Davies conditions, Eq.~(\ref{eq:MS_BD}), for  the mode functions take the form

$$v_k(T) \bigg \vert_{T \to \infty}  \to  \f{1}{\sqrt{2k}} \, e^{ i T} = \sqrt{T}  \, C_1  H_\nu^{(1)}(T) \bigg \vert_{T \to \infty} ~,$$
which yields 
$$C_1 = \f{1}{\sqrt{2k}}  \, \sqrt{\f{\pi}{2}} \, e^{i \l(  \nu + \f{1}{2}  \r) \f{\pi}{2}}  ~,~~~{\rm and}~~~  C_2 =0~,$$
and hence the final expression for the mode functions becomes 

\beq
v_k (T) = e^{i \l(  \nu + \f{1}{2}  \r) \f{\pi}{2}} \, \sqrt{\f{\pi}{2}}  \, \f{1}{\sqrt{2k}} \, \sqrt{T} \,  H_\nu^{(1)}(T)  \, .
\label{eq_MS_mode_fun_Hankel}
\eeq

\subsection{In terms of Bessel functions}
\label{sec:app:MS_analyt_sol_bessel}

The general solution to the Bessel equation, Eq.~(\ref{eq:MS_modes_nu_Bessel}), in terms  of the  Bessel functions  of the first kind of order $\pm \nu$ is given by

\beq
F(T) = \sqrt{\f{\pi}{2}} \, \f{1}{\sin{(\pi\nu)}}  \, \l[ \, C_{+} \, J_{-\nu}(T) \, +   \, C_{-} \, J_{\nu}(T) \, \r] ~,
\label{eq:Bessel_sol_Bessel_1st}
\eeq
where  the coefficients $C_{+}$ and $C_{-}$ are again  to be fixed by initial/boundary conditions. Hence the solution to MS Eq.~(\ref{eq:MS_modes}) can be written as

\beq
 v_k (T) =  \sqrt{\f{\pi}{2}} \, \f{1}{\sin{(\pi\nu)}}  \, \sqrt{T}  \, \l[  \,  C_{+} \, J_{-\nu}(T) \, +   \, C_{-} \, J_{\nu}(T)  \, \r] \, .
\label{eq:MS_sol_Bessel_1st}
\eeq

Imposing Bunch-Davies initial conditions, Eq.~(\ref{eq:MS_BD}), we get

$$C_{+} = -i \, \f{1}{\sqrt{2k}} \, e^{i \l(  \nu + \f{1}{2}  \r) \f{\pi}{2}}  ~,~~~{\rm and}~~~  C_{-} = i \, \f{1}{\sqrt{2k}} \,e^{i \l(  \nu + \f{1}{2}  \r) \f{\pi}{2}} \, e^{-i \pi \nu }  ~,$$
and hence the final expression for the mode functions becomes 

\beq
 v_k (T) = - i  \, \sqrt{\f{\pi}{2}} \,   \f{e^{i \l(  \nu + \f{1}{2}  \r) \f{\pi}{2}} }{ \sin{(\pi\nu)}} \, \f{1}{\sqrt{2k}} \, \sqrt{T} \l[ \, J_{-\nu}(T) \, -  e^{-i \pi \nu } \, J_{\nu}(T)  \,   \r]  ~.
\label{eq_MS_mode_fun_Bessel}
\eeq
 With the help of Eq.~(\ref{eq:Hank_Bess12}), we see by equating Eqs.~(\ref{eq:MS_sol_Hankel}) and (\ref{eq:MS_sol_Bessel_1st}), that the relation between the Hankel coefficients $ \lbrace C_1, \, C_2 \rbrace $ and Bessel coefficients $ \lbrace  C_{+}, \,  C_{-}  \rbrace $ is given by

\beq
C_1 = i \sqrt{\f{\pi}{2}}  \, \l[ \,   \f{C_{+} + e^{-i2\pi\nu} C_{-}}{1-e^{-i2\pi\nu} }    \, \r] \, ; ~~~
C_2 = i \sqrt{\f{\pi}{2}} \,  \l[ \,   \f{C_{+} + e^{i\pi\nu} C_{-}}{1-e^{-i2\pi\nu} }    \, \r] e^{-i2\pi\nu}  \, . 
\label{eq:ceff_Hank_Bess12}
\eeq

In this work, we express analytical solutions of the MS equation in terms of Hankel functions.
However, our results can alternatively be easily expressed in terms of the Bessel functions by using Eqs.~(\ref{eq:Hank_Bess12}) and (\ref{eq:ceff_Hank_Bess12}).

\section{Super-Hubble expansion of the noise matrix elements}
\label{app:noise_elements_expanded}

 The full expressions for the noise matrix elements are given in Eqs.~(\ref{eq:Sigma_phi_phi_gen_T})-(\ref{eq:Sigma_pi_pi_gen_T}). Since we need to evaluate $\Sigma_{ij}$ in the super-Hubble limit with $T=\sigma \ll 1$, here we provide expressions for the noise terms, derived from the mode functions, $v_k(T)$, in Eq.~(\ref{eq:MS_sol_SR}), as a series expansion in $T$, up to ${\cal O}(T^4)$, for constant $\nu$:

\begin{eqnarray}
    \Sigma_{\phi \phi} &=&  2^{2 \l( \nu - \f{3}{2} \r)} \, \l[ \f{\Gamma(\nu)}{\Gamma(3/2)} \r]^2\left(\frac{H}{2\pi}\right)^2 T^{2\l(-\nu + \f{3}{2}\r)} \left[ 1 + \f{1}{2(-1+\nu)} \, T^2  \right.\nonumber \\ &~& \left.  +
    \frac{\nu-3/2}{8(-2+\nu)(-1+\nu)^2} \, T^4  \right.\nonumber \\ &~& \left.  +  \l( 2   +  \frac{T^2}{(1-\nu^2)} \r)  \f{2^{-2\nu} \Gamma[-\nu]}{\Gamma[\nu]}  \cos{(\pi \nu)} \, T^{2\nu} + 
 {\cal O}(T^{4+2\nu}) \right]  \label{eq:Sig_phiphi_expand} \,,
\end{eqnarray}

\begin{eqnarray}
    {\rm Re}(\Sigma_{\pi \phi}) &=&  - 2^{2 \l( \nu - \f{3}{2} \r)} \, \l[ \f{\Gamma(\nu)}{\Gamma(3/2)} \r]^2\left(\frac{H}{2\pi}\right)^2 T^{2\l(-\nu + \f{3}{2}\r)} \left[ \left(\f{3}{2} - \nu\right) + \f{5 - 2\nu}{4(-1+\nu)} \, T^2  \right.\nonumber \\ &~& \left. + \frac{-7+2\nu}{16(-2+\nu)(-1+\nu)^2} \l(\f{3}{2} - \nu \r) \, T^4  
     \right.\nonumber \\ &~& \left. - \l( 3  +
     \frac{5}{2(-1+\nu^2)} T^2  \r)  \frac{2^{-2\nu}}{\Gamma(1+\nu)^2 }  \pi \nu \cot{(\pi \nu)} \, T^{2\nu} +  {\cal O}(T^{4+2\nu}) \right] \,,
\label{eq:Sig_piphi_expand}
\end{eqnarray}

\begin{eqnarray}
     \Sigma_{\pi \pi} &=&  2^{2 \l( \nu - \f{3}{2} \r)} \, \l[ \f{\Gamma(\nu)}{\Gamma(3/2)} \r]^2\left(\frac{H}{2\pi}\right)^2 T^{2\l(-\nu + \f{3}{2}\r)} \left[ \l(-\nu + \f{3}{2}\r)^2 + \l(-\nu + \f{3}{2}\r) \l(\f{-2\nu + 7}{4(-\nu+1)}\r) \, T^2  \right.\nonumber \\ &~& \left.  +
     \l(\f{131-2\nu(83-34\nu+4\nu^2)}{64 (-\nu+2)(-1+\nu)^2}\r) \, T^4    \right.\nonumber \\ &~& \left.  -
     \l( \f{-4\nu^2+9}{2} + \f{21+4\nu^2}{4(-1+\nu^2)} \, T^2 \r) \frac{2^{-2\nu} }{\Gamma[1+\nu]^2} \pi \nu  \cot{(\pi \nu)} \, T^{2\nu}
     + {\cal O}(T^{4+2\nu}) \right] \label{eq:Sig_pipi_expand} \,.
\end{eqnarray}

The above expressions for $\Sigma_{ij}$ are valid for any value of $\nu$, and accurately reproduce the de Sitter results for $\nu = 3/2$.  We have verified that there are no higher order corrections in the dS limit.

\section{Functional form of $z^{\prime \prime}/z$ during sharp transitions}
\label{app:eta_to_nu}

In this Appendix we derive the analytic expressions for $z^{\prime \prime}/z$ for the instantaneous transitions that we use in Sec.~\ref{sec:SI_noise_analytical}. Under the quasi-dS approximation, $\epsilon_H \simeq 0$, the effective mass term in the MS Eq.~(\ref{eq:MS_modes}) becomes
\beq
 \f{z''}{z} \, \tau^2 \simeq 2 - 3\, \eta_H + \eta_H^2 + \tau \, \f{{\rm d}\eta_H}{{\rm d}\tau} \, ,
\label{eq:meff_qdS_eta}
\eeq
where this final form of $z''/z$ depends upon the expression for the second slow-roll  parameter $\eta_H(\tau)$. In the following, we will assume that $\eta_H$ is piece-wise constant and makes instantaneous, but finite transitions. We will begin with the simplest case where $\eta_H$ makes only one transition and later generalise this to the case of two or more  successive transitions.

\subsection{Single instantaneous transition}
\label{sec:eta_to_nu_single}

Suppose that during inflation, the second SR parameter makes a sharp transition from $\eta_H = \eta_1 \to \eta_2$ at time $\tau = \tau_1$. In this case $\eta_H$ can be written as 

\beq
\eta_H (\tau) = \eta_1 + \l( \eta_2 - \eta_1 \r) \, \Theta(\tau-\tau_1)~.
\label{eq:eta_singleT}
\eeq
Using Eq.~(\ref{eq:meff_qdS_eta}), we obtain the following expression for $z''/z$ 

\beq
\f{z''}{z} \, \tau^2 =  {\cal A} \, \tau \, \delta_D(\tau-\tau_1) +  \nu_1^2 - \f{1}{4} + \l( \nu_2^2 - \nu_1^2 \r) \, \Theta(\tau-\tau_1) \, ,
\label{eq:meff_qdS_eta_singleT}
\eeq
with ${\cal A} = \eta_2 - \eta_1$ and $\nu_{1,2}^2 = (9/4)  - 3\, \eta_{1,2} + \eta_{1,2}^2$.  For $\eta_1 = 0$ and $\eta_2 = +3$, the system  reduces to the case of  pure de Sitter SR $\to$ USR transition, where $\nu_1^2 = \nu_2^2 = 9/4$ and ${\cal A} = +3$.

\subsection{Two successive instantaneous transitions}

If  the second SR parameter makes another sharp transition from $\eta_H = \eta_2 \to \eta_3$ at time $\tau = \tau_2$, then $\eta_H$  can be written as 

\beq
\eta_H (\tau) = \eta_1 + \l( \eta_2 - \eta_1 \r) \,  \Theta(\tau-\tau_1) + \l( \eta_3 - \eta_2 \r) \,  \Theta(\tau-\tau_2) \, .
\label{eq:eta_doubleT}
\eeq
Once again using Eq.~(\ref{eq:meff_qdS_eta}), we now obtain 
\beq
\f{z''}{z} \, \tau^2 =  \nu_1^2 - \f{1}{4} + \l( \nu_2^2 - \nu_1^2 \r) \, \Theta(\tau-\tau_1) +  \l( \nu_3^2 - \nu_2^2 \r) \, \Theta(\tau-\tau_2)  + {\cal A}_1 \, \tau \, \delta_D(\tau-\tau_1) +  {\cal A}_2 \, \tau \, \delta_D(\tau-\tau_2)   \, ,
\label{eq:meff_qdS_eta_doubleT_gen}
\eeq
with ${\cal A}_{1,2} = \eta_{2,3} - \eta_{1,2}$ and $\nu_{1,2,3}^2 = (9/4)  - 3\, \eta_{1,2,3} + \eta_{1,2,3}^2$.

\subsection{Generalising to multiple instantaneous transitions}

If the inflaton potential exhibits a number of tiny features/modulations, then the second SR parameter $\eta_H$ might undergo a number of successive transitions before the end of inflation. Assuming each transition to be instantaneous for ease of analytical treatment, we can write the following general expression for $\eta_H$
\beq
\eta_H (\tau) = \eta_1 + \l( \eta_2 - \eta_1 \r) \,  \Theta(\tau-\tau_1) + \l( \eta_3 - \eta_2 \r) \,  \Theta(\tau-\tau_2) + ... + \l( \eta_{n+1} - \eta_n \r) \,  \Theta(\tau-\tau_n) \, ,
\label{eq:eta_multiT}
\eeq
where `$n$' is the total number of instantaneous  transitions occurring at times $\tau_1, \tau_2, ...., \tau_n$. In this case, the expression for $z''/z$  becomes
\beq
\begin{split}
 \f{z''}{z} \, \tau^2 =  \nu_1^2 - \f{1}{4} + \l( \nu_2^2 - \nu_1^2 \r) \, \Theta(\tau-\tau_1) +  \l( \nu_3^2 - \nu_2^2 \r) \, \Theta(\tau-\tau_2)  + ... + \l( \nu_{n+1}^2 - \nu_n^2 \r) \, \Theta(\tau-\tau_n)     \\ 
 + {\cal A}_1 \, \tau \, \delta_D(\tau-\tau_1) +  {\cal A}_2 \, \tau \, \delta_D(\tau-\tau_2)  + ... + {\cal A}_n \, \tau \, \delta_D(\tau-\tau_n)  \, ,
\label{eq:meff_qdS_eta_multiT_gen}
\end{split}
\eeq
where ${\cal A}_{n} = \eta_{n+1} - \eta_{n}$ and $\nu_{n}^2 = (9/4)  - 3\, \eta_{n} + \eta_{n}^2$. 

In compact notation, we can write the general expressions for $\eta_H$ and $z''/z$ for `$n$' successive instantaneous transitions as
\ber
 \eta_H (\tau) &=&  \eta_1  \, + \,  \sum_{i = 1}^{n}  \, \l[ \, \eta_{i+1} - \eta_{i} \, \r] \, \Theta(\tau-\tau_i)  \, ,\\ \label{eq:eta_multiT_compact}
  \f{z''}{z} \, \tau^2 &=& \nu_1^2 - \f{1}{4}   \, + \,  \sum_{i = 1}^{n}  \, \l[ \, \nu_{i+1}^2 - \nu_{i}^2  \, \r] \, \Theta(\tau-\tau_i) \, + \, {\cal A}_i \, \tau \, \delta_D(\tau-\tau_i)  \, ,
\label{eq:meff_qdS_eta_multiT_compact}
\eer
where $ {\cal A}_{i} = \eta_{i+1} - \eta_{i} \, , ~~~ \nu_{i}^2 = (9/4)  - 3\, \eta_{i} + \eta_{i}^2  $\, .

\section{Noise matrix elements for two successive  instantaneous transitions}
\label{app:Sig_2_trans}

In this Appendix, we present  the full calculations  for the mode functions in a closed form for the case of two successive instantaneous transitions during inflation. This generalises the results presented in Case 1 (Pure dS limit) and Case 2 (transition between two different values $\nu_1 \to \nu_2$) of Sec.~\ref{sec:SI_noise_analytical}.

\subsection{Pure dS limit}

For two successive instantaneous transitions SR $\to$  USR $\to$ SR in the pure dS limit\footnote{ See Ref.~\cite{Ivanov:1994pa} for an extension of the  Starobinsky model \cite{Starobinsky:1992ts} featuring two successive transitions in the context of  PBH formation.},  the effective mass term in the MS equation takes the form

\beq
 \f{z''}{z} \, \tau^2=  {\cal A} \, \tau \, \l[ \,  \delta_D(\tau-\tau_1) - \delta_D(\tau-\tau_2) \, \r] \, +  \, 2 \, ,
\label{eq:meff_jump_dS_dS_dS}
 \eeq
 where $\tau_1$ is the transition time from SR to USR and $\tau_2$ is the transition time from USR back to  SR. ${\cal A} = +3 $ is the strength of the transition, as discussed before 

The MS (complex) mode functions  generalise those derived in Eq.~(\ref{eq:vk_jump_dS_dS}) and are given by
\begin{flalign}
 v_k (T)  \equiv  \begin{cases}   v_k^{E}(T) = \f{1}{\sqrt{2k}} \, \l( 1 + \f{i}{T} \r) \, e^{i \, T}  \, , &  T_2 < T_1 < T ~ ,\\ 
 v_k^{I}(T) = \f{1}{\sqrt{2k}} \, \l[  \, C_1^I \, \l( 1 + \f{i}{T} \r) \, e^{i \, T}  \, +   \,  C_2^I \, \l( 1 - \f{i}{T} \r) \, e^{-i \, T}  \, \r] \, , & T_2 < T < T_1 ~, \\
 v_k^{L}(T) = \f{1}{\sqrt{2k}} \, \l[  \, C_1^L \, \l( 1 + \f{i}{T} \r) \, e^{i \, T}  \, +   \,  C_2^L \, \l( 1 - \f{i}{T} \r) \, e^{-i \, T}  \, \r] \, , &  T < T_2 < T_1 ~,
 \end{cases} 
\label{eq:vk_jump_dS_dS_dS}
\end{flalign}
and the derivatives of the mode functions  generalise those derived in Eq.~(\ref{eq:vkT_jump_dS_dS}) and are given by 
\begin{flalign}
 \f{{\rm d}v_k}{{\rm d}T}  \equiv  \begin{cases}   \f{{\rm d}v_k^E}{{\rm d}T}  = \f{1}{\sqrt{2k}} \, \l[ -\f{1}{T} + i \, \l( 1 - \f{1}{T^2}  \r) \r] \, e^{i \, T}  \, , & T_2 < T_1 < T ~ ,\\ 
 \f{{\rm d}v_k^I}{{\rm d}T}  = \f{1}{\sqrt{2k}} \, \l[  \, C_1^I \, \l[ -\f{1}{T} + i \, \l( 1 - \f{1}{T^2}  \r) \r] \, e^{i \, T}  \, +   \,   C_2^I  \, \l[ -\f{1}{T} - i \, \l( 1 - \f{1}{T^2}  \r) \r] \, e^{-i \, T}  \, \r] \, , & T_2 < T < T_1  ~, \\
 \f{{\rm d}v_k^L}{{\rm d}T}  = \f{1}{\sqrt{2k}} \, \l[  \, C_1^L  \, \l[ -\f{1}{T} + i \, \l( 1 - \f{1}{T^2}  \r) \r] \, e^{i \, T}  \, +   \, C_2^L \, \l[ -\f{1}{T} - i \, \l( 1 - \f{1}{T^2}  \r) \r] \, e^{-i \, T}  \, \r] \, , &  T < T_2 < T_1 ~. \end{cases} 
\label{eq:vkT_jump_dS_dS+_dS}
\end{flalign}

Here the superscripts `$E$', `$I$' and `$L$' stand for early, intermediate and late respectively. After implementing the Israel junction matching conditions, we obtain the final expressions for the mode functions  

\begin{flalign}
  C_1^{L} = \f{ d_3 \, b_4 - d_4 \, b_3  }{ a_3 \, b_4 - a_4 \, b_3   } ~,~~~   C_2^{L} = \f{ d_4 \, a_3 - d_3 \, a_4  }{ a_3 \, b_4 - a_4 \, b_3   } ~, 
\end{flalign}

where 
\begin{align}
a_3 &=   T_2 + i  \,, & \\
b_3 &=   \l( T_2 - i \r) \, e^{ -i \, 2T_2 } \,,  & \\
a_4 &=  \l(  {\cal A} -1 \r) \, T_2 + i \,   \l(   T_2^2  - 1 + {\cal A} \r) \,, & \\
b_4 &=  \ \l(  {\cal A} -1 \r) \, T_2 - i \,   \l(   T_2^2  - 1 + {\cal A} \r) \, e^{ -i \, 2T_2 } \,, & \\ 
d_3 &=   \l(  T_2 + i \r) \, C_1^I +  \l( T_2 - i \r) \, e^{ -i \, 2T_2 } \,  C_2^I \,, & \\  
d_4 &=  \l[  T_2 - i \, \l(  T_2^2 -1 \r) \r] \, C_1^I + \l[  T_2 + i \, \l(  T_2^2 -1 \r) \r] \, C_2^I~,
\end{align}
and the intermediate transition coefficients $C_1^I$ and  $C_2^I$ are just $\alpha_k$ and $\beta_k$, respectively,  derived previously in Eqs.~(\ref{eq:alphak_dS_dS}) and (\ref{eq:betak_dS_dS}).

\subsection{$\nu_1 \to \nu_2 \to \nu_3$}

In case of two successive instantaneous transitions from  $\nu_1 \to \nu_2 \to \nu_3$,  the effective mass term in the MS equation takes the form

\beq
\f{z''}{z} \, \tau^2 =  \nu_1^2 - \f{1}{4} + \l( \nu_2^2 - \nu_1^2 \r) \, \Theta(\tau-\tau_1) + \l( \nu_3^2 - \nu_2^2 \r) \, \Theta(\tau-\tau_2)   +  {\cal A}_1 \, \tau \, \delta_D(\tau-\tau_1) +  {\cal A}_2 \, \tau \, \delta_D(\tau-\tau_2)   \, ,
\label{eq:meff_jump_nu1_n2_nu3}
\eeq
where $\tau_1$ is the transition time from $\nu_1 \to \nu_2$ and $\tau_2$ is the transition time from $\nu_2 \to \nu_3$. ${\cal A}_1 , \, {\cal A}_2 $ are  the strengths of the first and second transitions  respectively.

The MS (complex) mode functions and their derivatives  generalise those derived in Eqs.~(\ref{eq:vk_jump_nu1_nu2}) and (\ref{eq:vkT_jump_nu1_nu2}) to yield
\begin{flalign}
 v_k (T)  \equiv  \begin{cases}   v_k^E(T) = \f{1}{\sqrt{2k}} \, \sqrt{\f{\pi}{2}}    \, \sqrt{T} \,  H_{\nu_1}^{(1)}(T)  \, e^{i \l(  \nu_1 + \f{1}{2}  \r) \f{\pi}{2}} \, , &  T_2 < T_1 < T~ , \\ 
 v_k^{I}(T) = \sqrt{T} \, \l[  \, C_1^I \, H_{\nu_2}^{(1)}(T)  \, +   \,  C_2^I \, H_{\nu_2}^{(2)}(T) \, \r] \, , & T_2 < T < T_1 ~, \\
 v_k^{L}(T) = \sqrt{T} \, \l[  \, C_1^L \, H_{\nu_3}^{(1)}(T)  \, +   \,  C_2^L \, H_{\nu_3}^{(2)}(T) \, \r] \, , & T < T_2 < T_1   ~,
 \end{cases} 
\label{eq:vk_jump_nu1_nu2_nu3}
\end{flalign}

\begin{small}
\begin{flalign}
 \f{{\rm d}v_k}{{\rm d}T}   \equiv   \begin{cases}  \f{1}{\sqrt{2k}}  \sqrt{\f{\pi}{2}}   e^{i \l(  \nu_1 + \f{1}{2}  \r) \f{\pi}{2}}   \f{1}{\sqrt{T}}  \l[  \l( \f{1}{2} - \nu_1 \r) \,  H_{\nu_1}^{(1)}(T) + T \,  H_{\nu_1-1}^{(1)}(T)  \r] \, , &  T_2 < T_1 < T ~, \\     \f{1}{\sqrt{T}}  \l[  C_1^{I}  \l(  \l( \f{1}{2} - \nu_2 \r)   H_{\nu_2}^{(1)}(T) + T \,  H_{\nu_2-1}^{(1)}(T)  \r)   +    C_2^{I} \l(  \l( \f{1}{2} - \nu_2 \r)   H_{\nu_2}^{(2)}(T) + T \,  H_{\nu_2-1}^{(2)}(T)  \r)  \r] \, , &   T_2 < T < T_1 ~, \\
  \f{1}{\sqrt{T}}  \l[  C_1^{L}  \l(  \l( \f{1}{2} - \nu_3 \r)   H_{\nu_3}^{(1)}(T) + T \,  H_{\nu_3-1}^{(1)}(T)  \r)   +    C_2^{L} \l(  \l( \f{1}{2} - \nu_3 \r)   H_{\nu_3}^{(2)}(T) + T \,  H_{\nu_3-1}^{(2)}(T)  \r)  \r] \, , & T < T_2 < T_1  \end{cases}  
  \label{eq:vkT_jump_nu1_nu2_nu3}
\end{flalign}
\end{small}
The superscripts `$E$', `$I$' and `$L$' again stand for early, intermediate and late respectively. After implementing the Israel junction matching conditions, we find the final expressions for the mode functions  
\begin{flalign}
   C_1^{L} = \f{ d_3 \, b_4 - d_4 \, b_3  }{ a_3 \, b_4 - a_4 \, b_3   } ~,~~~   C_2^{L} = \f{ d_4 \, a_3 - d_3 \, a_4  }{ a_3 \, b_4 - a_4 \, b_3   } ~, 
\end{flalign}
where 
\begin{align}
a_3 &=   H_{\nu_3}^{(1)}(T_2) \,,  &\\
b_3 &=   H_{\nu_3}^{(2)}(T_2) \,,  &\\
a_4 &=  \l( \, \f{1}{2} - \nu_3    - {\cal A}_2 \, \r)  \,  H_{\nu_3}^{(1)}(T_2) + T_2 \,  H_{\nu_3 - 1}^{(1)}(T_2) \,, & \\
b_4 &=  \l( \, \f{1}{2} - \nu_3    - {\cal A}_2 \, \r)  \,  H_{\nu_3}^{(2)}(T_2) + T_2 \,  H_{\nu_3 - 1}^{(2)}(T_2)  \,, & \\ 
d_3 &=      H_{\nu_2}^{(1)}(T_2)  \, C_1^{I} +    H_{\nu_2}^{(2)}(T_2) \, C_2^{I} \,, & \\  
d_4 &=    \l[ \l( \f{1}{2} - \nu_2 \r)  H_{\nu_2}^{(1)}(T_2) + T_2  \, H_{\nu_2 - 1}^{(1)}(T_2)   \r] \, C_1^{I} +  \l[ \l( \f{1}{2} - \nu_2 \r)   H_{\nu_2}^{(2)}(T_2) + T_2 \,  H_{\nu_2 - 1}^{(2)}(T_2)  \r]  C_2^{I} \,,
\end{align}
and the intermediate transition coefficients $C_1^I$ and  $C_2^I$ are given by

\begin{flalign}
  C_1^{I} = \f{ d_1 \, b_2 - d_2 \, b_1  }{ a_1 \, b_2 - a_2 \, b_1   } ~,~~~   C_2^{I} = \f{ d_2 \, a_1 - d_1 \, a_2  }{ a_1 \, b_2 - a_2 \, b_1   } ~, 
\end{flalign}
where 
\begin{align}
a_1 &=   H_{\nu_2}^{(1)}(T_1)  \,, &\\
b_1 &=   H_{\nu_2}^{(2)}(T_1) \,,  &\\
a_2 &=  \l( \, \f{1}{2} - \nu_2    - {\cal A}_1 \, \r)  \,  H_{\nu_2}^{(1)}(T_1) + T_1 \,  H_{\nu_2 - 1}^{(1)}(T_1) \,, & \\
b_2 &=  \l( \, \f{1}{2} - \nu_2    - {\cal A}_1 \, \r)  \,  H_{\nu_2}^{(2)}(T_1) + T_1 \,  H_{\nu_2 - 1}^{(2)}(T_1)  \,, & \\ 
d_1 &=  \f{1}{\sqrt{2k}} \, \sqrt{\f{\pi}{2}}  \,   e^{i \l(  \nu_1 + \f{1}{2}  \r) \f{\pi}{2}} \, H_{\nu_1}^{(1)}(T_1)  \,, & \\  
d_2 &=    \f{1}{\sqrt{2k}} \, \sqrt{\f{\pi}{2}}  \,   e^{i \l(  \nu_1 + \f{1}{2}  \r) \f{\pi}{2}} \, \l[ \, \l( \f{1}{2} - \nu_1 \r)  \,  H_{\nu_1}^{(1)}(T_1) + T_1 \,  H_{\nu_1 - 1}^{(1)}(T_1)    \r]  \,.
\end{align}


\begin{thebibliography}{69}




\bibitem{Bertone:2016nfn}
G.~Bertone and D.~Hooper,
``History of dark matter,''
Rev. Mod. Phys. \textbf{90} (2018) no.4, 045002
[arXiv:1605.04909 [astro-ph.CO]].


\bibitem{Peebles:2017bzw}
P.~J.~E.~Peebles,
``Growth of the nonbaryonic dark matter theory,''
Nature Astron. \textbf{1} (2017) no.3, 0057
[arXiv:1701.05837 [astro-ph.CO]].


\bibitem{Green:2021jrr}
A.~M.~Green,
``Dark matter in astrophysics/cosmology,''
SciPost Phys. Lect. Notes \textbf{37} (2022), 1
[arXiv:2109.05854 [hep-ph]].


\bibitem{Zeldovich:1967lct}
Y.~B.~Zel'dovich and I.~D.~Novikov,
``The Hypothesis of Cores Retarded during Expansion and the Hot Cosmological Model,''
Soviet Astron. AJ (Engl. Transl. ), \textbf{10} (1967), 602.

\bibitem{Hawking:1971ei}
S.~Hawking,
``Gravitationally collapsed objects of very low mass,''
Mon. Not. Roy. Astron. Soc. \textbf{152} (1971), 75.

\bibitem{Carr:1974nx}
B.~J.~Carr and S.~W.~Hawking,
``Black holes in the early Universe,''
Mon. Not. Roy. Astron. Soc. \textbf{168} (1974), 399-415


\bibitem{Carr:1975qj}
B.~J.~Carr,
``The Primordial black hole mass spectrum,''
Astrophys. J. \textbf{201} (1975), 1-19


\bibitem{Novikov:1979aa}
 I.~D.~Novikov, A.~G. Polnarev, A.~A. Starobinsky and Ya.~B. Zeldovich, ``Primordial Black Holes,'' Astron. Astroph. 80, (1979) 104-109.

\bibitem{Sasaki:2018dmp}
M.~Sasaki, T.~Suyama, T.~Tanaka and S.~Yokoyama,
``Primordial black holes\textemdash{}perspectives in gravitational wave astronomy,''
Class. Quant. Grav. \textbf{35} (2018) no.6, 063001
[arXiv:1801.05235 [astro-ph.CO]].



\bibitem{Chapline:1975ojl}
G.~F.~Chapline,
``Cosmological effects of primordial black holes,''
Nature \textbf{253} (1975) no.5489, 251-252

\bibitem{Meszaros:1975ef}
P.~Meszaros,
``Primeval black holes and galaxy formation,''
Astron. Astrophys. \textbf{38} (1975), 5-13


\bibitem{Ivanov:1994pa}
P.~Ivanov, P.~Naselsky and I.~Novikov,
``Inflation and primordial black holes as dark matter,''
Phys. Rev. D \textbf{50} (1994), 7173-7178



\bibitem{Carr:2016drx}
B.~Carr, F.~Kuhnel and M.~Sandstad,
``Primordial Black Holes as Dark Matter,''
Phys. Rev. D \textbf{94} (2016) no.8, 083504
[arXiv:1607.06077 [astro-ph.CO]].


\bibitem{Green:2020jor}
A.~M.~Green and B.~J.~Kavanagh,
``Primordial Black Holes as a dark matter candidate,''
J. Phys. G \textbf{48}, no.4, 043001 (2021)
[arXiv:2007.10722 [astro-ph.CO]].




\bibitem{Carr:2020xqk}
B.~Carr and F.~Kuhnel,
``Primordial Black Holes as Dark Matter: Recent Developments,''
Ann. Rev. Nucl. Part. Sci. \textbf{70} (2020), 355-394
[arXiv:2006.02838 [astro-ph.CO]].




\bibitem{LIGOScientific:2016aoc}
B.~P.~Abbott \textit{et al.} [LIGO Scientific and Virgo],
``Observation of Gravitational Waves from a Binary Black Hole Merger,''
Phys. Rev. Lett. \textbf{116} (2016) no.6, 061102
[arXiv:1602.03837 [gr-qc]].


\bibitem{Bird:2016dcv}
S.~Bird, I.~Cholis, J.~B.~Mu\~noz, Y.~Ali-Ha\"\i{}moud, M.~Kamionkowski, E.~D.~Kovetz, A.~Raccanelli and A.~G.~Riess,
``Did LIGO detect dark matter?,''
Phys. Rev. Lett. \textbf{116} (2016) no.20, 201301
[arXiv:1603.00464 [astro-ph.CO]].


\bibitem{Clesse:2016vqa}
S.~Clesse and J.~Garc\'\i{}a-Bellido,
``The clustering of massive Primordial Black Holes as Dark Matter: measuring their mass distribution with Advanced LIGO,''
Phys. Dark Univ. \textbf{15} (2017), 142-147
[arXiv:1603.05234 [astro-ph.CO]].


\bibitem{Sasaki:2016jop}
M.~Sasaki, T.~Suyama, T.~Tanaka and S.~Yokoyama,
``Primordial Black Hole Scenario for the Gravitational-Wave Event GW150914,''
Phys. Rev. Lett. \textbf{117} (2016) no.6, 061101
[erratum: Phys. Rev. Lett. \textbf{121} (2018) no.5, 059901]
[arXiv:1603.08338 [astro-ph.CO]].

\bibitem{Starobinsky:1980te}
A.~A.~Starobinsky,
``A New Type of Isotropic Cosmological Models Without Singularity,''
Phys. Lett. B \textbf{91} (1980), 99-102


\bibitem{Guth:1980zm}
A.~H.~Guth,
``The Inflationary Universe: A Possible Solution to the Horizon and Flatness Problems,''
Phys. Rev. D \textbf{23} (1981), 347-356


\bibitem{Linde:1981mu}
A.~D.~Linde,
``A New Inflationary Universe Scenario: A Possible Solution of the Horizon, Flatness, Homogeneity, Isotropy and Primordial Monopole Problems,''
Phys. Lett. B \textbf{108} (1982), 389-393

\bibitem{Albrecht:1982wi}
A.~Albrecht and P.~J.~Steinhardt,
``Cosmology for Grand Unified Theories with Radiatively Induced Symmetry Breaking,''
Phys. Rev. Lett. \textbf{48} (1982), 1220-1223


\bibitem{Linde:1983gd}
A.~D.~Linde,
``Chaotic Inflation,''
Phys. Lett. B \textbf{129} (1983), 177-181


\bibitem{Baumann:2009ds}
D.~Baumann,
``TASI Lectures on Inflation,''
[arXiv:0907.5424 [hep-th]].

\bibitem{Mukhanov:1981xt}
V.~F.~Mukhanov and G.~V.~Chibisov,
``Quantum Fluctuations and a Nonsingular Universe,''
JETP Lett. \textbf{33} (1981), 532-535

\bibitem{Guth:1982ec}
A.~H.~Guth and S.~Y.~Pi,
``Fluctuations in the New Inflationary Universe,''
Phys. Rev. Lett. \textbf{49} (1982), 1110-1113

\bibitem{Starobinsky:1982ee}
A.~A.~Starobinsky,
``Dynamics of Phase Transition in the New Inflationary Universe Scenario and Generation of Perturbations,''
Phys. Lett. B \textbf{117} (1982), 175-178

\bibitem{Hawking:1982cz}
S.~W.~Hawking,
``The Development of Irregularities in a Single Bubble Inflationary Universe,''
Phys. Lett. B \textbf{115} (1982), 295

\bibitem{Mukhanov:1990me}
V.~F.~Mukhanov, H.~A.~Feldman and R.~H.~Brandenberger,
``Theory of cosmological perturbations. Part 1. Classical perturbations. Part 2. Quantum theory of perturbations. Part 3. Extensions,''
Phys. Rept. \textbf{215} (1992), 203-333.

\bibitem{Baumann:2018muz}
D.~Baumann,
``Primordial Cosmology,''
PoS \textbf{TASI2017}, 009 (2018)
[arXiv:1807.03098 [hep-th]].


\bibitem{Planck:2018nkj}
N.~Aghanim \textit{et al.} [Planck],
``Planck 2018 results. I. Overview and the cosmological legacy of Planck,''
Astron. Astrophys. \textbf{641} (2020), A1
[arXiv:1807.06205 [astro-ph.CO]].

\bibitem{Planck:2018vyg}
N.~Aghanim \textit{et al.} [Planck],
``Planck 2018 results. VI. Cosmological parameters,''
Astron. Astrophys. \textbf{641} (2020), A6
[erratum: Astron. Astrophys. \textbf{652} (2021), C4]
[arXiv:1807.06209 [astro-ph.CO]].


\bibitem{Tegmark:2004qd}
M.~Tegmark,
``What does inflation really predict?,''
JCAP \textbf{04} (2005), 001
[arXiv:astro-ph/0410281 [astro-ph]].







\bibitem{Planck:2018jri}
Y.~Akrami \textit{et al.} [Planck],
``Planck 2018 results. X. Constraints on inflation,''
Astron. Astrophys. \textbf{641} (2020), A10
[arXiv:1807.06211 [astro-ph.CO]].

\bibitem{BICEP:2021xfz}
P.~A.~R.~Ade \textit{et al.} [BICEP and Keck],
``Improved Constraints on Primordial Gravitational Waves using Planck, WMAP, and BICEP/Keck Observations through the 2018 Observing Season,''
Phys. Rev. Lett. \textbf{127} (2021) no.15, 151301
[arXiv:2110.00483 [astro-ph.CO]].


\bibitem{Ferrante:2022mui}
G.~Ferrante, G.~Franciolini, A.~Iovino, Junior. and A.~Urbano,
``Primordial non-gaussianity up to all orders: theoretical aspects and implications for primordial black hole models,''
[arXiv:2211.01728 [astro-ph.CO]].

\bibitem{Gow:2022jfb}
A.~D.~Gow, H.~Assadullahi, J.~H.~P.~Jackson, K.~Koyama, V.~Vennin and D.~Wands,
``Non-perturbative non-Gaussianity and primordial black holes,''
[arXiv:2211.08348 [astro-ph.CO]].

\bibitem{Escriva:2022duf}
A.~Escriv\`a, F.~Kuhnel and Y.~Tada,
``Primordial Black Holes,''
[arXiv:2211.05767 [astro-ph.CO]].


\bibitem{DeLuca:2022rfz}
V.~De Luca and A.~Riotto,
``A note on the abundance of primordial black holes: Use and misuse of the metric curvature perturbation,''
Phys. Lett. B \textbf{828}, 137035 (2022)
[arXiv:2201.09008 [astro-ph.CO]].

\bibitem{Biagetti:2021eep}
M.~Biagetti, V.~De Luca, G.~Franciolini, A.~Kehagias and A.~Riotto,
``The formation probability of primordial black holes,''
Phys. Lett. B \textbf{820} (2021), 136602
[arXiv:2105.07810 [astro-ph.CO]].

\bibitem{Starobinsky:1986fx}
A.~A.~Starobinsky,
``STOCHASTIC DE SITTER (INFLATIONARY) STAGE IN THE EARLY UNIVERSE,''
Lect. Notes Phys. \textbf{246}, 107-126 (1986)
doi:10.1007/3-540-16452-9\_6


\bibitem{Salopek:1990jq}
D.~S.~Salopek and J.~R.~Bond,
``Nonlinear evolution of long wavelength metric fluctuations in inflationary models,''
Phys. Rev. D \textbf{42}, 3936-3962 (1990)
doi:10.1103/PhysRevD.42.3936



\bibitem{Salopek:1990re}
D.~S.~Salopek and J.~R.~Bond,
``Stochastic inflation and nonlinear gravity,''
Phys. Rev. D \textbf{43}, 1005-1031 (1991)
doi:10.1103/PhysRevD.43.1005





\bibitem{Starobinsky:1994bd}
A.~A.~Starobinsky and J.~Yokoyama,
``Equilibrium state of a selfinteracting scalar field in the De Sitter background,''
Phys. Rev. D \textbf{50}, 6357-6368 (1994)
[arXiv:astro-ph/9407016 [astro-ph]].

\bibitem{Fujita:2013cna}
T.~Fujita, M.~Kawasaki, Y.~Tada and T.~Takesako,
``A new algorithm for calculating the curvature perturbations in stochastic inflation,''
JCAP \textbf{12}, 036 (2013)
[arXiv:1308.4754 [astro-ph.CO]].


\bibitem{Fujita:2014tja}
T.~Fujita, M.~Kawasaki and Y.~Tada,
``Non-perturbative approach for curvature perturbations in stochastic $\delta N$ formalism,''
JCAP \textbf{10}, 030 (2014)
[arXiv:1405.2187 [astro-ph.CO]].



\bibitem{Vennin:2015hra}
V.~Vennin and A.~A.~Starobinsky,
Eur. Phys. J. C \textbf{75}, 413 (2015)
[arXiv:1506.04732].


\bibitem{Celoria:2021vjw}
M.~Celoria, P.~Creminelli, G.~Tambalo and V.~Yingcharoenrat,
``Beyond perturbation theory in inflation,''
JCAP \textbf{06}, 051 (2021)
doi:10.1088/1475-7516/2021/06/051
[arXiv:2103.09244 [hep-th]].

\bibitem{Cohen:2022clv}
T.~Cohen, D.~Green and A.~Premkumar,
``Large Deviations in the Early Universe,''
[arXiv:2212.02535 [hep-th]].

\bibitem{Hooshangi:2021ubn}
S.~Hooshangi, M.~H.~Namjoo and M.~Noorbala,
``Rare events are nonperturbative: Primordial black holes from heavy-tailed distributions,''
Phys. Lett. B \textbf{834} (2022), 137400
[arXiv:2112.04520 [astro-ph.CO]].

\bibitem{Achucarro:2021pdh}
A.~Achucarro, S.~Cespedes, A.~C.~Davis and G.~A.~Palma,
``The hand-made tail: non-perturbative tails from multifield inflation,''
JHEP \textbf{05} (2022), 052
[arXiv:2112.14712 [hep-th]].

\bibitem{Cai:2021zsp}
Y.~F.~Cai, X.~H.~Ma, M.~Sasaki, D.~G.~Wang and Z.~Zhou,
``One small step for an inflaton, one giant leap for inflation: A novel non-Gaussian tail and primordial black holes,''
Phys. Lett. B \textbf{834} (2022), 137461
[arXiv:2112.13836 [astro-ph.CO]].

\bibitem{Ezquiaga:2022qpw}
J.~M.~Ezquiaga, J.~Garc\'\i{}a-Bellido and V.~Vennin,
``Could ''El Gordo'' be hinting at primordial quantum diffusion?,''
[arXiv:2207.06317 [astro-ph.CO]].


\bibitem{Cai:2022erk}
Y.~F.~Cai, X.~H.~Ma, M.~Sasaki, D.~G.~Wang and Z.~Zhou,
``Highly non-Gaussian tails and primordial black holes from single-field inflation,''
[arXiv:2207.11910 [astro-ph.CO]].



\bibitem{Pattison:2017mbe}
C.~Pattison, V.~Vennin, H.~Assadullahi and D.~Wands, ``Quantum diffusion during inflation and primordial black holes,'' 
JCAP \textbf{10}, 046 (2017)
[arXiv:1707.00537].


\bibitem{Ezquiaga:2018gbw}
J.~M.~Ezquiaga and J.~Garc\'\i{}a-Bellido,
``Quantum diffusion beyond slow-roll: implications for primordial black-hole production,''
JCAP \textbf{08} (2018), 018
[arXiv:1805.06731 [astro-ph.CO]].

\bibitem{Biagetti:2018pjj}
M.~Biagetti, G.~Franciolini, A.~Kehagias and A.~Riotto,
``Primordial Black Holes from Inflation and Quantum Diffusion,''
JCAP \textbf{07} (2018), 032
[arXiv:1804.07124 [astro-ph.CO]].


\bibitem{Ezquiaga:2019ftu}
J.~M.~Ezquiaga, J.~Garc\'\i{}a-Bellido and V.~Vennin,
``The exponential tail of inflationary fluctuations: consequences for primordial black holes,''
JCAP \textbf{03} (2020), 029
[arXiv:1912.05399 [astro-ph.CO]].


\bibitem{Firouzjahi:2018vet}
H.~Firouzjahi, A.~Nassiri-Rad and M.~Noorbala,
``Stochastic Ultra Slow Roll Inflation,''
JCAP \textbf{01} (2019), 040
[arXiv:1811.02175 [hep-th]].

\bibitem{Pattison:2019hef}
C.~Pattison, V.~Vennin, H.~Assadullahi and D.~Wands,
``Stochastic inflation beyond slow roll,''
JCAP \textbf{07}, 031 (2019)
[arXiv:1905.06300 [astro-ph.CO]].







\bibitem{Ballesteros:2020sre}
G.~Ballesteros, J.~Rey, M.~Taoso and A.~Urbano,
``Stochastic inflationary dynamics beyond slow-roll and consequences for primordial black hole formation,''
JCAP \textbf{08} (2020), 043
[arXiv:2006.14597 [astro-ph.CO]].


\bibitem{Vennin:2020kng}
V.~Vennin,
``Stochastic inflation and primordial black holes,''
[arXiv:2009.08715 [astro-ph.CO]].


\bibitem{Ando:2020fjm}
K.~Ando and V.~Vennin,
``Power spectrum in stochastic inflation,''
JCAP \textbf{04}, 057 (2021)
[arXiv:2012.02031 [astro-ph.CO]].




\bibitem{De:2020hdo}
A.~De and R.~Mahbub,
``Numerically modeling stochastic inflation in slow-roll and beyond,''
Phys. Rev. D \textbf{102}, no.12, 123509 (2020)
[arXiv:2010.12685 [astro-ph.CO]].

\bibitem{Figueroa:2020jkf}
D.~G.~Figueroa, S.~Raatikainen, S.~Rasanen and E.~Tomberg,
``Non-Gaussian Tail of the Curvature Perturbation in Stochastic Ultra slow-Roll Inflation: Implications for Primordial Black Hole Production,''
Phys. Rev. Lett. \textbf{127}, no.10, 101302 (2021)
[arXiv:2012.06551 [astro-ph.CO]].

\bibitem{Cruces:2021iwq}
D.~Cruces and C.~Germani,
Phys. Rev. D \textbf{105}, no.2, 023533 (2022)
[arXiv:2107.12735].

\bibitem{Rigopoulos:2021nhv}
G.~Rigopoulos and A.~Wilkins,
JCAP \textbf{12}, no.12, 027 (2021)
[arXiv:2107.05317].


\bibitem{Pattison:2021oen}
C.~Pattison, V.~Vennin, D.~Wands and H.~Assadullahi,
``Ultra-slow-roll inflation with quantum diffusion,''
JCAP \textbf{04}, 080 (2021)
[arXiv:2101.05741 [astro-ph.CO]].





\bibitem{Tomberg:2021xxv}
E.~Tomberg,
``A numerical approach to stochastic inflation and primordial black holes,''
J. Phys. Conf. Ser. \textbf{2156}, no.1, 012010 (2021)
[arXiv:2110.10684 [astro-ph.CO]].


\bibitem{Figueroa:2021zah}
D.~G.~Figueroa, S.~Raatikainen, S.~Rasanen and E.~Tomberg,
``Implications of stochastic effects for primordial black hole production in ultra-slow-roll inflation,''
[arXiv:2111.07437 [astro-ph.CO]].




\bibitem{Tada:2021zzj}
Y.~Tada and V.~Vennin,
``Statistics of coarse-grained cosmological fields in stochastic inflation,''
JCAP \textbf{02}, no.02, 021 (2022)
[arXiv:2111.15280 [astro-ph.CO]].



\bibitem{Mahbub:2022osb}
R.~Mahbub and A.~De,
``Smooth coarse-graining and colored noise dynamics in stochastic inflation,''
[arXiv:2204.03859 [astro-ph.CO]].


\bibitem{Jackson:2022unc}
J.~H.~P.~Jackson, H.~Assadullahi, K.~Koyama, V.~Vennin and D.~Wands,
``Numerical simulations of stochastic inflation using importance sampling,''
JCAP \textbf{10} (2022), 067
[arXiv:2206.11234 [astro-ph.CO]].

\bibitem{Starobinsky:1985ibc}
A.~A.~Starobinsky,
``Multicomponent de Sitter (Inflationary) Stages and the Generation of Perturbations,''
JETP Lett. \textbf{42} (1985), 152-155

\bibitem{Sasaki:1995aw}
M.~Sasaki and E.~D.~Stewart,
``A General analytic formula for the spectral index of the density perturbations produced during inflation,''
Prog. Theor. Phys. \textbf{95} (1996), 71-78
[arXiv:astro-ph/9507001 [astro-ph]].

\bibitem{Lyth:2004gb}
D.~H.~Lyth, K.~A.~Malik and M.~Sasaki,
``A General proof of the conservation of the curvature perturbation,''
JCAP \textbf{05} (2005), 004
[arXiv:astro-ph/0411220 [astro-ph]].


\bibitem{Wands:2000dp}
D.~Wands, K.~A.~Malik, D.~H.~Lyth and A.~R.~Liddle,
``A New approach to the evolution of cosmological perturbations on large scales,''
Phys. Rev. D \textbf{62} (2000), 043527
[arXiv:astro-ph/0003278 [astro-ph]].

\bibitem{Lyth:2005fi}
D.~H.~Lyth and Y.~Rodriguez,
``The Inflationary prediction for primordial non-Gaussianity,''
Phys. Rev. Lett. \textbf{95} (2005), 121302
[arXiv:astro-ph/0504045 [astro-ph]].

\bibitem{Grain:2017dqa}
J.~Grain and V.~Vennin,
``Stochastic inflation in phase space: Is slow-roll a stochastic attractor?,''
JCAP \textbf{05}, 045 (2017)
doi:10.1088/1475-7516/2017/05/045
[arXiv:1703.00447 [gr-qc]].


\bibitem{MCG_ST_FPE_22}
S.~S.~Mishra, E. J. Copeland and A. M. Green,
``Primordial black holes and stochastic inflation beyond slow-roll: II - solutions of the adjoint Fokker-Planck equation,''
(in preparation).




\bibitem{Liddle:2003as}
A.~R.~Liddle and S.~M.~Leach,
``How long before the end of inflation were observable perturbations produced?,''
Phys. Rev. D \textbf{68} (2003), 103503
[arXiv:astro-ph/0305263 [astro-ph]].

\bibitem{Motohashi:2017kbs}
H.~Motohashi and W.~Hu,
``Primordial Black Holes and Slow-Roll Violation,''
Phys. Rev. D \textbf{96} (2017) no.6, 063503
doi:10.1103/PhysRevD.96.063503
[arXiv:1706.06784 [astro-ph.CO]].


\bibitem{Tsamis:2003px}
N.~C.~Tsamis and R.~P.~Woodard,
``Improved estimates of cosmological perturbations,''
Phys. Rev. D \textbf{69} (2004), 084005
[arXiv:astro-ph/0307463 [astro-ph]].

\bibitem{Kinney:2005vj}
W.~H.~Kinney,
``Horizon crossing and inflation with large eta,''
Phys. Rev. D \textbf{72} (2005), 023515
[arXiv:gr-qc/0503017 [gr-qc]].

\bibitem{Byrnes:2018txb}
C.~T.~Byrnes, P.~S.~Cole and S.~P.~Patil,
``Steepest growth of the power spectrum and primordial black holes,''
JCAP \textbf{06}, 028 (2019)
[arXiv:1811.11158 [astro-ph.CO]].

\bibitem{Karam:2022nym}
A.~Karam, N.~Koivunen, E.~Tomberg, V.~Vaskonen and H.~Veerm\"ae,
``Anatomy of single-field inflationary models for primordial black holes,''
[arXiv:2205.13540 [astro-ph.CO]].



\bibitem{Winitzki:1999ve}
S.~Winitzki and A.~Vilenkin,
``Effective noise in stochastic description of inflation,''
Phys. Rev. D \textbf{61} (2000), 084008
[arXiv:gr-qc/9911029 [gr-qc]].

\bibitem{Andersen:2021lii}
J.~O.~Andersen, M.~Eriksson and A.~Tranberg,
``Stochastic inflation from quantum field theory and the parametric dependence of the effective noise amplitude,''
JHEP \textbf{02}, 121 (2022)
[arXiv:2111.14503 [hep-ph]].

\bibitem{Polarski:1995jg}
D.~Polarski and A.~A.~Starobinsky,
``Semiclassicality and decoherence of cosmological perturbations,''
Class. Quant. Grav. \textbf{13}, 377-392 (1996)
doi:10.1088/0264-9381/13/3/006
[arXiv:gr-qc/9504030 [gr-qc]].


\bibitem{Kiefer:1998qe}
C.~Kiefer, D.~Polarski and A.~A.~Starobinsky,
``Quantum to classical transition for fluctuations in the early universe,''
Int. J. Mod. Phys. D \textbf{7}, 455-462 (1998)
doi:10.1142/S0218271898000292
[arXiv:gr-qc/9802003 [gr-qc]].

\bibitem{Kiefer:2008ku}
C.~Kiefer and D.~Polarski,
``Why do cosmological perturbations look classical to us?,''
Adv. Sci. Lett. \textbf{2}, 164-173 (2009)
doi:10.1166/asl.2009.1023
[arXiv:0810.0087 [astro-ph]].



\bibitem{SDE_Gardiner}
C.W. Gardiner, ``Handbook of Stochastic Methods for Physics, Chemistry and the Natural
Sciences'', Springer, Berlin, Germany (2004).

\bibitem{SDE_Evans}
L. Evans, 
``Introduction to Stochastic Differential Equations'', American Mathematical Society (2013).

\bibitem{Creminelli:2008es}
P.~Creminelli, S.~Dubovsky, A.~Nicolis, L.~Senatore and M.~Zaldarriaga,
``The Phase Transition to Slow-roll Eternal Inflation,''
JHEP \textbf{09}, 036 (2008)
[arXiv:0802.1067 [hep-th]].

\bibitem{Rudelius:2019cfh}
T.~Rudelius,
``Conditions for (No) Eternal Inflation,''
JCAP \textbf{08}, 009 (2019)
[arXiv:1905.05198 [hep-th]].

\bibitem{Ahmadi:2022lsm}
N.~Ahmadi, M.~Noorbala, N.~Feyzabadi, F.~Eghbalpoor and Z.~Ahmadi,
``Quantum Diffusion in Sharp Transition to Non-Slow-Roll Phase,''
JCAP \textbf{08} (2022), 078
[arXiv:2207.10578 [gr-qc]].






\bibitem{Bunch:1978yq} 
  T.~S.~Bunch and P.~C.~W.~Davies,
  ``Quantum Field Theory in de Sitter Space: Renormalization by Point Splitting,''
  Proc.\ Roy.\ Soc.\ Lond.\ A {\bf 360}, 117 (1978).

\bibitem{KKLT}
S.~Kachru, R.~Kallosh, A.~D.~Linde and S.~P.~Trivedi,
``De Sitter vacua in string theory,''
Phys. Rev. D \textbf{68}, 046005 (2003)
[arXiv:hep-th/0301240 [hep-th]].



\bibitem{KKLMMT} 
  S.~Kachru, R.~Kallosh, A.~D.~Linde, J.~M.~Maldacena, L.~P.~McAllister and S.~P.~Trivedi,
  ``Towards inflation in string theory,''
  JCAP {\bf 0310}, 013 (2003)
  [hep-th/0308055].

\bibitem{Kallosh_Linde_CMB_targets1}
R.~Kallosh and A.~Linde,
``CMB targets after the latest Planck data release,''
Phys. Rev. D \textbf{100}, no.12, 123523 (2019)
[arXiv:1909.04687 [hep-th]].

\bibitem{inf_encyclo}
J.~Martin, C.~Ringeval and V.~Vennin,
``Encyclop\ae{}dia Inflationaris,''
Phys. Dark Univ. \textbf{5-6} (2014), 75-235
[arXiv:1303.3787 [astro-ph.CO]].

\bibitem{Mishra:2022ijb}
S.~S.~Mishra and V.~Sahni,
``Canonical and Non-canonical Inflation in the light of the recent BICEP/Keck results,''
[arXiv:2202.03467 [astro-ph.CO]].

\bibitem{Bhatt:2022mmn}
S.~S.~Bhatt, S.~S.~Mishra, S.~Basak and S.~N.~Sahoo,
``Numerical simulations of inflationary dynamics: slow-roll and beyond,''
[arXiv:2212.00529 [gr-qc]].



\bibitem{Motohashi:2014ppa}
H.~Motohashi, A.~A.~Starobinsky and J.~Yokoyama,
``Inflation with a constant rate of roll,''
JCAP \textbf{09} (2015), 018
[arXiv:1411.5021 [astro-ph.CO]].

\bibitem{Mishra:2019pzq}
S.~S.~Mishra and V.~Sahni,
`Primordial Black Holes from a tiny bump/dip in the Inflaton potential,''
JCAP \textbf{04} (2020), 007
[arXiv:1911.00057 [gr-qc]].

\bibitem{Wands:1998yp}
D.~Wands,
``Duality invariance of cosmological perturbation spectra,''
Phys. Rev. D \textbf{60}, 023507 (1999)
[arXiv:gr-qc/9809062 [gr-qc]].

\bibitem{Deruelle:1995kd}
N.~Deruelle and V.~F.~Mukhanov,
``On matching conditions for cosmological perturbations,''
Phys. Rev. D \textbf{52}, 5549-5555 (1995)
[arXiv:gr-qc/9503050 [gr-qc]].





\bibitem{Starobinsky:1992ts}
A.~A.~Starobinsky,
``Spectrum of adiabatic perturbations in the universe when there are singularities in the inflation potential,''
JETP Lett. \textbf{55}, 489-494 (1992)



\bibitem{Joy:2007na}
M.~Joy, V.~Sahni and A.~A.~Starobinsky,
``A New Universal Local Feature in the Inflationary Perturbation Spectrum,''
Phys. Rev. D \textbf{77} (2008), 023514
[arXiv:0711.1585 [astro-ph]].



\bibitem{Hazra:2014goa}
D.~K.~Hazra, A.~Shafieloo, G.~F.~Smoot and A.~A.~Starobinsky,
``Wiggly Whipped Inflation,''
JCAP \textbf{08} (2014), 048
[arXiv:1405.2012 [astro-ph.CO]].


\bibitem{Hazra:2021eqk}
D.~K.~Hazra, D.~Paoletti, I.~Debono, A.~Shafieloo, G.~F.~Smoot and A.~A.~Starobinsky,
``Inflation story: slow-roll and beyond,''
JCAP \textbf{12} (2021) no.12, 038
[arXiv:2107.09460 [astro-ph.CO]].


\bibitem{Tomberg:2022mkt}
E.~Tomberg,
``Numerical stochastic inflation constrained by frozen noise,''
[arXiv:2210.17441 [astro-ph.CO]].

\bibitem{Tomberg:2023kli}
E.~Tomberg,
``Stochastic constant-roll inflation and primordial black holes,''
[arXiv:2304.10903 [astro-ph.CO]].

\bibitem{Press:1973iz}
W.~H.~Press and P.~Schechter,
``Formation of galaxies and clusters of galaxies by selfsimilar gravitational condensation,''
Astrophys. J. \textbf{187} (1974), 425-438


\bibitem{Germani:2018jgr}
C.~Germani and I.~Musco,
``Abundance of Primordial Black Holes Depends on the Shape of the Inflationary Power Spectrum,''
Phys. Rev. Lett. \textbf{122} (2019) no.14, 141302
[arXiv:1805.04087 [astro-ph.CO]].



\bibitem{Kristiano:2022maq}
J.~Kristiano and J.~Yokoyama,
``Ruling Out Primordial Black Hole Formation From Single-Field Inflation,''
[arXiv:2211.03395 [hep-th]].


\bibitem{Inomata:2022yte}
K.~Inomata, M.~Braglia and X.~Chen,
``Questions on calculation of primordial power spectrum with large spikes: the resonance model case,''
[arXiv:2211.02586 [astro-ph.CO]].


\bibitem{Choudhury:2023vuj}
S.~Choudhury, M.~R.~Gangopadhyay and M.~Sami,
``No-go for the formation of heavy mass Primordial Black Holes in Single Field Inflation,''
[arXiv:2301.10000 [astro-ph.CO]].

\bibitem{Choudhury:2023jlt}
S.~Choudhury, S.~Panda and M.~Sami,
``No-go for PBH formation in EFT of single field inflation,''
[arXiv:2302.05655 [astro-ph.CO]].

\bibitem{Kristiano:2023scm}
J.~Kristiano and J.~Yokoyama,
``Response to criticism on ''Ruling Out Primordial Black Hole Formation From Single-Field Inflation'': A note on bispectrum and one-loop correction in single-field inflation with primordial black hole formation,''
[arXiv:2303.00341 [hep-th]].

\bibitem{Riotto:2023gpm}
A.~Riotto,
``The Primordial Black Hole Formation from Single-Field Inflation is Still Not Ruled Out,''
[arXiv:2303.01727 [astro-ph.CO]].

\bibitem{Riotto:2023hoz}
A.~Riotto,
``The Primordial Black Hole Formation from Single-Field Inflation is Not Ruled Out,''
[arXiv:2301.00599 [astro-ph.CO]].

\bibitem{Firouzjahi:2023aum}
H.~Firouzjahi,
``One-loop Corrections in Power Spectrum in Single Field Inflation,''
[arXiv:2303.12025 [astro-ph.CO]].

\bibitem{Maldacena:2002vr}
J.~M.~Maldacena,
``Non-Gaussian features of primordial fluctuations in single field inflationary models,''
JHEP \textbf{05}, 013 (2003)
[arXiv:astro-ph/0210603 [astro-ph]].

\bibitem{book_nist_gov}
 B.~Schneider, B.~Miller, B.~aunders, ``The NIST Digital Library of Mathematical Functions: A 21st Century Source of Information on the Special Functions of Mathematical Physics'', Physics Today (2018), [online], https://doi.org/10.1063/PT.3.3846.
 
 
 
\bibitem{Birrell:1982ix}
N.~D.~Birrell and P.~C.~W.~Davies,
``Quantum Fields in Curved Space,''
Cambridge Univ. Press, 1984,
ISBN 978-0-521-27858-4, 978-0-521-27858-4
doi:10.1017/CBO9780511622632.




\end{thebibliography}
\end{document}